\documentclass[10pt,letterpaper,times]{article}
\usepackage[fleqn]{amsmath}
\usepackage{amssymb,fancyhdr,graphicx}
\usepackage{setspace}
\usepackage{breqn}
\usepackage{IEEEtrantools}
\DeclareGraphicsExtensions{.pdf,.png,.jpg}

\usepackage[top=3cm, bottom=3cm, left=2cm, right=3cm]{geometry} 
\title{An Adaptive Statistical Non-uniform Quantizer for Detail Wavelet Components in Lossy JPEG2000 Image Compression}
\author{Madhur Srivastava \thanks{M. Srivastava is with the Department of Biomedical Engineering, Cornell University, Ithaca, NY, 14850 USA e-mail: ms2736@cornell.edu}\and Satish K. Singh \thanks{S.K. Singh is with the Indian Institute of Information Technology, Allahabad,UP, 211012 India e-mail: sk.singh@iiita.ac.in} \and Prasanta K. Panigrahi \thanks{P.K. Panigrahi is with the Indian Institute of Science Education and Research - Kolkata, Nadia, WB, 741252 India e-mail: pprasanta@iiserkol.ac.in}}
\date{}

\begin{document}
\maketitle

\doublespacing
\section*{Abstract}
The paper presents a non-uniform quantization method for the Detail components in the JPEG2000 standard. Incorporating the fact that the coefficients lying towards the ends of the histogram plot of each Detail component represent the structural information of an image, the quantization step sizes become smaller at they approach the ends of the histogram plot. The variable quantization step sizes are determined by the actual statistics of the wavelet coefficients. Mean and standard deviation are the two statistical parameters used iteratively to obtain the variable step sizes. Moreover, the mean of the coefficients lying within the step size is chosen as the quantized value, contrary to the deadzone uniform quantizer which selects the midpoint of the quantization step size as the quantized value. The experimental results of the deadzone uniform quantizer and the proposed non-uniform quantizer are objectively compared by using Mean-Squared Error (MSE) and Mean Structural Similarity Index Measure (MSSIM), to evaluate the quantization error and reconstructed image quality, respectively. Subjective analysis of the reconstructed images is also carried out. Through the objective and subjective assessments, it is shown that the non-uniform quantizer performs better than the deadzone uniform quantizer in the perceptual quality of the reconstructed image, especially at low bitrates. More importantly, unlike the deadzone uniform quantizer, the non-uniform quantizer accomplishes better visual quality with a few quantized values.

\vspace{2cm}
\textbf{Keywords:} Mean and standard deviation, quantization, discrete wavelet transform, image compression, image processing, human visual system, JPEG2000 standard.
\section{Introduction}

The JPEG2000 standard (hereinafter: standard) is a state-of-art compression for diverse applications and platforms, embedded into a single system and a single compressed bit stream \cite{Acharya_2005,Skodras_2001,Taubman_2002a}. Apart from providing higher compression efficiency and improved quality of an image compared to baseline JPEG, additional features like the multiresolution representation, embedded coding, signal to noise ratio scalability, region of interest coding and error robustness are included in the standard \cite{Skodras_2001,Taubman_2002a,Rabbani_2002}.

One major reason for better performance of the standard over baseline JPEG and other image compression methods is the introduction of the discrete wavelet transform (DWT) into the standard. The perfect reconstruction of DWT has enabled the lossy and lossless compression into a unified system. Moreover, DWT possesses multiresolution capability which naturally allows this property in the standard. Further, DWT provides high energy compaction resulting in better compression ratio. Also, DWT removes blocking affects due to higher decorrelation of the image because it is applied to a complete image \cite{Rabbani_2002}.

Another important factor that results in the improved performance is the embedded coding of DWT coefficients, which is accomplished at present by using a uniform quantizer \cite{Rabbani_2002, Marcellin_2002}. Fig. 1 displays the fundamental block diagram of the standard. The quantization process contributes most to the lossy compression because the quantization step size controls the compression ratio and bitrate. For example, in any uniform quantizer, large step sizes will result in higher compression compared to small step sizes. Other encoding steps like DWT and entropy coding as well as choice of color space also contribute to the higher compression, but unlike quantization, there is no direct objective correlation between the variation in these encoding steps and the amount of compression that can be achieved. Their influence and control are limited. Therefore, the right choice of quanizer and quantization steps size is important to obtain maximum compression for the perceived quality of an image.

The Part I-II of the standard employs a deadzone uniform quantizer for compression, which incorporates Shannon's rate-distortion (R-D) theory \cite{Shannon_1959,Berger_1998} to select quantization step sizes for each subband. Mean-Squared Error (MSE) is the most common measure used to optimize R-D for the given data. However, in the case of images (and also videos), MSE may not relate to the perceived quality of image \cite{Watson_1993,Wu_2005} because it does not incorporate Human Visual System (HVS) into its calculations. HVS is the factor that determines the visual quality of image. To incorporate HVS, the standard allows the assignment of a weighting factor to each subband depending upon its visual significance in the image. This technique is called the visual frequency weighting \cite{JPEG2000_TechRep}. Further, the visual frequency weighting is categorized into the fixed visual weighting \cite{JPEG2000_TechRep} and visual progressive weighting \cite{Li_1999}. In the fixed visual weighting, the weights are computed based on the viewing distance of the reconstructed image. Whereas, for the visual progressive weighting, the weights are selected from a table of weights which corresponds to a specific bitrate. It enables the inclusion of common visual factor in both the quantization as well as embedding process for effective compression. Moreover, the distortion-adaptive visual progessive weighting is also present in the standard \cite{Zeng_2002}, which considers visual distortions due to suprathresholds before choosing an appropriate weight for a subband.

\graphicspath{{UntitledFolder/}}
\begin{figure}[t]
\vspace{-11mm}
  \centering
  \includegraphics[width=8.5cm]{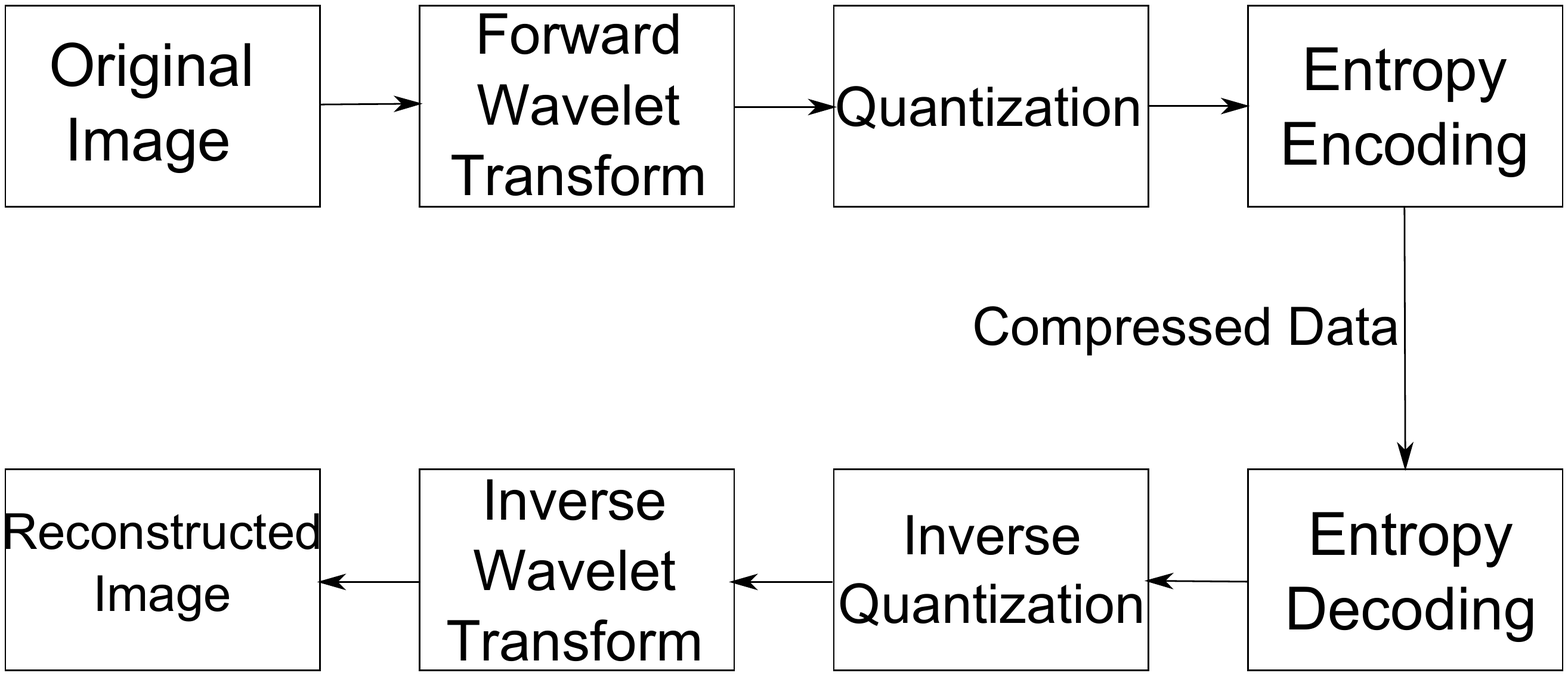}
\vspace{-14mm}
  \caption{Block Diagram of the JPEG2000 Standard.}

\end{figure}

The standard allows the use and implementation of HVS algorithms to produce perceptually superior lossy images. Many perceptual based quantization schemes have been proposed that incorporated HVS in the lossy image compression. Nadenau and Reichel \cite{Nadenau_1999} analyzed the contrast sensitivity curves for different color spaces, and accordingly, calculated the visual weighting factors for each subband. In their book \cite{Taubman_2002b}, Taubman and Marcellin suggested a visual masking technique for the standard. Zeng et al. \cite{Zeng_2000} applied a masking function to retain the significance of the image edges. Watson et al. \cite{Watson_1997} characterized the visibility thresholds by modelling uniform noise using the subband level, orientation, and display resolution. This mathematical model tried to precapture the quantization distortion. Liu et al. \cite{Liu_2006} incorporated the visibility thresholds, measured from the uniform noise modelling, into the standard. Using phychophysical experiments, Larabi et al. \cite{Larabi_2009} applied visual thresholds on each subband to obtain visually lossless compressed images for the digital cinema. It was shown that there is necessity to increase the maximum bitrate i.e., 250 Mbps, to achieve it. Similarly, Ramos and Hemami \cite{Ramos_2000,Ramos_2001} proposed a method to select the quantization step sizes using the visibility thresholds through phychophysical experiments for the low bitrate wavelet based image compression. Subsequently, incorporating the previous results \cite{Ramos_2000,Ramos_2001}, Chandler and Hemami \cite{Chandler_2005} proposed a unified contrast-based quantization scheme which provided competitive visual quality at high bitrates and improved visual quality at low bitrates. Gaubatz et al. \cite{Gaubatz_2005a,Gaubatz_2005b,Gaubatz_2006} proposed a spatially selective quantization scheme for the wavelet-based image compression that performed similar to the standard in visual quality. Introducing adaptive quantization step sizes for a subband, Albanesi and Guerrini \cite{Albanesi_2003} utilized the contrast sensivity and luminance masking to calculate the visibility thresholds independently for each component i.e., Approximation, Horizontal, Vertical, and Diagonal. In a similar work, Liu and Chou \cite{Liu_2008} captured perceptual redundancy as the noise detection thresholds for each component, and accordingly, adjusted the quantization step sizes for each component. In their two papers \cite{Sreelekha_2009, Sreelekha_2010}, Sreelekha and Sathidevi presented an adaptive quantization scheme which extended the luminance component visual modelling to the chrominance component. They used the contrast thresholds to initially eliminate insignificant components followed by a k-means clustering approach to quantize remaining coefficients. Wu et al. \cite{Wu_2006,Wu_2010} proposed a mechanism that used the visual pruning and human vision modelling to remove visually insignificant coefficients in the JPEG2000 compressed image. More recently, Oh et al. \cite{Oh_2013} calculated the visibility thresholds using a distortion model that considers the deadzone uniform quantizer properties and the statistical features of wavelets coefficients to obtain visually lossless image compression in the standard. 

Alternatively, Reichel et al. \cite{Reichel_2001} further investigated the reversible Integer-to-Integer Wavelet Transform (IWT) by testing various uniform quantization schemes. They concluded that the application of DWT for the lossy compression in the standard results in far superior image quality for the small number of quantization step sizes, compared to reversible IWT. However, in the case of high compression ratio, both the transforms performed similarly in terms of MSE and visual quality. Long et al. \cite{Long_2002} overcame the ineffective performance of reversible IWT and carried out the lossy image compression using 5/3 IWT and a uniform quantizer. The paper proposed two different selection of quantization step sizes for the uniform quantizer, both of them showing better performance in terms of PSNR. Reichel et al. independently proposed one of the quantization step sizes scheme presented in Long et al. paper. The two uniform quantizers can also be implemented in the standard with CDF 9/7 wavelet transformed image; however, their effects are unknown.

The current literature on visually lossy or lossless compression in the standard only produce better visual quality compared to the non-perceptual quantizers for a given bit rate. They lack the capability to directly correlate the quantization of wavelet coefficients with visual quality. Most importantly, in all the cases, the uniform quantization mechanism is used i.e., the uniform quantizer, deadzone uniform quantizer, and k-means clustering, to name a few. The perceptual quantization algorithms either re-adjusts the quantization step sizes and thresholds in the uniform quantizer, or reassigns the coefficient values through weights based on their visual significance.
   
Even though the perceptual thresholding using the deadzone uniform quantizer in the standard produces effective, efficient and better results, there are some major drawbacks when it is used in the \emph{Detail components} of the wavelet transformed image:

\begin{enumerate}
\item The uniform quantizer itself does not incorporate HVS. It relies on HVS algorithms for visually lossless and lossy compression.
\item Except for the case where coefficient values lie nearby zero, the deadzone uniform quantizer treats the information provided by each fixed step size in the image as equally important. This assumption is flawed because the Detail components represent the high frequency coefficients in the horizontal, vertical, and diagonal directions of the image, implying that the coefficients lying farther from the origin are more important. These high frequency coefficients capture the edge information in the images, and hence, depict the overall structure of the image. Quantization with the large step sizes using the deadzone uniform quantizer poses the problem of enhanced structural deformation. On the other hand, if smaller step sizes are used, then the overall bitrate is increased. 
\item The deadzone uniform quantizer has a deadzone region where all the coefficient values near zero, are assigned the quantized value zero. Although these coefficients are least important carrying minimal information, they do provide some information about the majority of coefficients that require necessary quantized value/s to represent them.
\item The number of the quantization step sizes required to maintain a certain quality of an image is not optimal because the same quality of image can be represented by a fewer number of step sizes.
\end{enumerate}

A non-uniform quantizer can potentially overcome the above disadvantages of the deadzone uniform quantizer. A non-uniform quantizer in conformity with HVS and the information represented by the Detail components can be natural match for HVS algorithms, especially in selecting thresholds for each step size. Moreover, the combination of the non-uniform quantizer and HVS algorithms are most likely to perform better than the current approach. To effectively replace the deadzone uniform quantizer, the non-uniform quantizer is required to have following qualities:
\begin{enumerate}
\item The quantized value within a quantization step size should have overall minimum error with the original coefficients.
\item The number of the quantization step sizes should be minimum for a given acceptable error. This would allow higher compression for a given visual quality of the compressed image.
\item The quantizer should be able to distinguish between the essential and non-essential coefficients, and accordingly, choose the appropriate step sizes for the coefficients.
\item The quantizer should be adaptive so that the step sizes can be instantaneously decided based on the actual wavelet coefficients.
\end{enumerate}

\begin{figure}[t]
\vspace{-11mm}
  \centering
  \includegraphics[width=10cm]{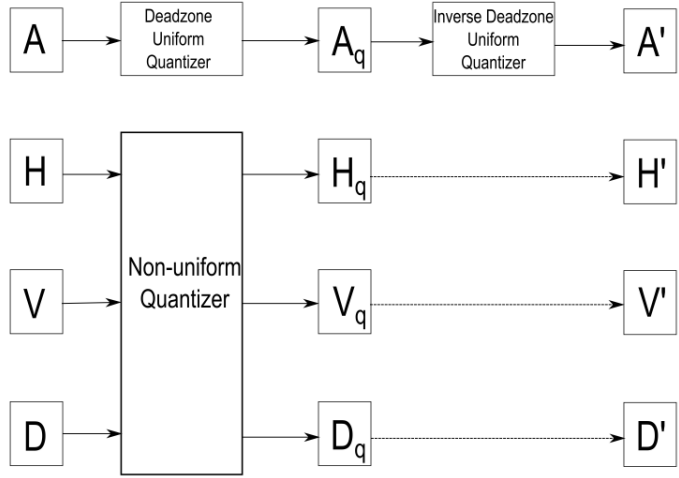}
\vspace{-2mm}
  \caption{Block Diagram of the Proposed Quantization Scheme.}

\end{figure}

This paper presents a non-uniform quantizer for the Detail components of a wavelet transformed image. It uses variable step sizes with the range of the step sizes reducing as they approach the end of the histogram plot of each Detail component. The non-linear quantizer was first used in \cite{Srivastava_2013b} for the image segmentation and object separation. Keeping the Approximation component unchanged, Srivastava and Panigrahi \cite{Srivastava_2013a} recently applied the presented non-uniform quantizer on the Detail components of the wavelet transformed images obtained from DB series wavelets. The results showed that the reconstructed images have the high Peak Signal to Noise Ratio (PSNR) and Mean Structural Similarity Index Measure (MSSIM) \cite{Wang_2004}, suggesting potential use in the standard. Building on the previous work \cite{Srivastava_2013a}, this paper applies the algorithm on the Detail components in the standard and compares it with the deadzone uniform quantizer results. The Approximation component is quantized with the deadzone uniform quantizer. The block diagram of the proposed method for the lossy image compression is shown in Fig. 2. 

The paper is organized as follows. Section 2 describes the deadzone uniform quantizer currently used in the standard, followed by section 3 which illustrates the non-uniform quantizer used in this paper. In section 4, the experimental results of the deadzone uniform and non-linear quantizer are examined and compared. Consequently, the discussion on the experimental results and the non-uniform quantizer's application in the standard is carried out in section 5. Lastly, section 6 concludes the paper with the future work. 

\section{Deadzone Uniform Quantizer}

The part I of the standard uses the uniform scalar quantizer having a deadzone around the origin, or the deadzone uniform quantizer. The quantization step size is same throughtout a subband, but step sizes vary among the subbands. The step size reduces towards the subbands that are representing higher decomposition levels. The quantized values for the wavelet coefficients are calculated as,

\begin{equation}
q_b(i,j)=sign(y_b(i,j))\left\lfloor\frac{\lvert y_b(i,j) \rvert}{\Delta_b}\right\rfloor
\end{equation}

\noindent where $q_b(i,j)$ is the quantized value, $y_b(i,j)$ is the input DWT coefficient, and $\Delta_b$ is the quantization step size. The subscript $b$ represents the subband. For deadzone region, the quantization size is $2\Delta_b$.

In the Part II of the standard, the deadzone region is devised to be flexible i.e., the deadzone region can be of variable length. The rest of the interval have fixed width $\Delta_b$. Similar to the Part I, the formula for calculating the quantized values is given by,

\begin{equation}
q_b(i,j)=sign(y_b(i,j))\left\lfloor\frac{\lvert y_b(i,j) + k\Delta_b \rvert}{\Delta_b}\right\rfloor
\end{equation}

\noindent where $k$ is the parameter which varies the deadzone step size. The quantization step size of the deadzone region is $2(1-k)\Delta_b$.

The quantization step size ($\Delta_b$) can be obtained from the following equation,

\begin{equation}
\Delta_b = 2^{R_b - \epsilon_b} \left(1 + \frac{\mu_b}{2^{11}} \right)
\end{equation}

\noindent where $R_b$ is the predicted bitdepth of wavelet coefficient at subband $b$, $\epsilon_b$ is the number of bits allocated to the exponent, and $\mu_b$ is the number of bits allocated to the mantessa. For irreversible wavelets (i.e., lossy compression), $\epsilon_b$ and $\mu_b$ are 5-bit and 11-bit integer, respectively.

In the deadzone uniform quantizer with fixed as well as variable deadzone width, the number of bits required to represent all the quantized step sizes is $max\, (\lceil \log_2 \lvert q(i,j)\rvert \rceil)$. For detailed information about the quantization in the standard, refer \cite{Marcellin_2002,Gonzalez_2003}.

\section{Non-uniform Quantizer}

The Detail components in a wavelet transformed image are the high frequency components of the image in the horizontal, vertical, and diagonal directions acquired from the low pass-high pass (LH), high pass-low pass (HL), and high pass-high pass (HH) filters, respectively. From human visual perspective, the high frequency components represent the structure of an image, mostly comprising visible edges in the image. The information about the structure and edges of the image is captured by the high coefficient values of the high frequency components. In other words, higher the value of the high frequency component, the higher information it contains about the structure and edges. Therefore, the high value coefficients need to be maximally preserved and should have minimum error during the quantization process. The coefficient values close to zero or within certain range around zero are least significant or perhaps insignificant in the context of human visual sensitivity and selectivity. Considering this fact, the deadzone uniform quantizer has a step size at and around zero two times (deadzone region) as large as the steps sizes for the rest of the high frequency coefficients. In lossy compression, the distortions that humans are unable to distinguish should be maximally exploited to increase image compression. Incorporating this factor, the presented non-uniform quantization algorithm starts with the large quantization step sizes close to zero, and then reduces the step sizes for every next set of coefficients to be quantized i.e., $\Delta_{b_1} > \Delta_{b_2} > .... > \Delta_{b_n}$, where $\Delta_{b_1}$ and $\Delta_{b_n}$ are the quantization step sizes for the coefficients with the lowest and highest magnitude, respectively.

The non-uniform quantizer calculates the varying step sizes with the help of mean and standard deviation ($std$). The formulae for obtaining boundaries of step sizes are given as,

\begin{equation}
B_L=\mu_L-\kappa_L\sigma_L
\end{equation}
\begin{equation}
B_R=\mu_R+\kappa_R\sigma_R
\end{equation}

\noindent where $\mu$ is the mean, $\sigma$ is the $std$, $\kappa$ is the skewness parameter, $B$ is the boundary of the quantization step size, $L$ and $R$ are the subscripts referring to the left and right part of the histogram plot from the mean of the coefficients, respectively. For example, $\mu_L$ and $\mu_R$ are the mean of coefficients lying the left and right side of the overall mean in the histogram plot.

As can be seen, the equations 4 and 5 select boundaries as the variation of coefficients from the mean in the left and right part of the histogram plot with the help of the standard deviation. This process naturally reduces the standard deviation values as we move towards the end of the histogram plot because there is a decrease in the number of coefficients with high magnitudes. The reduction in standard deviation would reduce the variation from the previous boundary point, resulting in smaller step size compared to the previous boundaries. Additionally, the real statistics of the coefficients would allow an adaptive step size based on the mean and standard deviation of the coefficients considered for the boundary determination. For instance, the step sizes would vary for different images with same resolution. Moreover, even for the same image, the step sizes belonging to the left and right part would vary unless the distribution of the coefficients is symmetric.

Here, $\kappa$ is similar to the $k$ of the deadzone uniform quantizer with the variable deadzone. It allows the non-uniform quantizer to further vary the step size. In general, $\kappa_L$ and $\kappa_R$ can have different values, but it is suggested that $\kappa_L=\kappa_R=1$ as it serves the purpose in the most cases. 

For the process of boundary selection, let $C$ be the coefficients of each Detail component and $n$ is the even number of quantization step sizes, in addition to the parameters mentioned above. The algorithm to obtain boundaries is given in the following steps:

\begin{enumerate}
\item Input $n$ and $C$
\item $\mu=mean(\forall C)$ and $\sigma=std(\forall C)$ 
\item $B_{L_1}=\mu-\kappa_L\sigma$ and $B_{R_1}=\mu+\kappa_R\sigma$
\item \textbf{Loop} $i=2$ to $\frac{n}{2}$ with unit increment.
\item $\mu_L=mean(\forall C \in [min(C),B_{L_{i-1}}))$ and $\mu_R=mean(\forall C \in [max(C),B_{R_{i-1}}))$
\item $\sigma_L=std(\forall C \in [min(C),B_{L_{i-1}}))$ and $\sigma_R=std(\forall C \in [max(C),B_{R_{i-1}}))$ 
\item $B_{L_i}=\mu_L-\kappa_1\sigma_L$ and $B_{R_i}=\mu_R+\kappa_2\sigma_R$
\item \textbf{end loop}
\item $B_{L_i}=min(C)$ and $B_{R_i}=max(C)$
\end{enumerate}

In the most cases, the number of quantization step sizes are even, so $n$ is even in the above boundary selection algorithm. However, the above algorithm can be modified to have odd number of quantization step sizes. For example, either the left or right part of the histogram plot can be neglected at any iteration of loop, merging that part with the coefficients to be considered in the next iteration.

The following two features of the above algorithm need to be noted:

\begin{enumerate}
\item The boundaries are selected based on the mean and standard deviation of the considered part of the Detail component histogram plot, making it adaptive.
\item For the higher number of quantization step sizes, only the leftmost and rightmost boundaries (excluding $min(C)$ and $max(C)$) are needed to find the next boundaries. Strictly speaking, none of the existing boundaries change with the increased number of step sizes; only new boundaries are added towards the ends of the histogram plot.
\end{enumerate}

After implementing the above algorithm, the boundaries of the non-uniform quantizer can be defined as,

\begin{equation}
B = \{B_{L_{\frac{n}{2}}}, ..., B_{L_1}, \mu, B_{R_1}, ..., B_{R_{\frac{n}{2}}}\}
\end{equation}

\noindent where, $B_{L_{\frac{n}{2}}} < ... < B_{L_1} < \mu < B_{R_1} < ... < B_{R_{\frac{n}{2}}}$

All the values lying within their respective boundaries are quantized to their mean. Mean is the centroid or the center of mass of the data, and hence, it minimizes the total error within a single quantization step. The same approach is applied in the deadzone uniform quantizer by considering the midpoint of the two boundaries as the quantized value, but the minimum error can only be obtained if the coefficients are uniformly distributed which is unlikely in most of the cases. The quantized values from the step size boundaries are calculated as, 

\begin{equation}
q_p= mean(\forall C \in [B_p, B_{p+1}))
\end{equation}

\noindent where $q_p$ is the quantized value, $p=\{[1,N),\, p\in Z\}$, and $N=\lvert B \rvert$.

\section{Experimental Results}

Experiments were conducted on various standard test images obtained from the USC-SIPI Image Database (http://sipi.usc.edu/database/). However, the results are only shown for \emph{Lenna}, \emph{Pepper}, and \emph{Baboon} images to include as much variety as possible in the limited space. As all the three test images are distinct, the effectiveness and the degree of effectiveness of the non-uniform quantizer over the deadzone uniform quantizer can be observed, tested, and evaluated. All the images were in gray scale having $512 \times 512$ dimensions. The experiment can easily be extended to the color images by applying the non-uniform quantization for each luminance-chrominance color channel. The goal here is to compare the two quantizers. The results of one channel will also be replicated by the other color channels.

The results are divided into three sections. In Section 4.1, the quantized values of the Approximation and Detail components obtained from the deadzone uniform quantizer and non-uniform quantizer for all the three test images are shown and discussed. The results of the non-uniform quantizer with different step sizes are also examined in this section. Section 4.2 carries out the objective evaluation of the non-uniform quantizer at various quantization step sizes using MSE and MSSIM, and simultaneously compare them with the deadzone uniform quantizer. Lastly, the reconstructed lossy images from the deadzone uniform quantizer and non-uniform quantizer are displayed and perceptually examined in section 4.3 for subjective evaluation.

\subsection{Assessment and Comparison of Quantized Values}

As shown in Fig. 2, in the proposed methodology, the deadzone uniform quantizer is applied on the Approximation component and the non-uniform quantizer is applied on each Detail component. Figs. 3-5 show the histogram plot of the original coefficients values of each component, followed by the histogram plot of their quantized values from both the quantizers at decomposition level 1, for the test images $Lenna$, $Pepper$, and $Baboon$, respectively. For comparison, the deadzone uniform quantization step size $\Delta_1$ is calculated using equation 3 with $R_1\, =\, 8$, $\mu_1\, = \, 8$, and $\epsilon_1\, = \, 7$. This results in the same quantized Approximation component for both the quantization schemes (see subfigs. b-d for each Fig. 3-5), allowing the comparison between the quantized values from the deadzone uniform quantizer and non-uniform quantizer for each Detail component. In other words, the comparison between the uniformly and non-uniformly quantized Detail component can only be conducted when the Approximation component is same for both of them. The number of quantization step sizes for the non-uniform quantizer is provided manually.

Additionally, the differences in the histogram plot of the quantized Detail component by the non-uniform quantizer at the quantization levels ($N$) 4 and 8 can be observed in Figs. 3-5. As mentioned in section 3, the increase of variable step sizes from 4 to 8 does not change the statistics of the quantized values which lie inside the leftmost and rightmost quantized value in the histogram plot. It is the leftmost and rightmost quantized values at $N=4$ that have been further divided into 6 quantized values at $N=8$. 

\begin{figure}[ht!]
\begin{minipage}[b]{0.24\linewidth}
  \centering
  \centerline{\includegraphics[width=\linewidth]{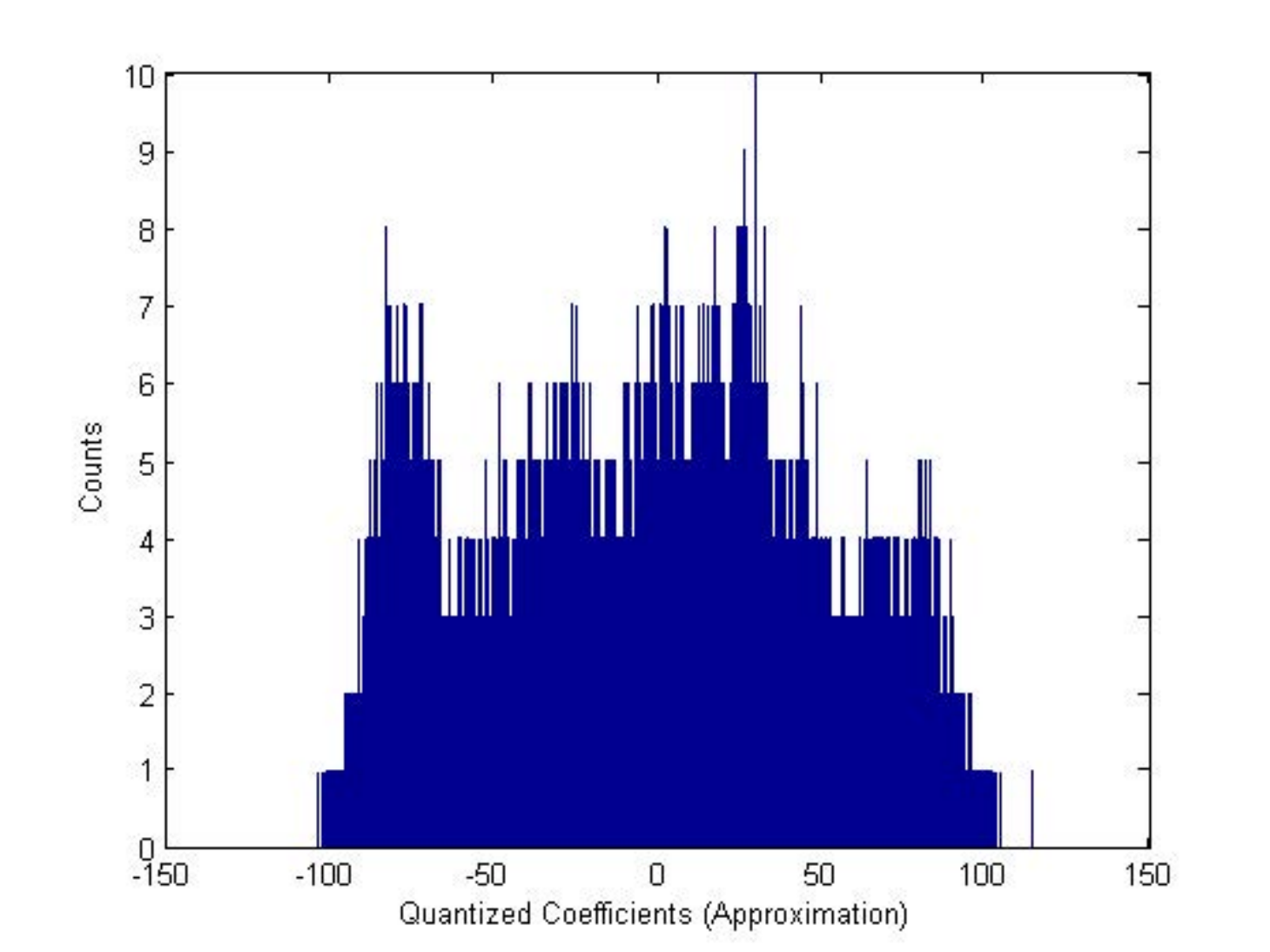}}
  \centerline{(a) Approximation}\medskip
\end{minipage}
\begin{minipage}[b]{0.24\linewidth}
  \centering
  \centerline{\includegraphics[width=\linewidth]{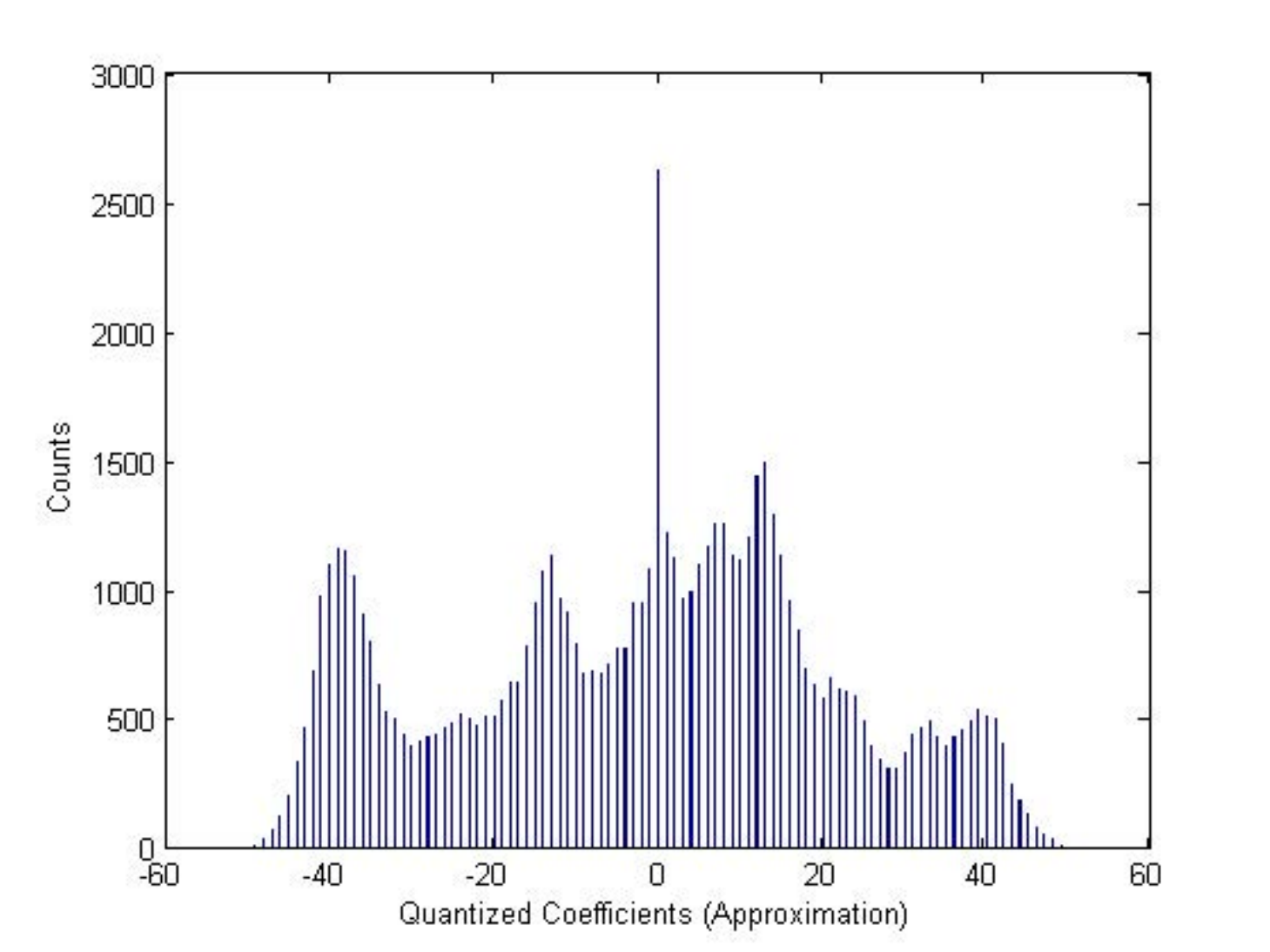}}
  \centerline{(b) UQ, $N=104$}\medskip
\end{minipage}
\begin{minipage}[b]{0.24\linewidth}
  \centering
  \centerline{\includegraphics[width=\linewidth]{Approximation_Lenna_e7m8.pdf}}
  \centerline{(c) UQ, $N=104$}\medskip
\end{minipage}
\begin{minipage}[b]{0.24\linewidth}
  \centering
  \centerline{\includegraphics[width=\linewidth]{Approximation_Lenna_e7m8.pdf}}
  \centerline{(d)  UQ, $N=104$}\medskip
\end{minipage}

\begin{minipage}[b]{0.24\linewidth}
  \centering
  \centerline{\includegraphics[width=\linewidth]{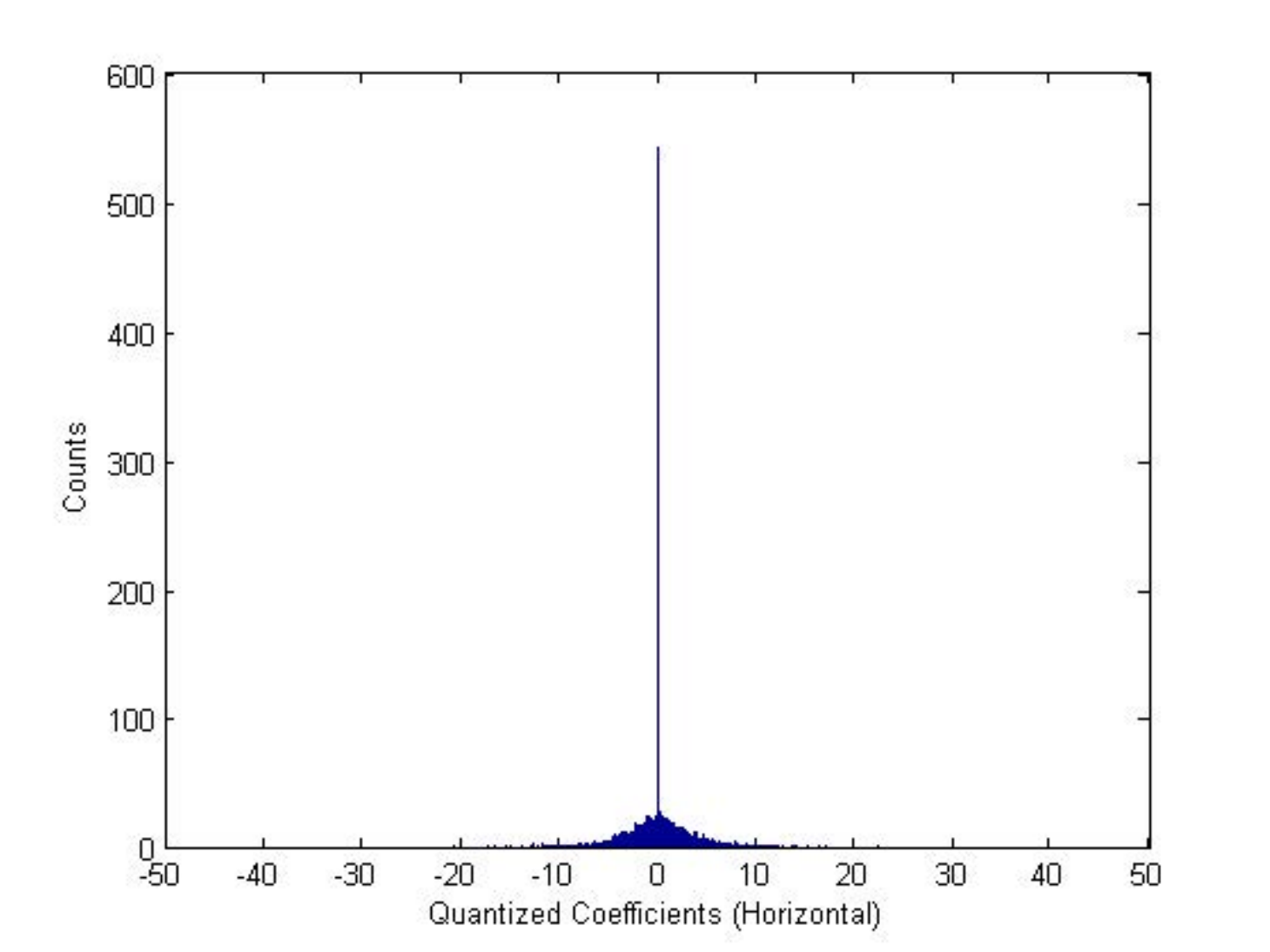}}
  \centerline{(e) Horizontal}\medskip
\end{minipage}
\begin{minipage}[b]{0.24\linewidth}
  \centering
  \centerline{\includegraphics[width=\linewidth]{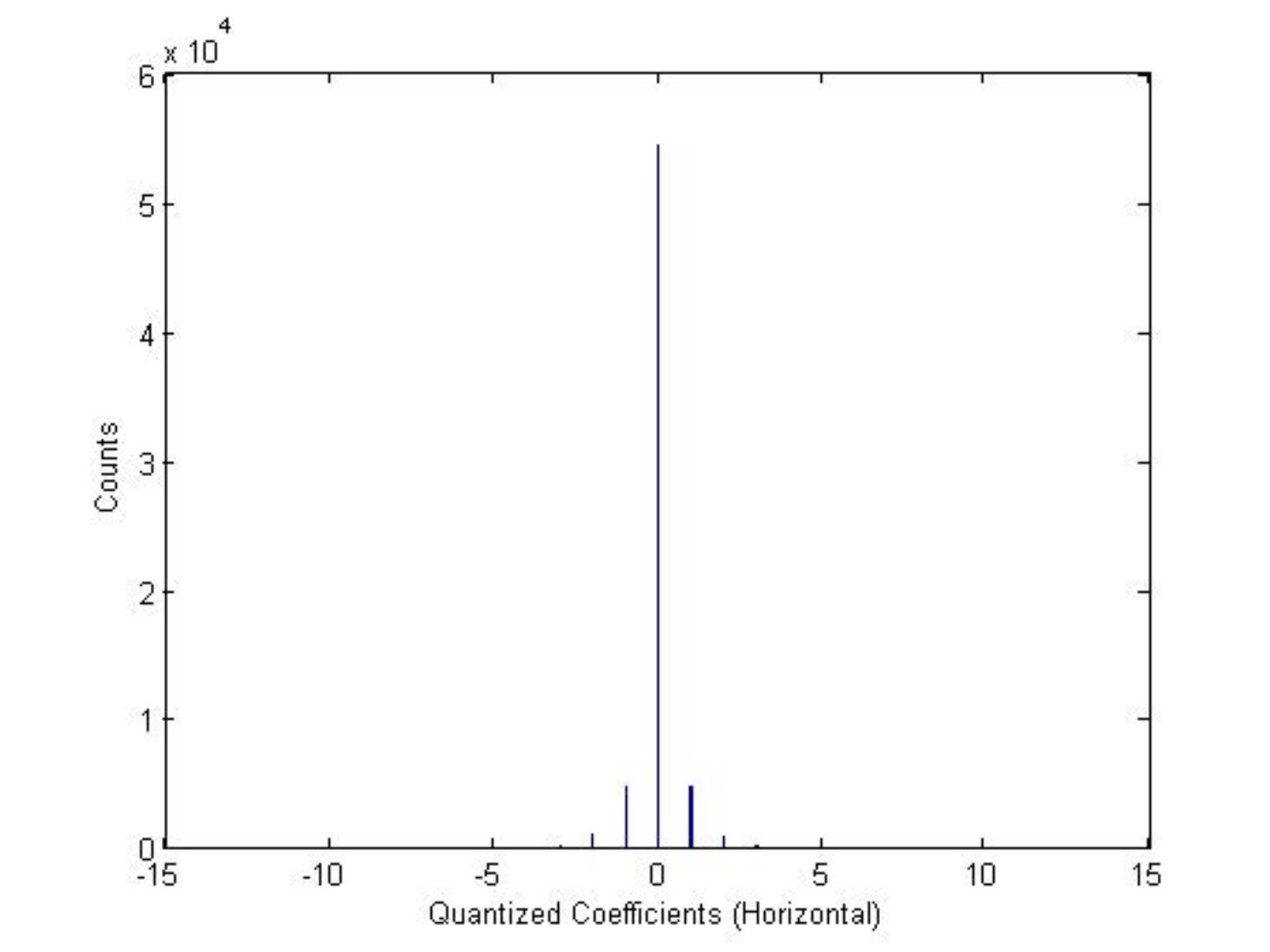}}
  \centerline{(f) UQ, $N=23$}\medskip
\end{minipage}
\begin{minipage}[b]{0.24\linewidth}
  \centering
  \centerline{\includegraphics[width=\linewidth]{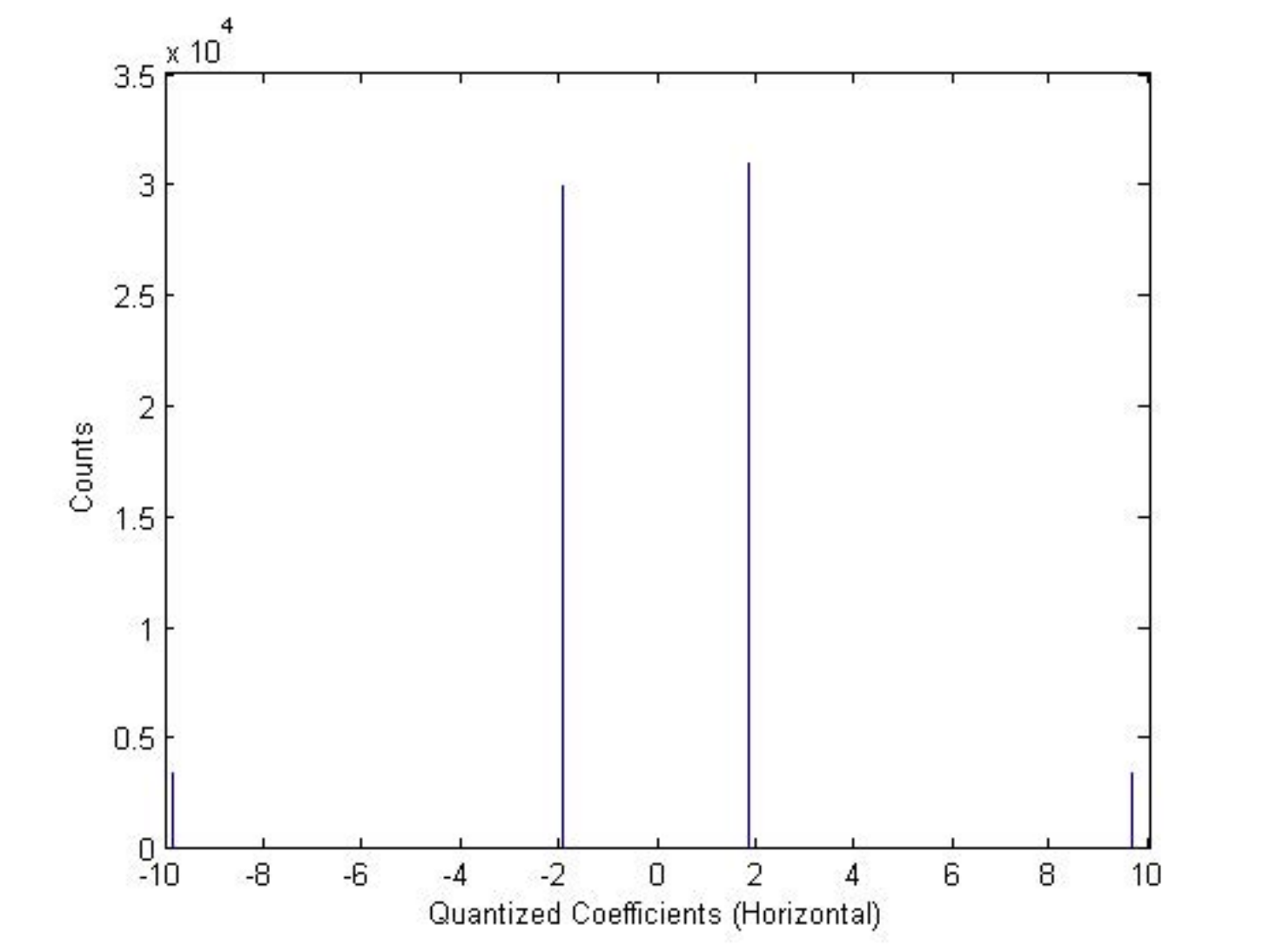}}
  \centerline{(g) NUQ, $N=4$}\medskip
\end{minipage}
\begin{minipage}[b]{0.24\linewidth}
  \centering
  \centerline{\includegraphics[width=\linewidth]{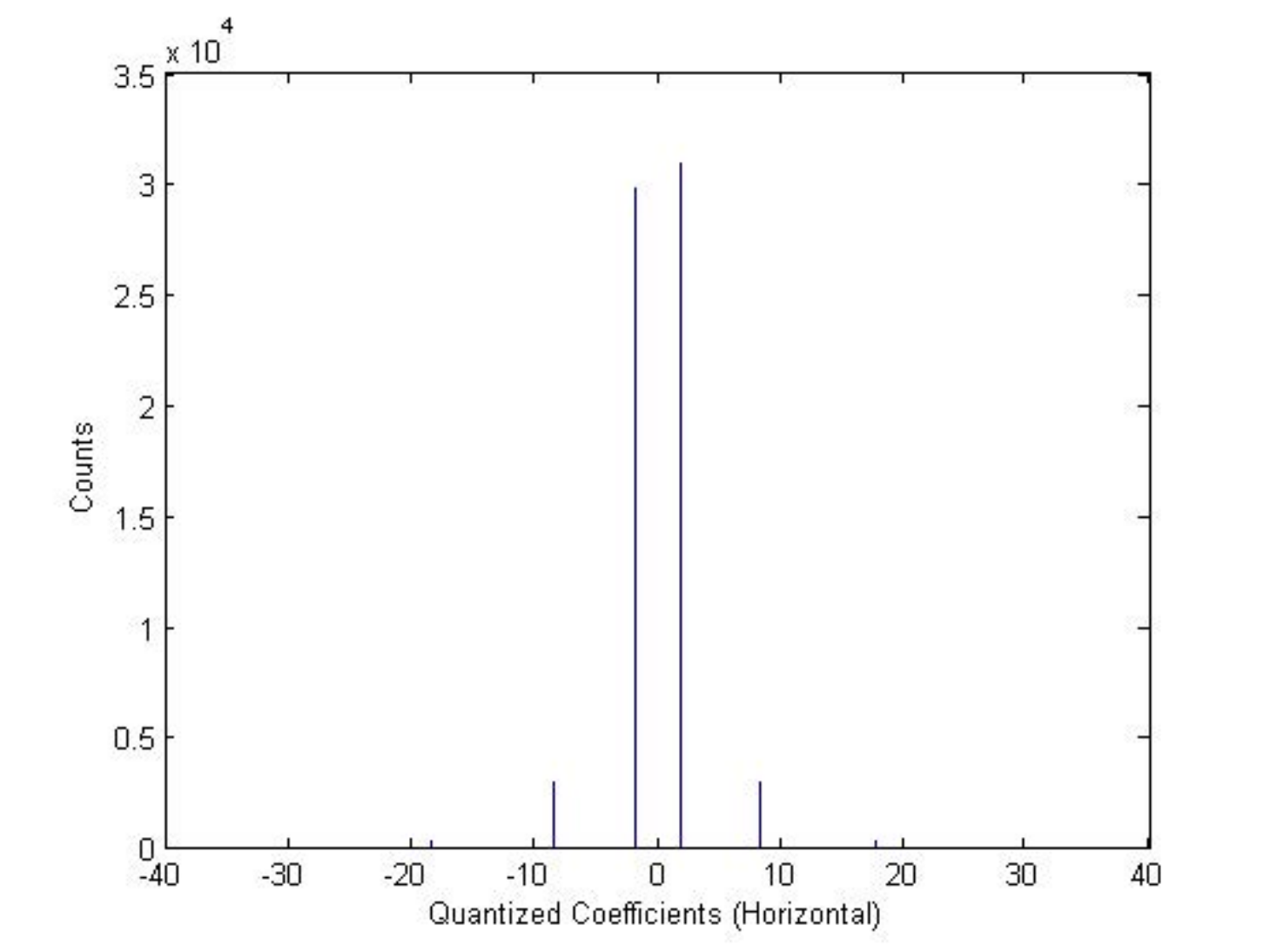}}
  \centerline{(h) NUQ, $N=8$}\medskip
\end{minipage}

\begin{minipage}[b]{0.24\linewidth}
  \centering
  \centerline{\includegraphics[width=\linewidth]{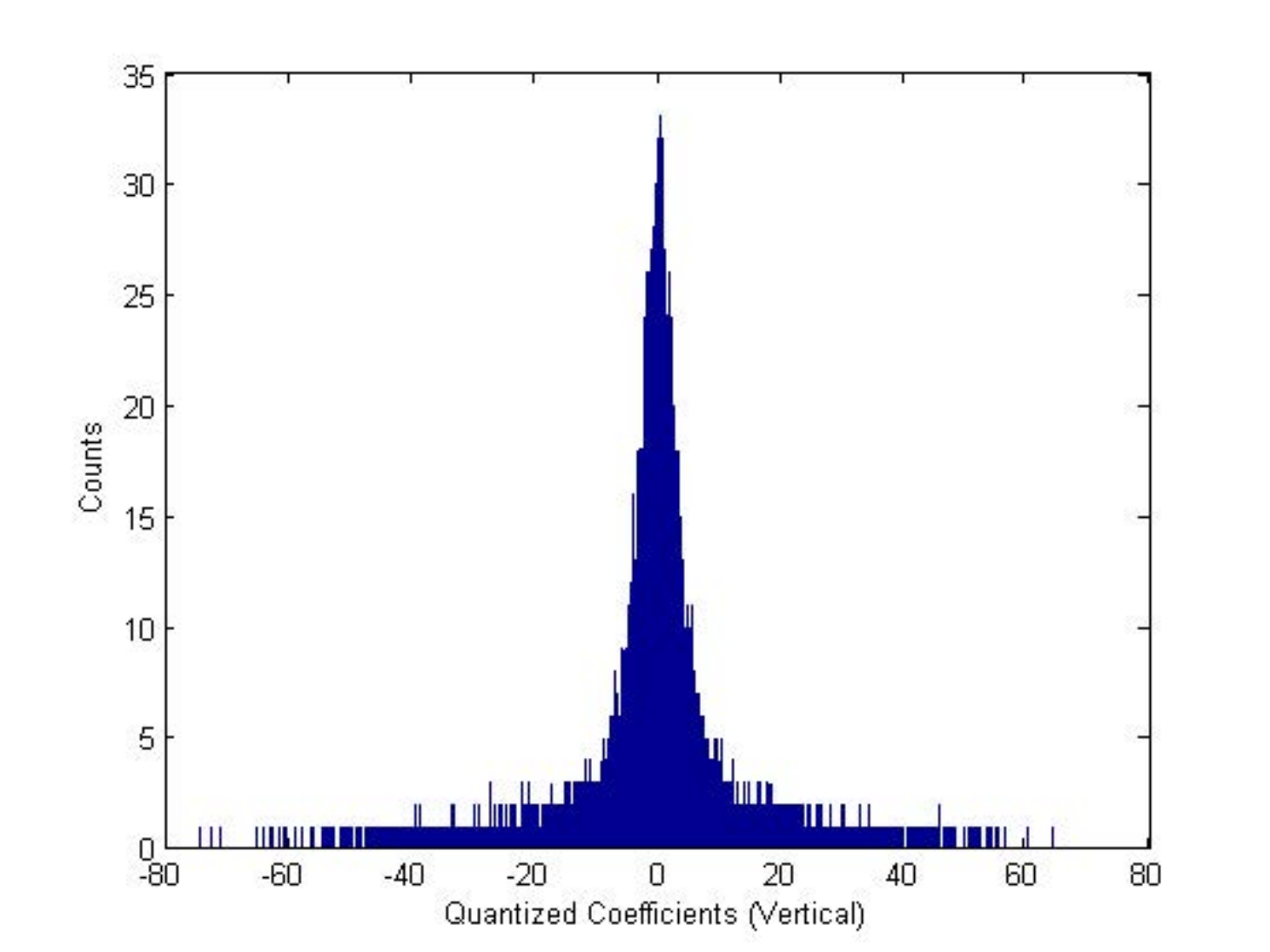}}
  \centerline{(i) Vertical}\medskip
\end{minipage}
\begin{minipage}[b]{0.24\linewidth}
  \centering
  \centerline{\includegraphics[width=\linewidth]{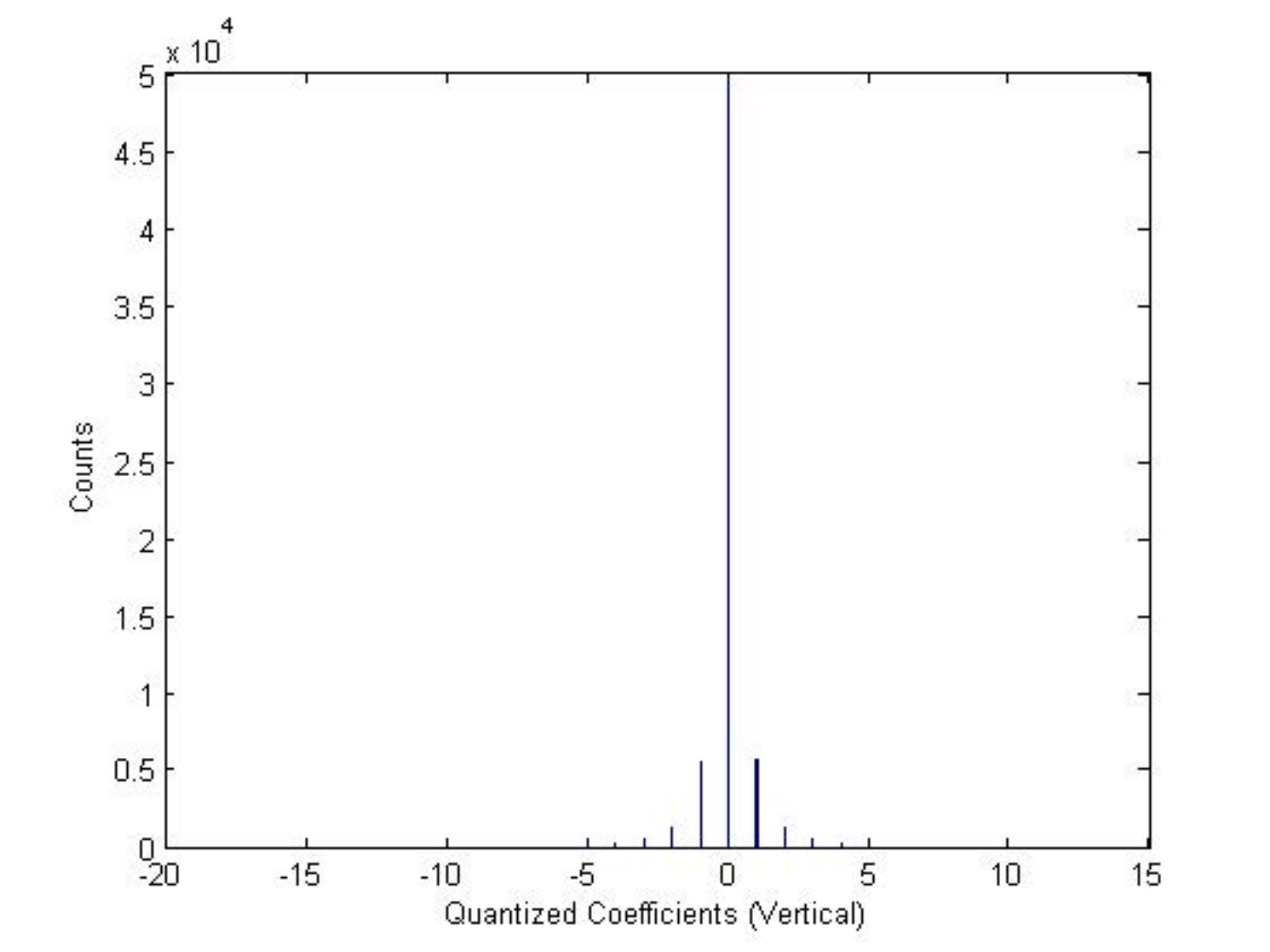}}
  \centerline{(j) UQ, $N=34$}\medskip
\end{minipage}
\begin{minipage}[b]{0.24\linewidth}
  \centering
  \centerline{\includegraphics[width=\linewidth]{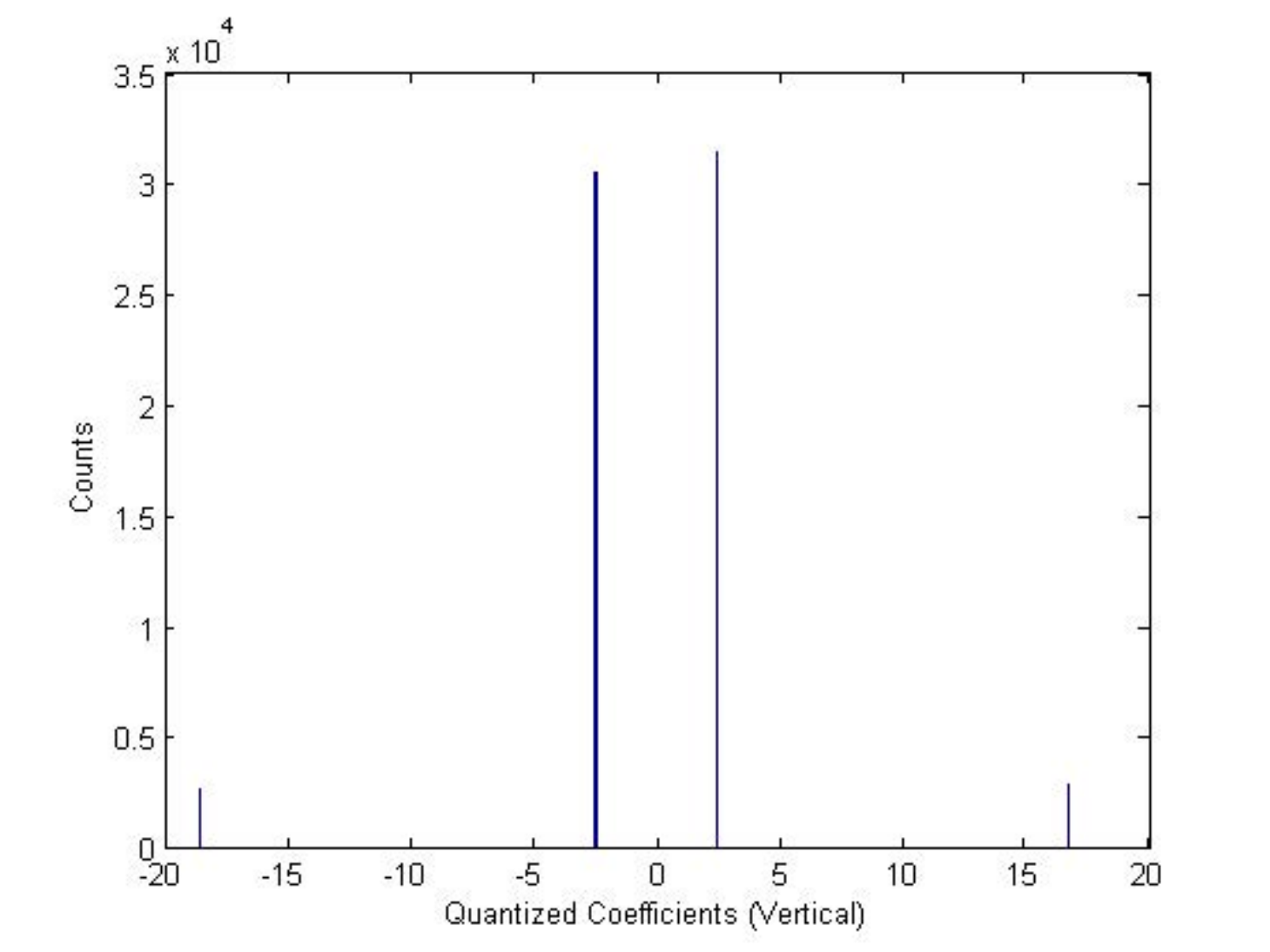}}
  \centerline{(k) NUQ, $N=4$}\medskip
\end{minipage}
\begin{minipage}[b]{0.24\linewidth}
  \centering
  \centerline{\includegraphics[width=\linewidth]{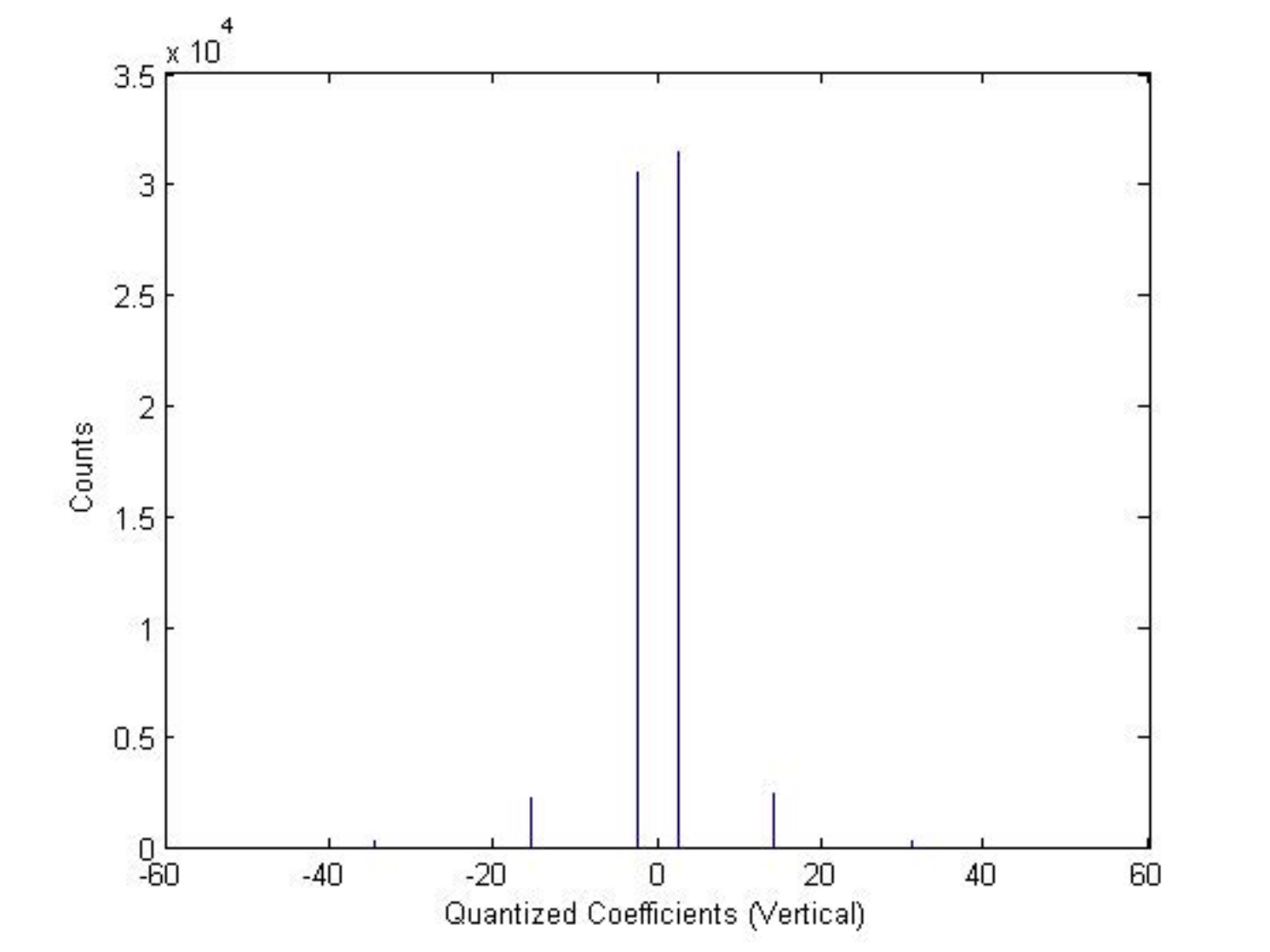}}
  \centerline{(l) NUQ, $N=8$}\medskip
\end{minipage}

\begin{minipage}[b]{0.24\linewidth}
  \centering
  \centerline{\includegraphics[width=\linewidth]{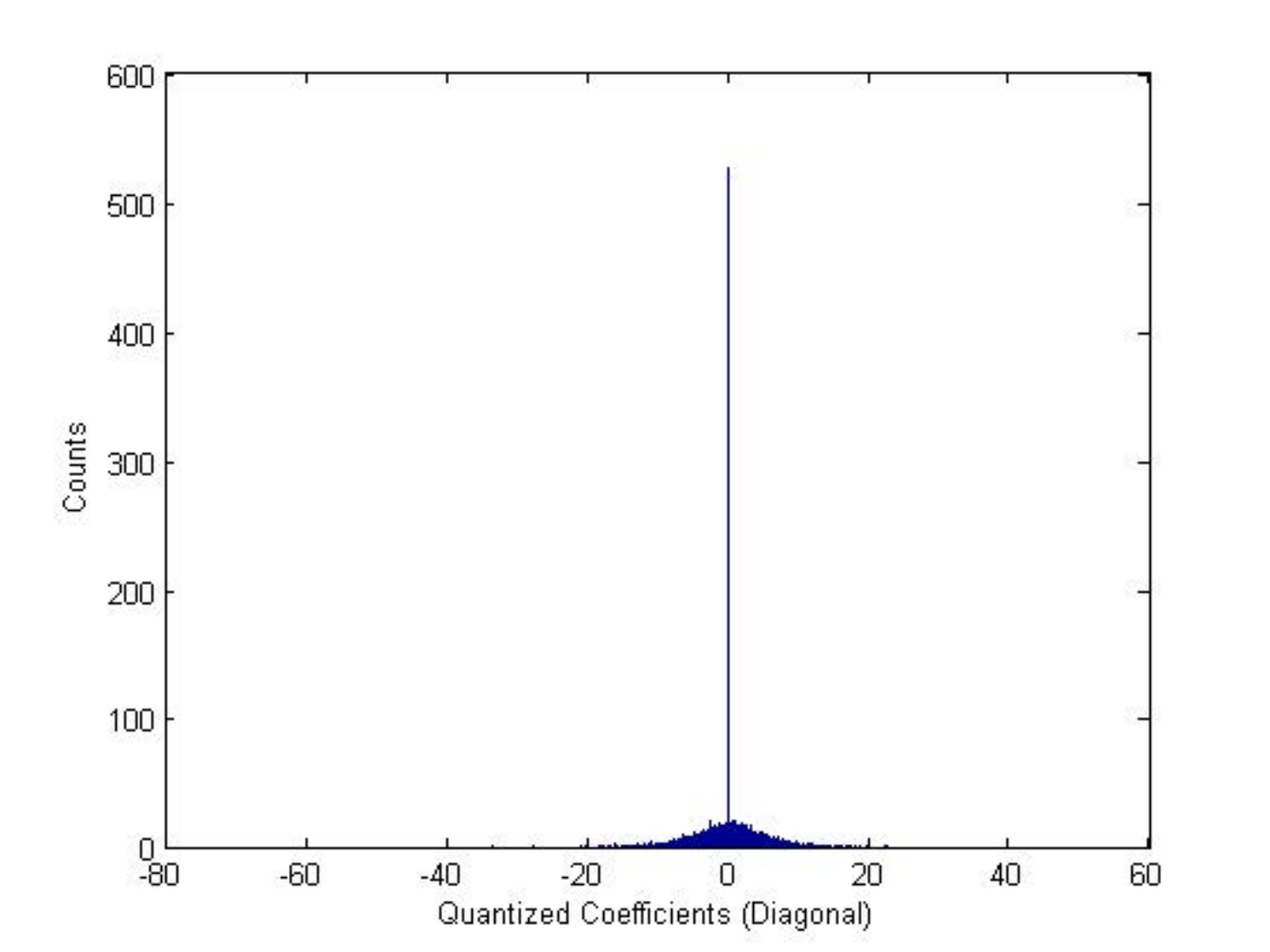}}
  \centerline{(m) Diagonal}\medskip
\end{minipage}
\begin{minipage}[b]{0.24\linewidth}
  \centering
  \centerline{\includegraphics[width=\linewidth]{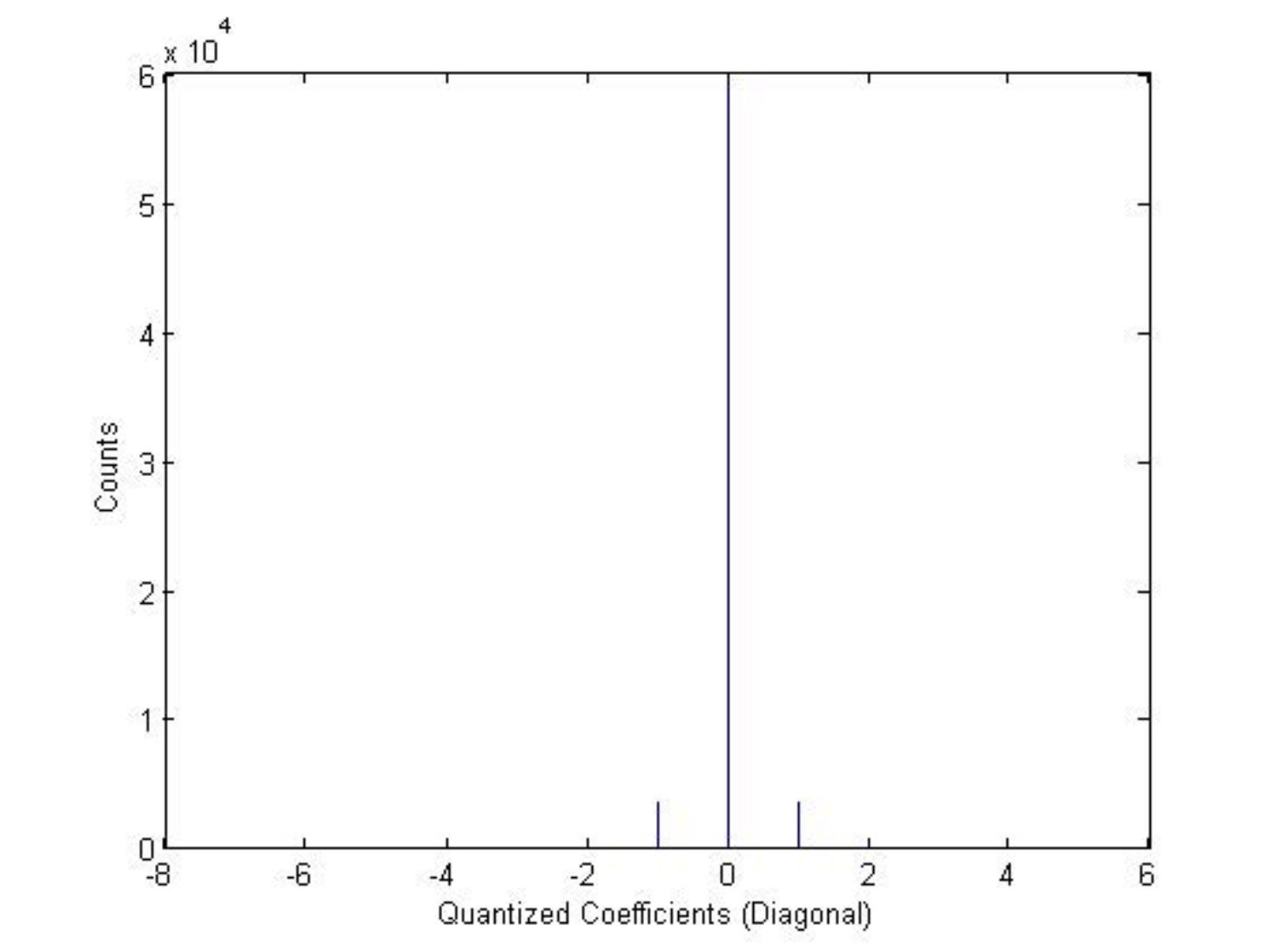}}
  \centerline{(n) UQ, $N=15$}\medskip
\end{minipage}
\begin{minipage}[b]{0.24\linewidth}
  \centering
  \centerline{\includegraphics[width=\linewidth]{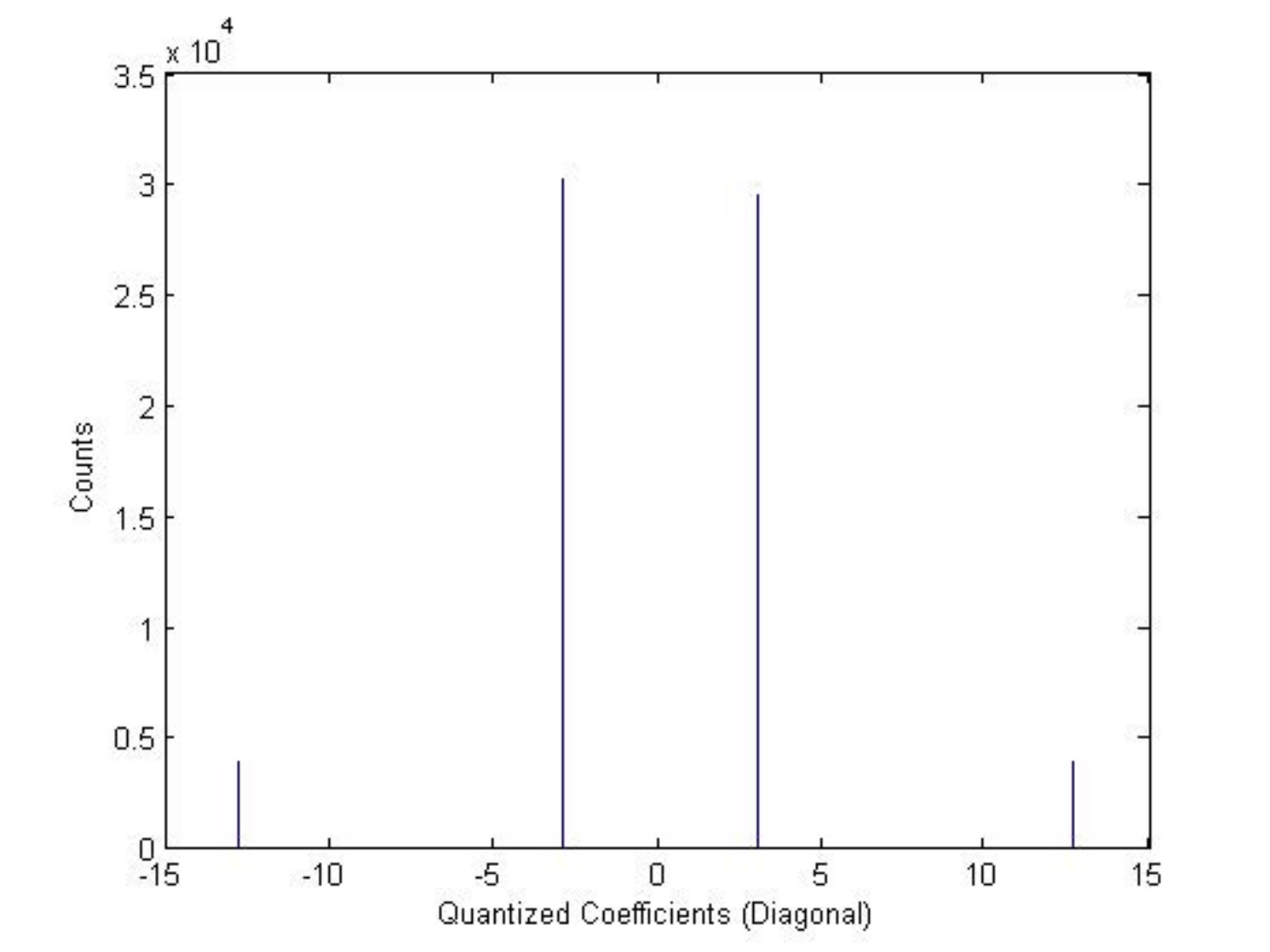}}
  \centerline{(o) NUQ, $N=4$}\medskip
\end{minipage}
\begin{minipage}[b]{0.24\linewidth}
  \centering
  \centerline{\includegraphics[width=\linewidth]{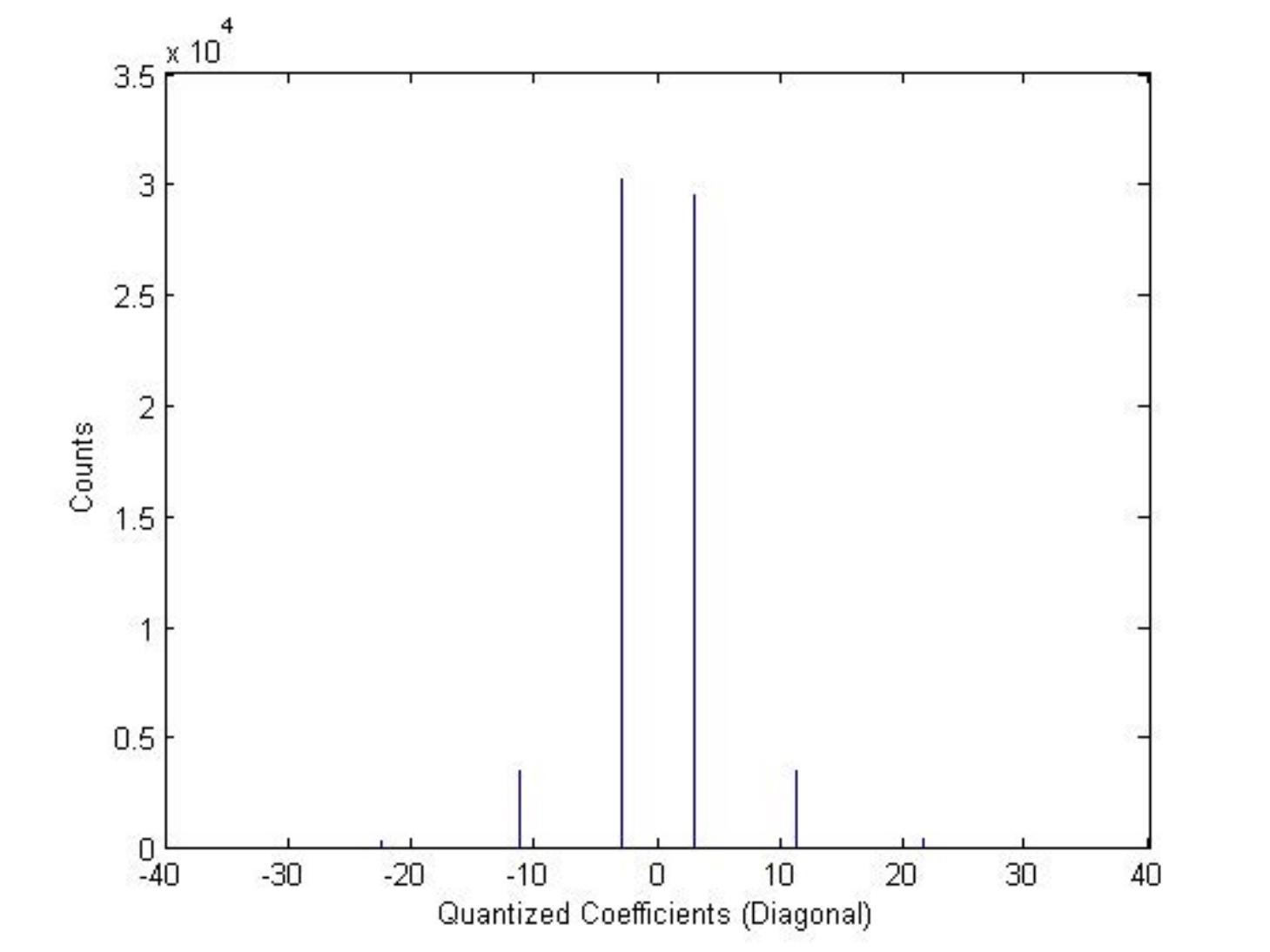}}
  \centerline{(p) NUQ, $N=8$}\medskip
\end{minipage}
 \caption{\emph{Lenna} ($512 \times 512$): (a,e,i,m) Histogram plots of the Approximation, Horizontal, Vertical, and Diagonal component, respectively; (b-d) Histogram plots of the Approximation component after the deadzone uniform quantization (UQ) and non-uniform quantization (NUQ); (f-h) Histogram plots of the Horizontal component after UQ and NUQ; (j-l) Histogram plots of the Vertical component after UQ and NUQ; (n-p) Histogram plots of the Diagonal component after UQ and NUQ. \textbf{\emph{N}} is the number of quantized values.}
\end{figure}

\begin{figure}[ht!]
\begin{minipage}[b]{0.24\linewidth}
  \centering
  \centerline{\includegraphics[width=\linewidth]{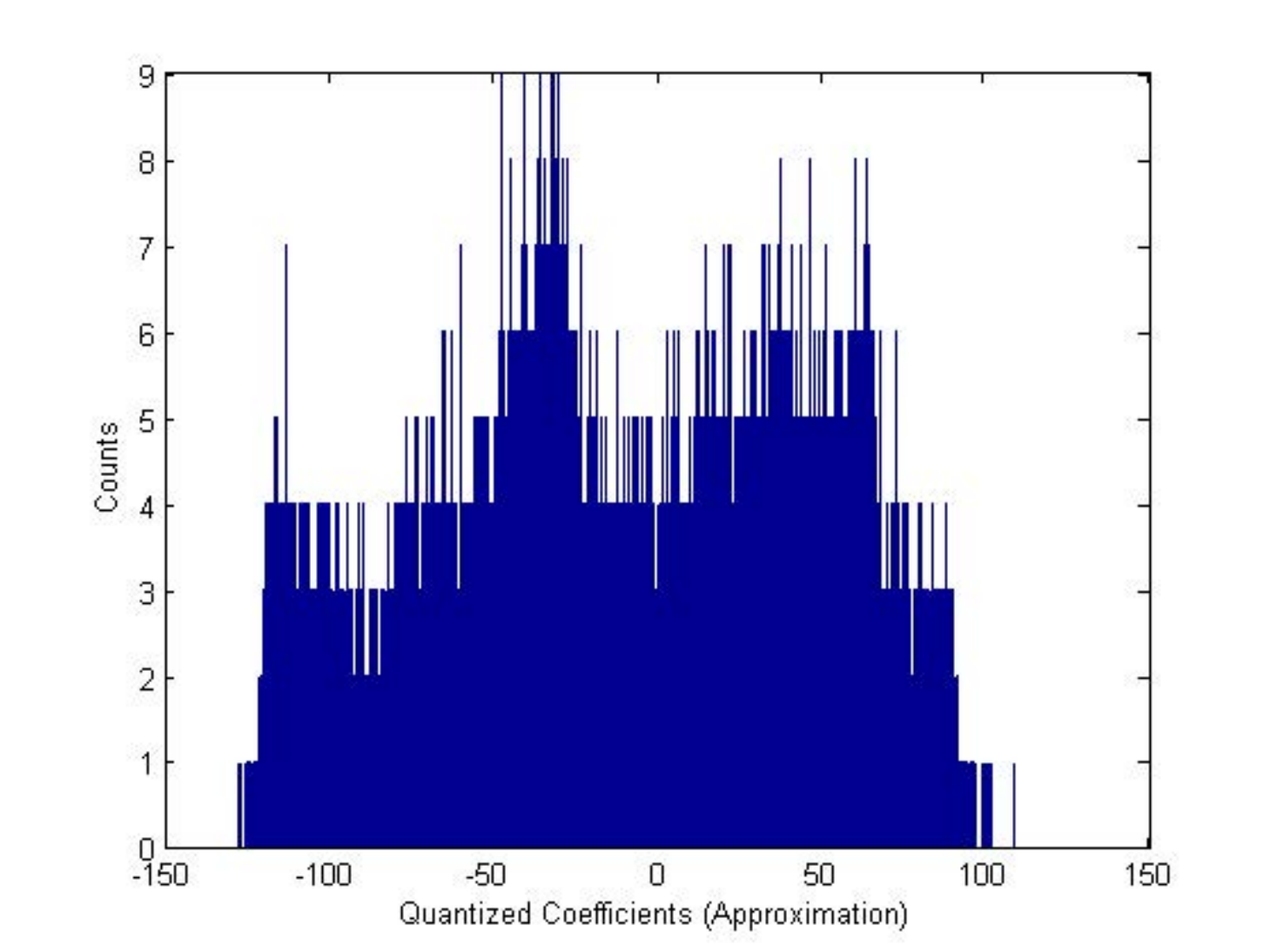}}
  \centerline{(a) Approximation}\medskip
\end{minipage}
\begin{minipage}[b]{0.24\linewidth}
  \centering
  \centerline{\includegraphics[width=\linewidth]{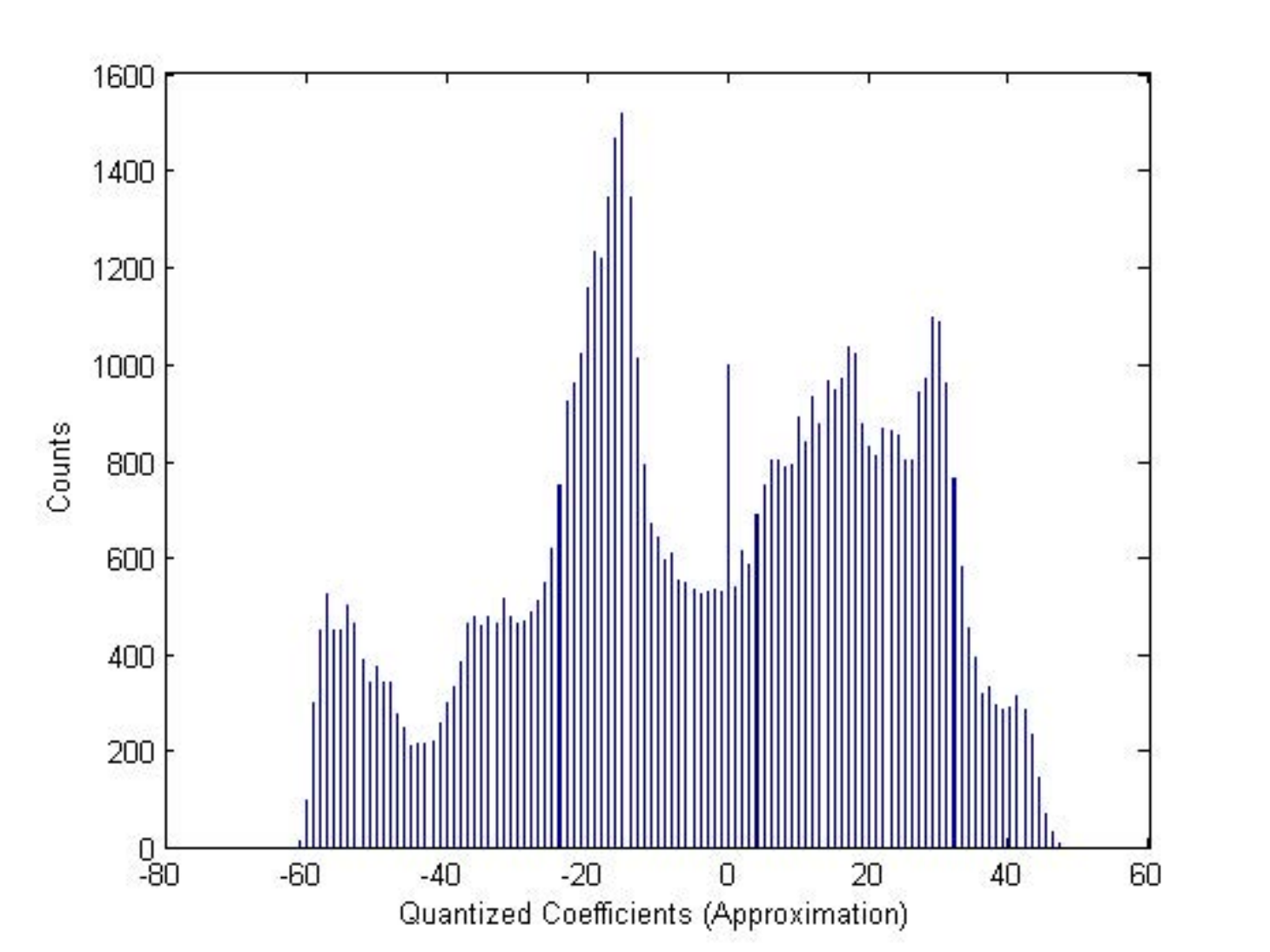}}
  \centerline{(b) UQ, $N=115$}\medskip
\end{minipage}
\begin{minipage}[b]{0.24\linewidth}
  \centering
  \centerline{\includegraphics[width=\linewidth]{Approximation_Pepper_e7m8.pdf}}
  \centerline{(c) UQ, $N=115$}\medskip
\end{minipage}
\begin{minipage}[b]{0.24\linewidth}
  \centering
  \centerline{\includegraphics[width=\linewidth]{Approximation_Pepper_e7m8.pdf}}
  \centerline{(d) UQ, $N=115$}\medskip
\end{minipage}

\begin{minipage}[b]{0.24\linewidth}
  \centering
  \centerline{\includegraphics[width=\linewidth]{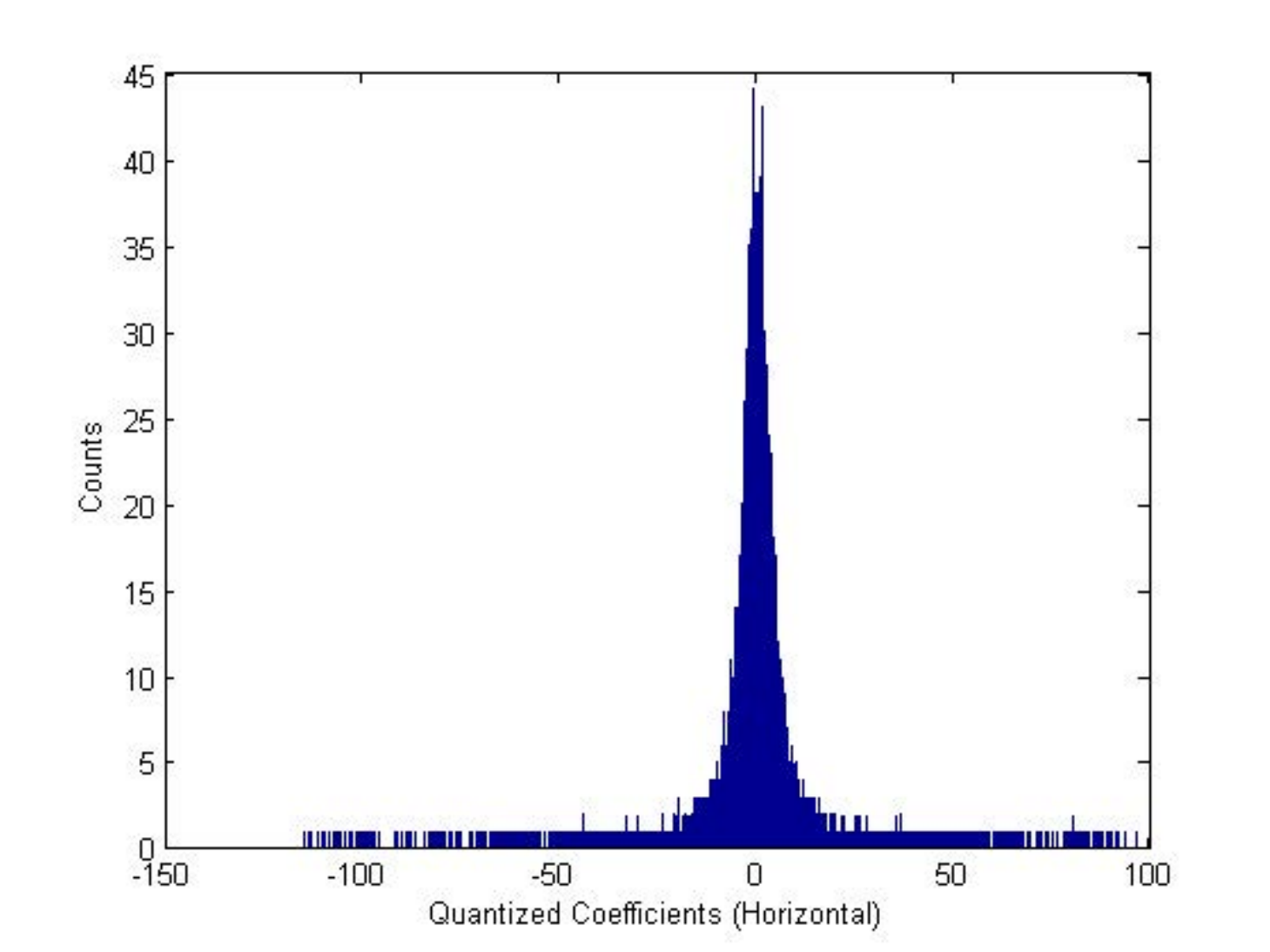}}
  \centerline{(e) Horizontal}\medskip
\end{minipage}
\begin{minipage}[b]{0.24\linewidth}
  \centering
  \centerline{\includegraphics[width=\linewidth]{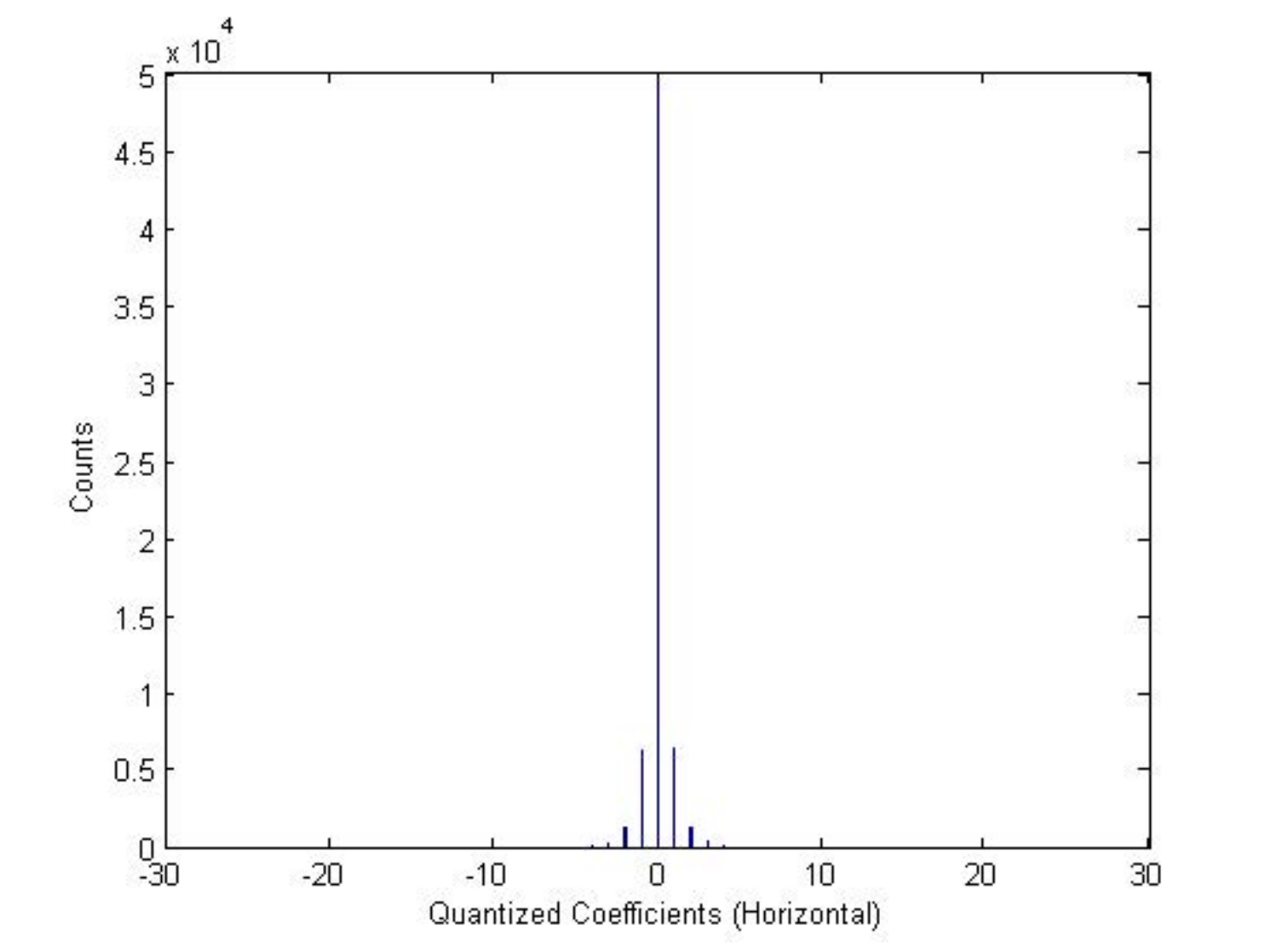}}
  \centerline{(f) UQ, $N=52$}\medskip
\end{minipage}
\begin{minipage}[b]{0.24\linewidth}
  \centering
  \centerline{\includegraphics[width=\linewidth]{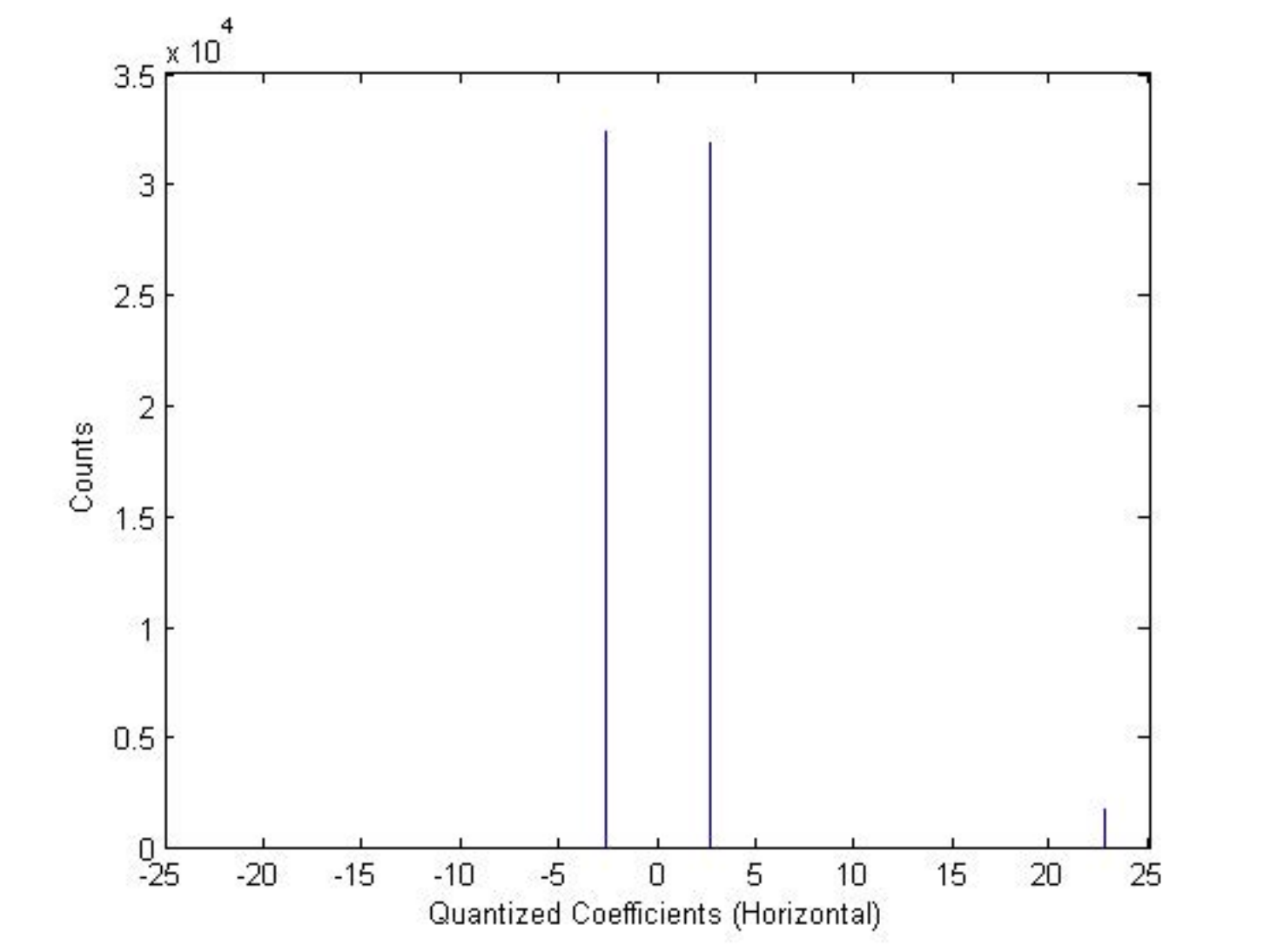}}
  \centerline{(g) NUQ, $N=4$}\medskip
\end{minipage}
\begin{minipage}[b]{0.24\linewidth}
  \centering
  \centerline{\includegraphics[width=\linewidth]{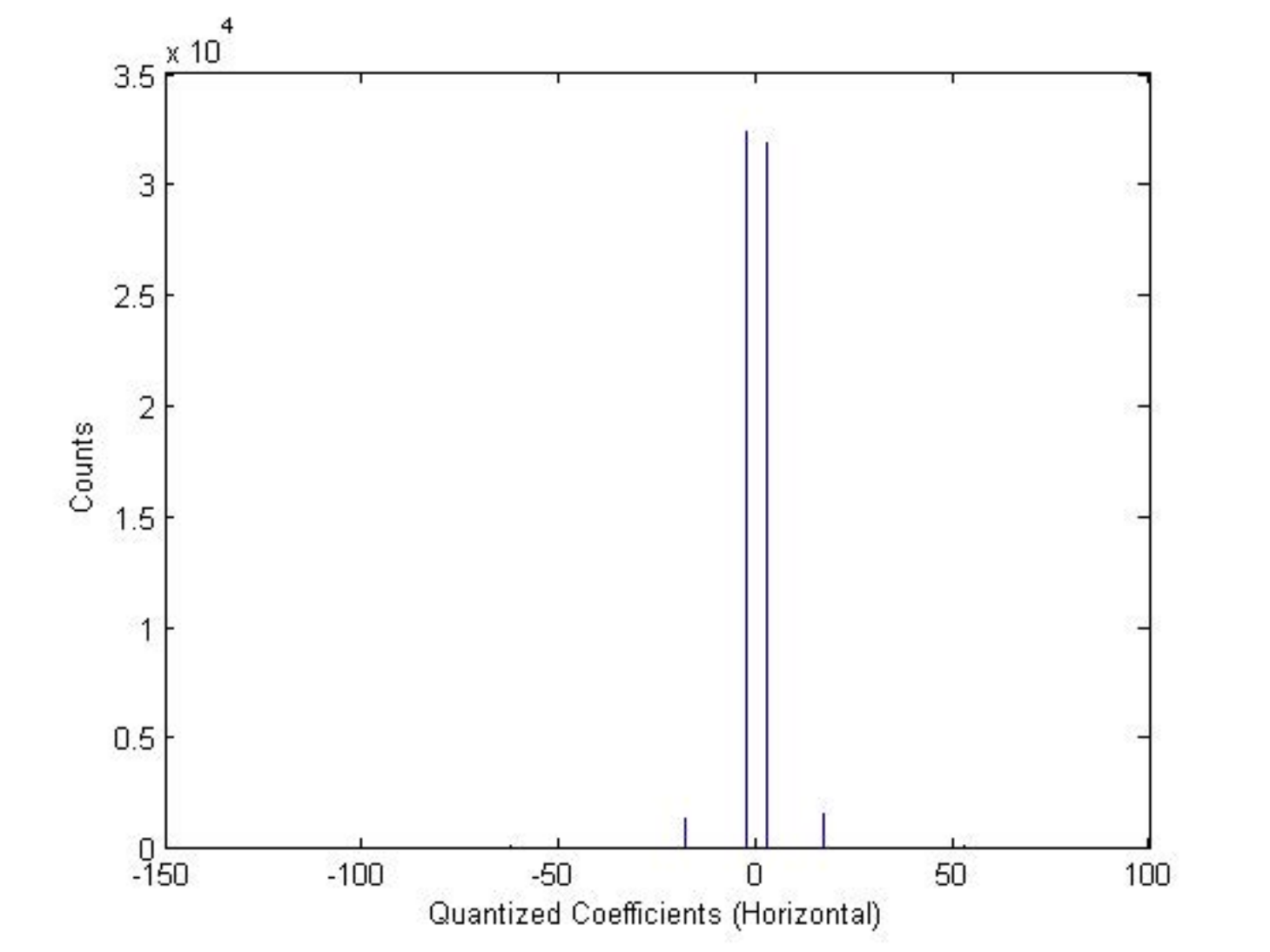}}
  \centerline{(h) NUQ, $N=8$}\medskip
\end{minipage}

\begin{minipage}[b]{0.24\linewidth}
  \centering
  \centerline{\includegraphics[width=\linewidth]{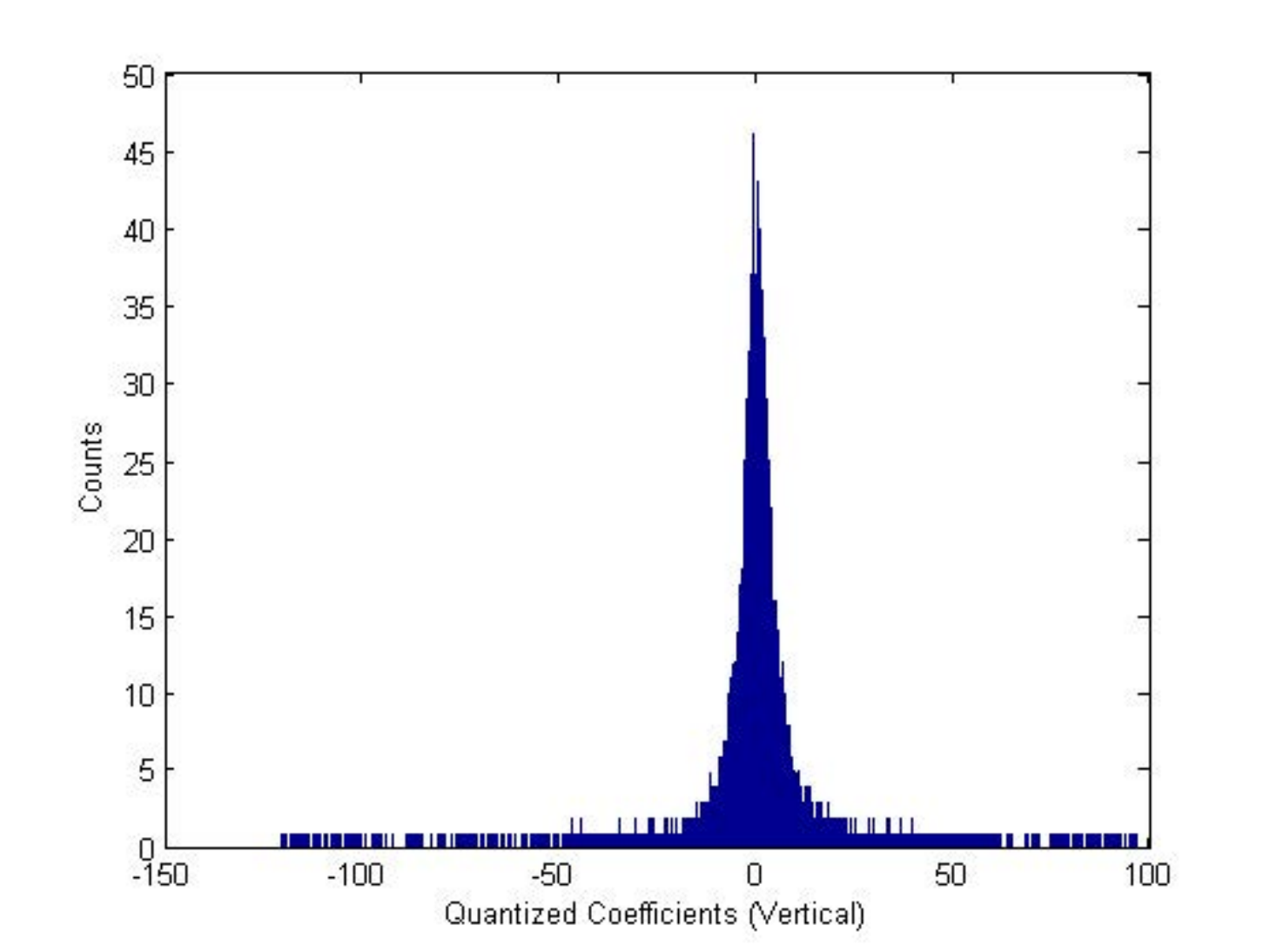}}
  \centerline{(i) Vertical}\medskip
\end{minipage}
\begin{minipage}[b]{0.24\linewidth}
  \centering
  \centerline{\includegraphics[width=\linewidth]{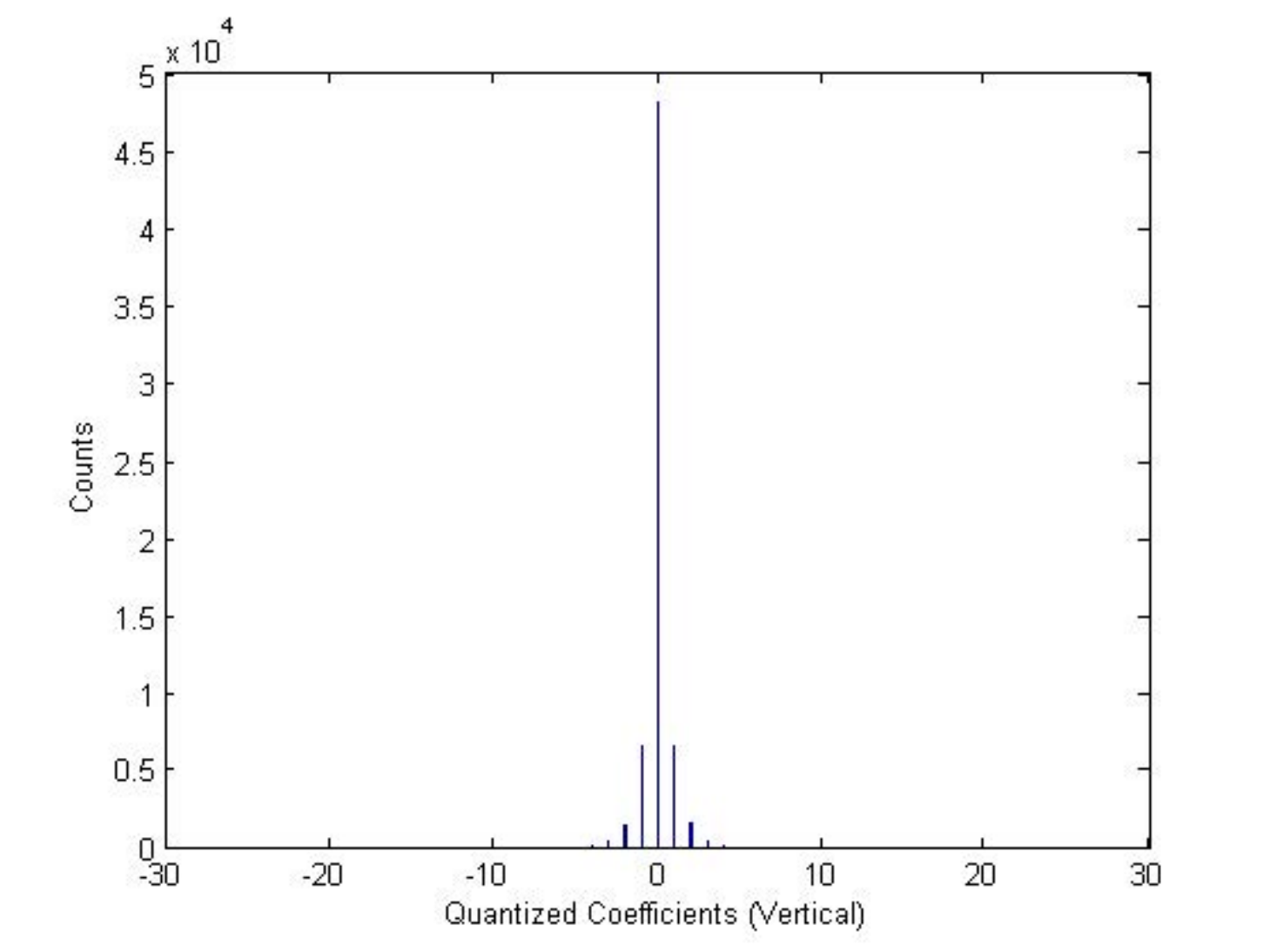}}
  \centerline{(j) UQ, $N=53$}\medskip
\end{minipage}
\begin{minipage}[b]{0.24\linewidth}
  \centering
  \centerline{\includegraphics[width=\linewidth]{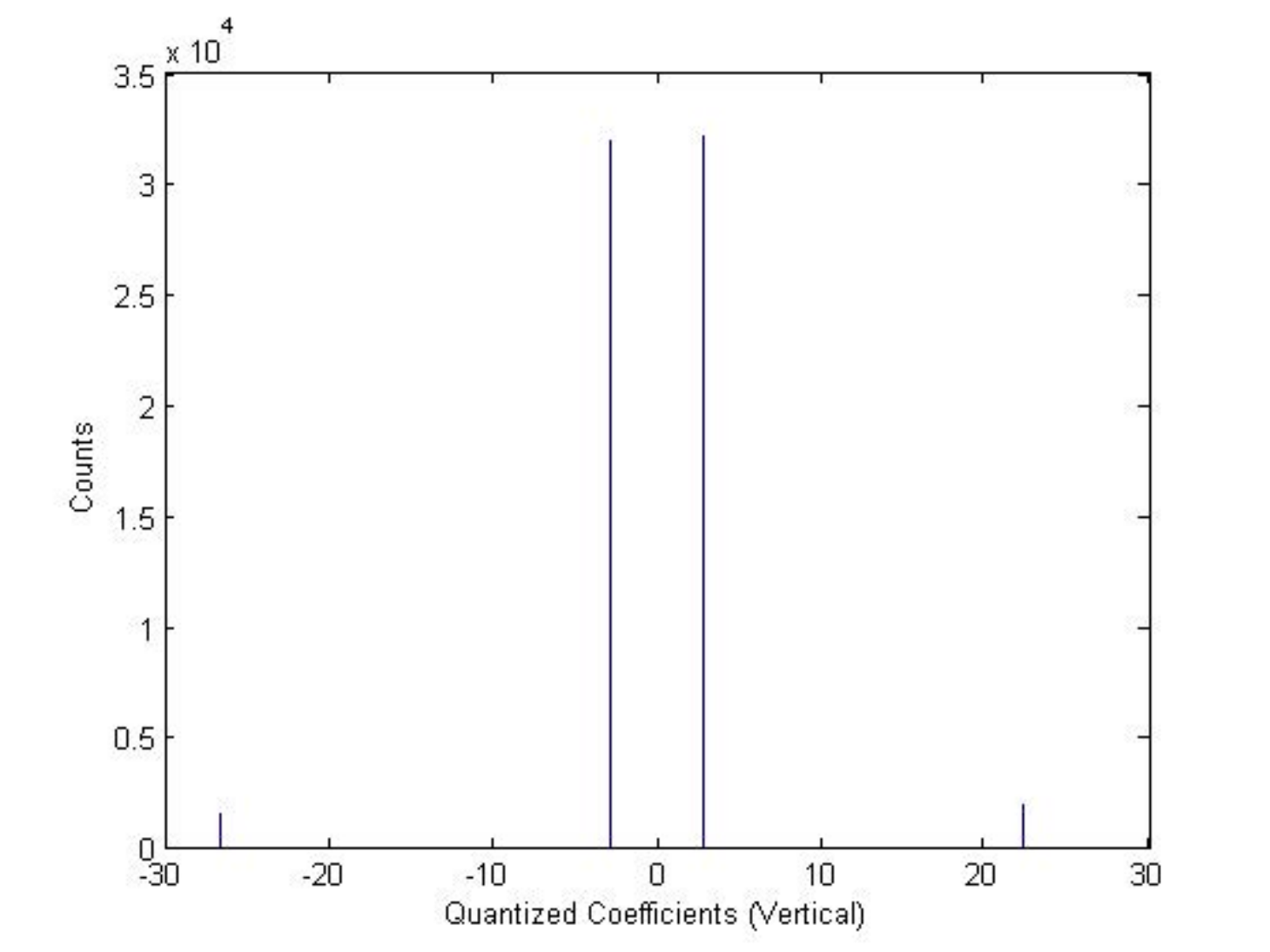}}
  \centerline{(k) NUQ, $N=4$}\medskip
\end{minipage}
\begin{minipage}[b]{0.24\linewidth}
  \centering
  \centerline{\includegraphics[width=\linewidth]{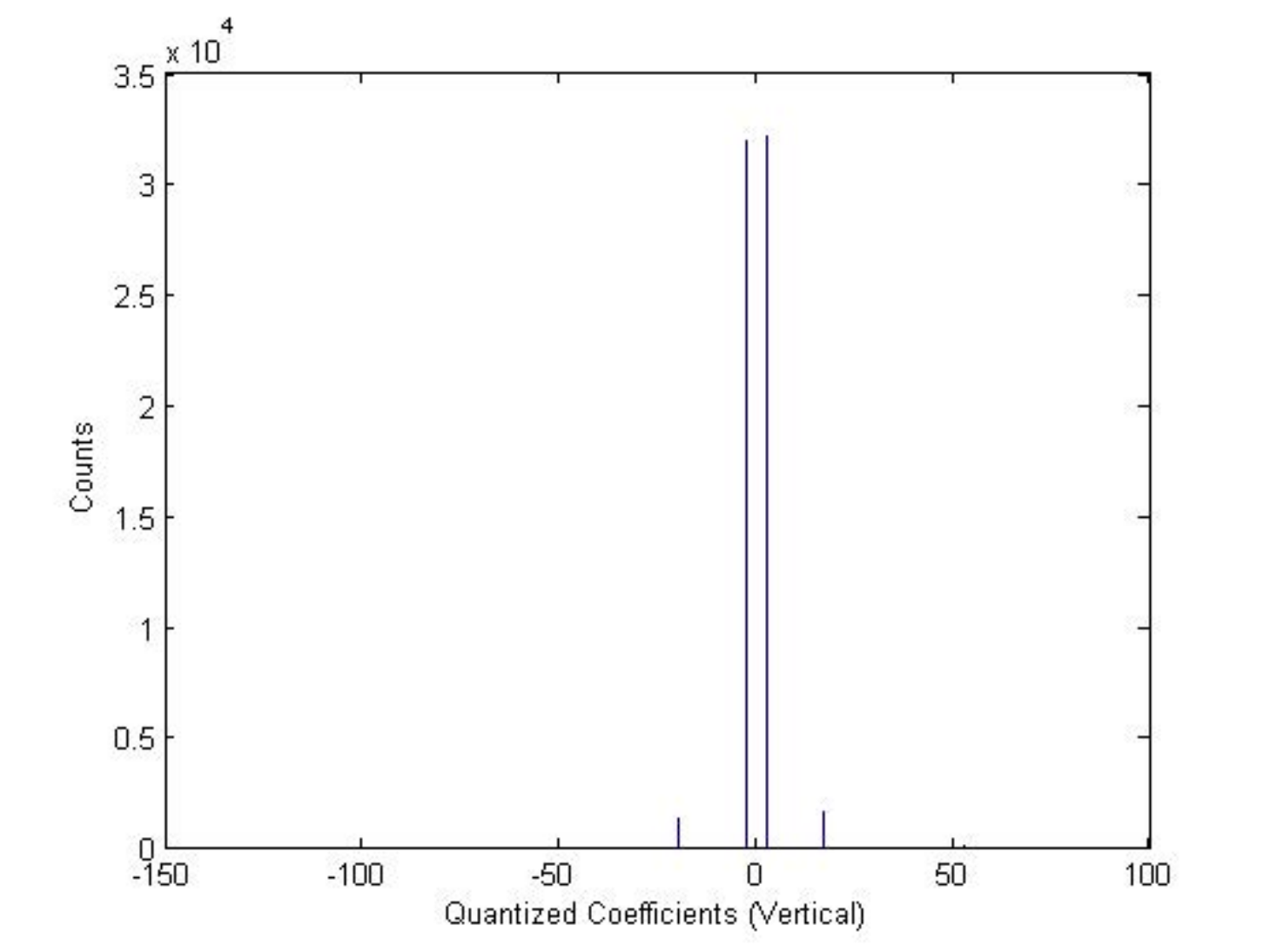}}
  \centerline{(l) NUQ, $N=8$}\medskip
\end{minipage}

\begin{minipage}[b]{0.24\linewidth}
  \centering
  \centerline{\includegraphics[width=\linewidth]{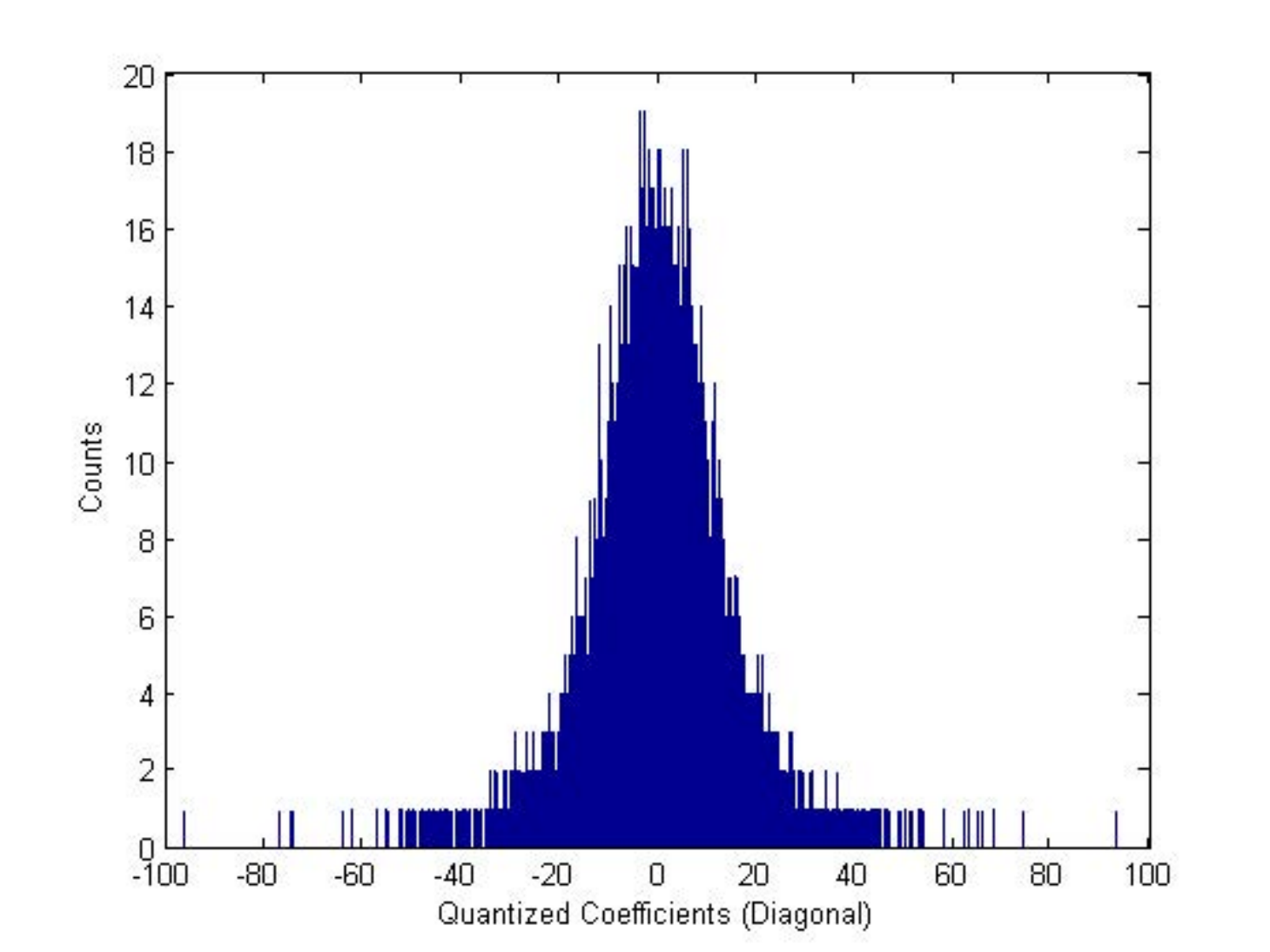}}
  \centerline{(m) Diagonal}\medskip
\end{minipage}
\begin{minipage}[b]{0.24\linewidth}
  \centering
  \centerline{\includegraphics[width=\linewidth]{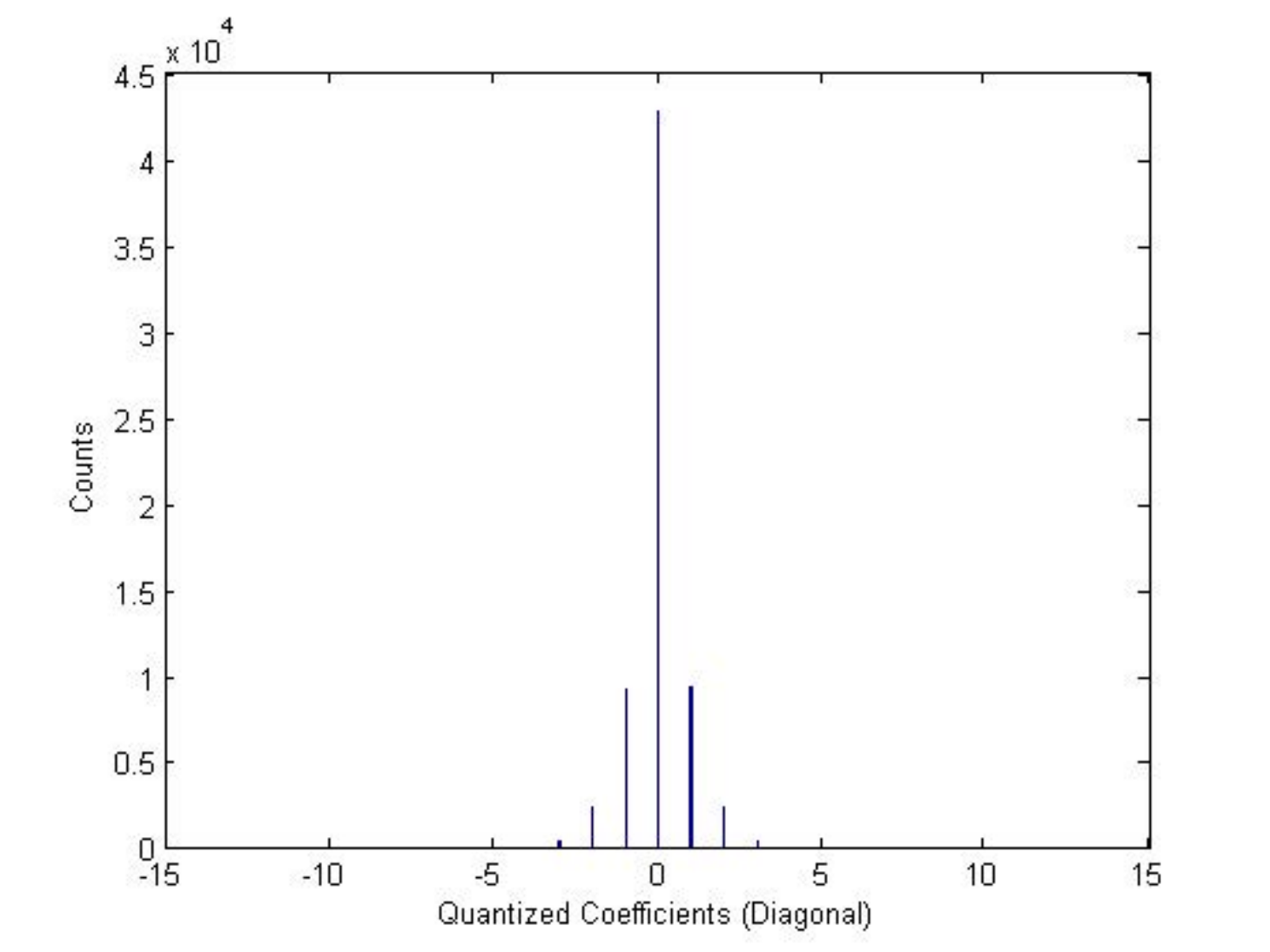}}
  \centerline{(n) UQ, $N=20$}\medskip
\end{minipage}
\begin{minipage}[b]{0.24\linewidth}
  \centering
  \centerline{\includegraphics[width=\linewidth]{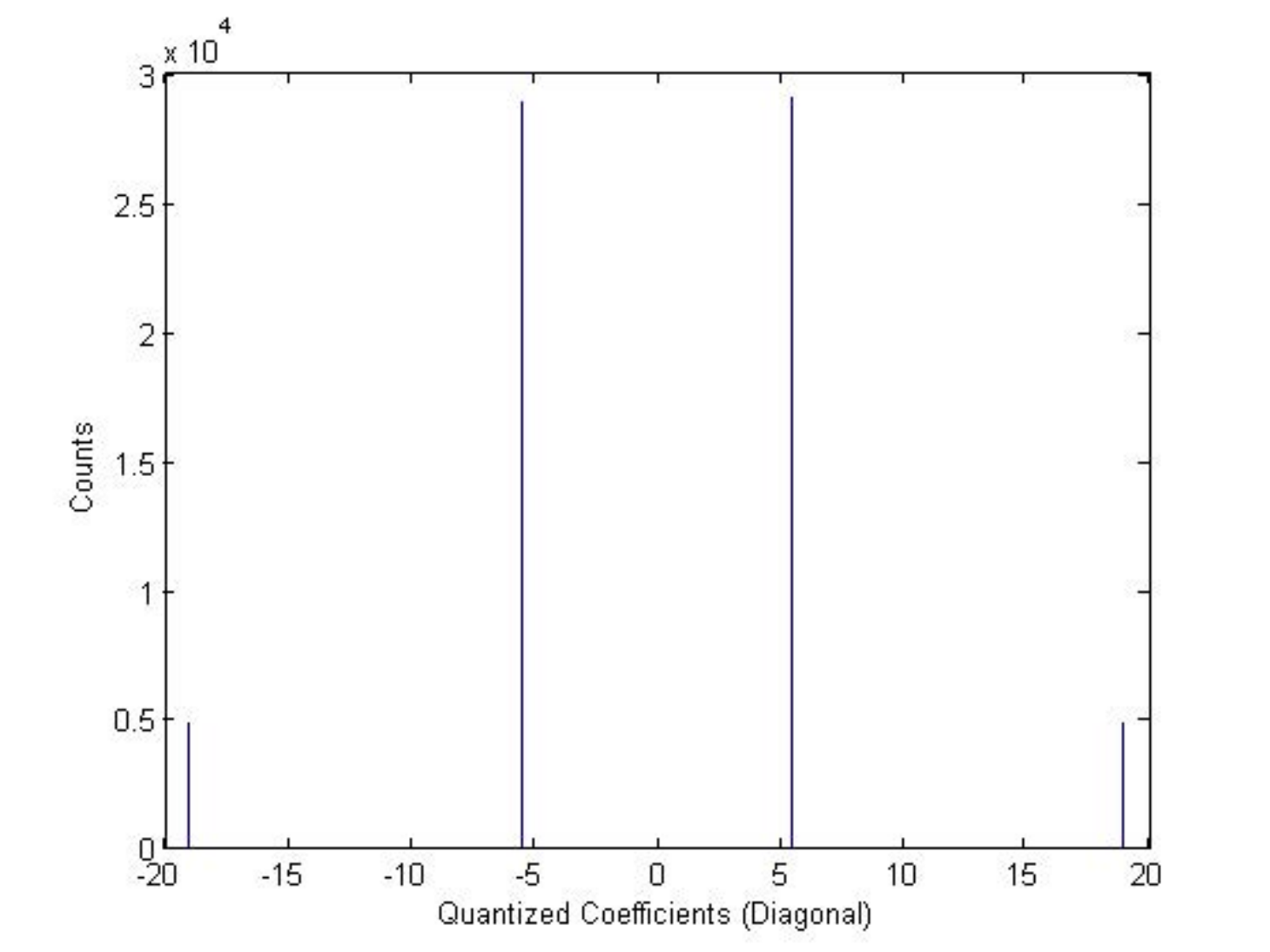}}
  \centerline{(o) NUQ, $N=4$}\medskip
\end{minipage}
\begin{minipage}[b]{0.24\linewidth}
  \centering
  \centerline{\includegraphics[width=\linewidth]{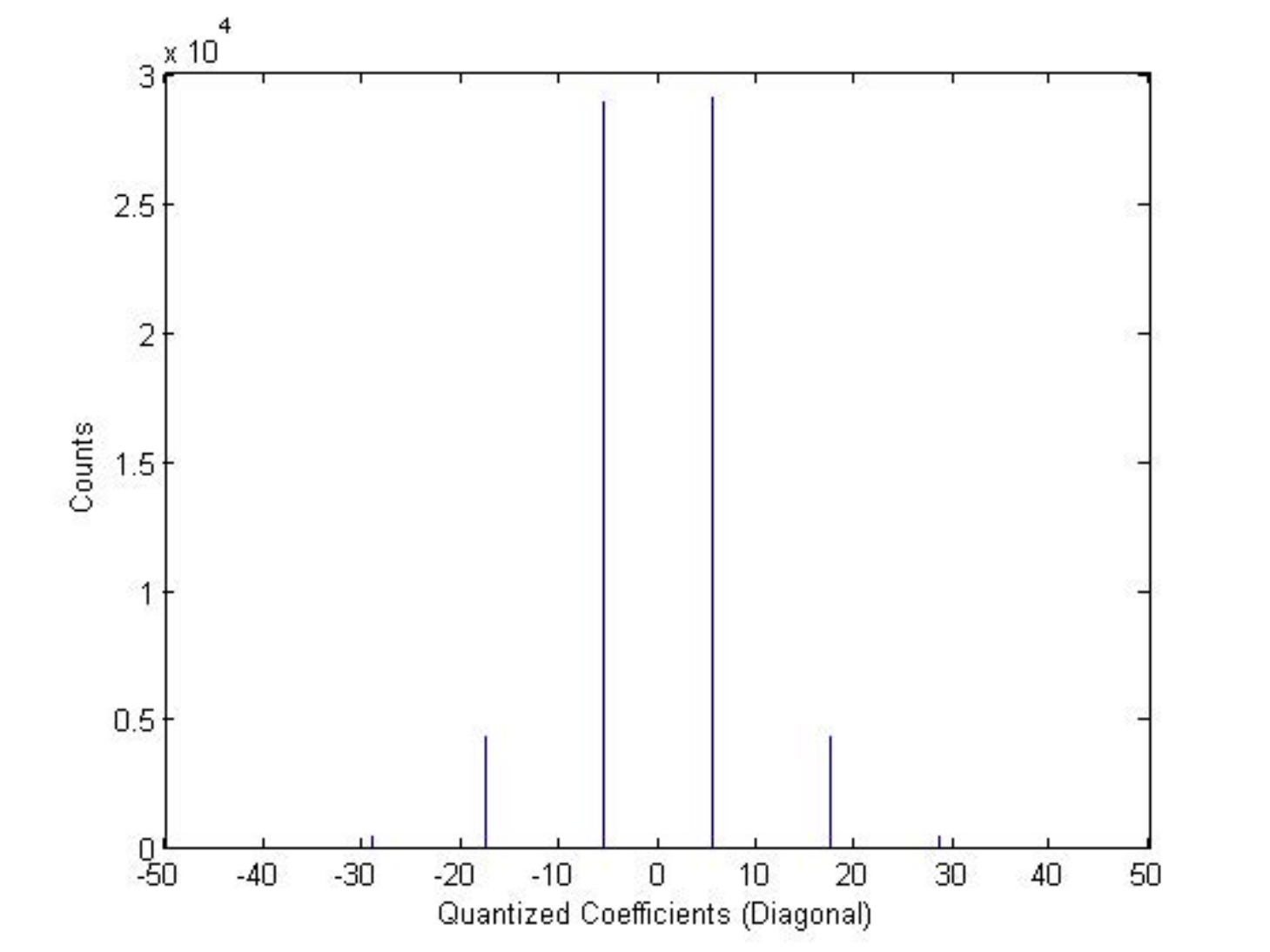}}
  \centerline{(p) NUQ, $N=8$}\medskip
\end{minipage}
 \caption{\emph{Pepper} ($512 \times 512$): (a,e,i,m) Histogram plots of the Approximation, Horizontal, Vertical, and Diagonal component, respectively; (b-d) Histogram plots of the Approximation component after the deadzone uniform quantization (UQ) and non-uniform quantization (NUQ); (f-h) Histogram plots of the Horizontal component after UQ and NUQ; (j-l) Histogram plots of the Vertical component after UQ and NUQ; (n-p) Histogram plots of the Diagonal component after UQ and NUQ. \textbf{\emph{N}} is the number of quantized values.}
\end{figure}

It can be seen in Figs. 3-5 (a,e,i,m) that the coefficient distribution in the histogram plot for the Approximation component and Detail components are very different. The Approximation component is more evenly distributed, whereas each Detail component has skewed distribution of the coefficients, centered to zero in their respective histogram plot. Only degree of skewness varies; Figs. 3 (e,m) are most skewed followed by Fig. 3(j) and Figs. 4(e,j), and then Fig. 4(m) and Figs. 5(e,j,m). The original histogram in each Detail component has a majority of coefficients in the vicinity of zero; only a few coefficients far from it. Based on their histogram distributions, separate quantization approaches for the Approximation component and Detail components are desirable. The deadzone uniform quantizer quantizes all the components with the same predefined fixed step size, leading to over quantization in the detail components and higher bitrates. Contrarily, the proposed methodology applies the non-uniform quantizer in each Detail component that maps all the coefficients into 4 and 8 quantized values using the variable quantization step sizes; see Figs. 3-5 (g,k,o) and Figs. 3-5 (h,l,p), respectively. The contribution of far away coefficients is low because they are extremely few in number. It also shows the capability and adaptability of the non-uniform quantizer to exploit the available statistics in choosing the appropriate step sizes. In this case, more importance is given to the coefficients lying in the vicinity of zero. However, in general, the degree of skewness should determine the number of quantization step sizes. Objective and subjective results for the same are discussed later in section 4.2 and 4.3. 

\begin{figure}[ht!]
\begin{minipage}[b]{0.24\linewidth}
  \centering
  \centerline{\includegraphics[width=\linewidth]{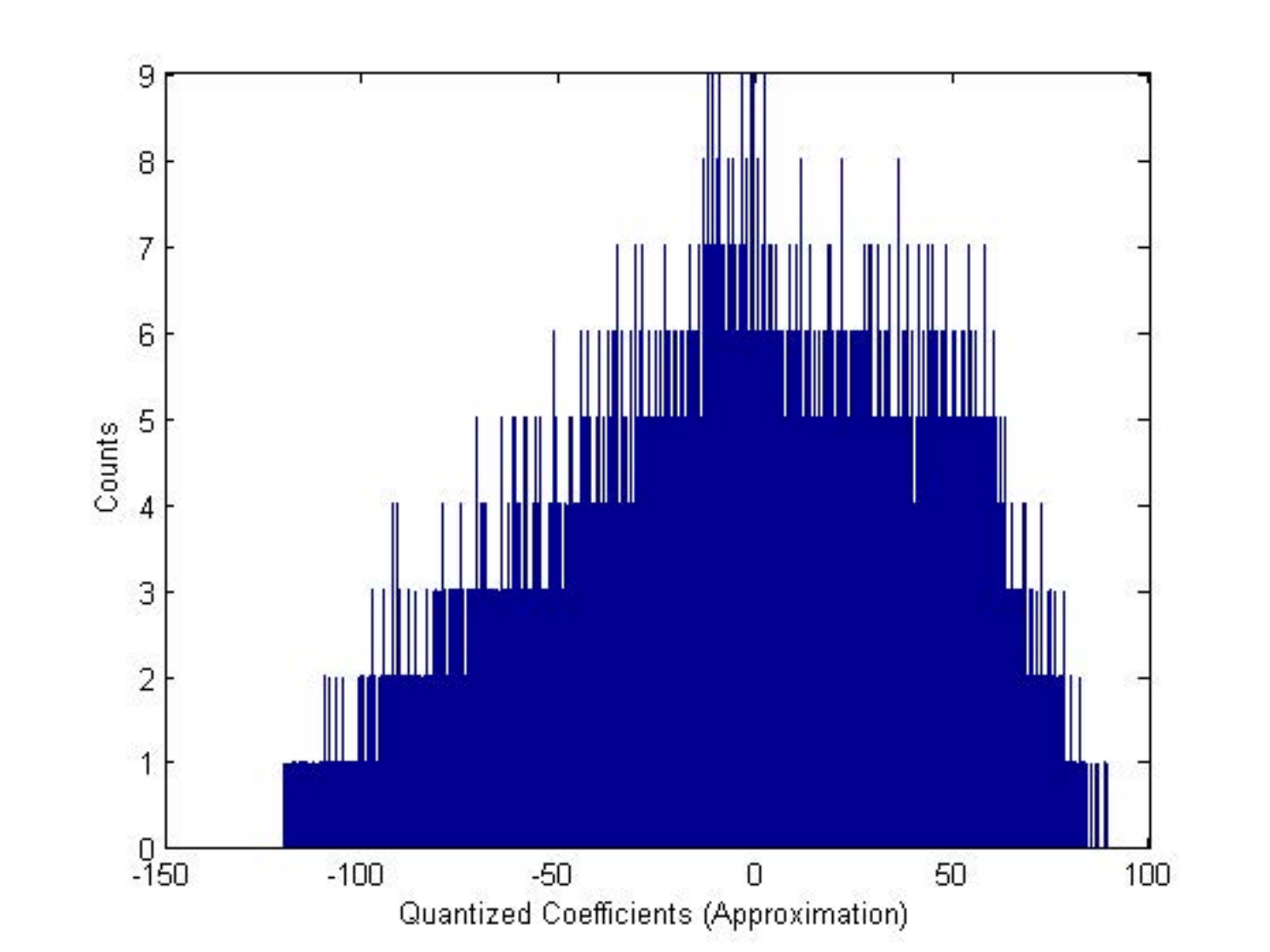}}
  \centerline{(a) Approximation}\medskip
\end{minipage}
\begin{minipage}[b]{0.24\linewidth}
  \centering
  \centerline{\includegraphics[width=\linewidth]{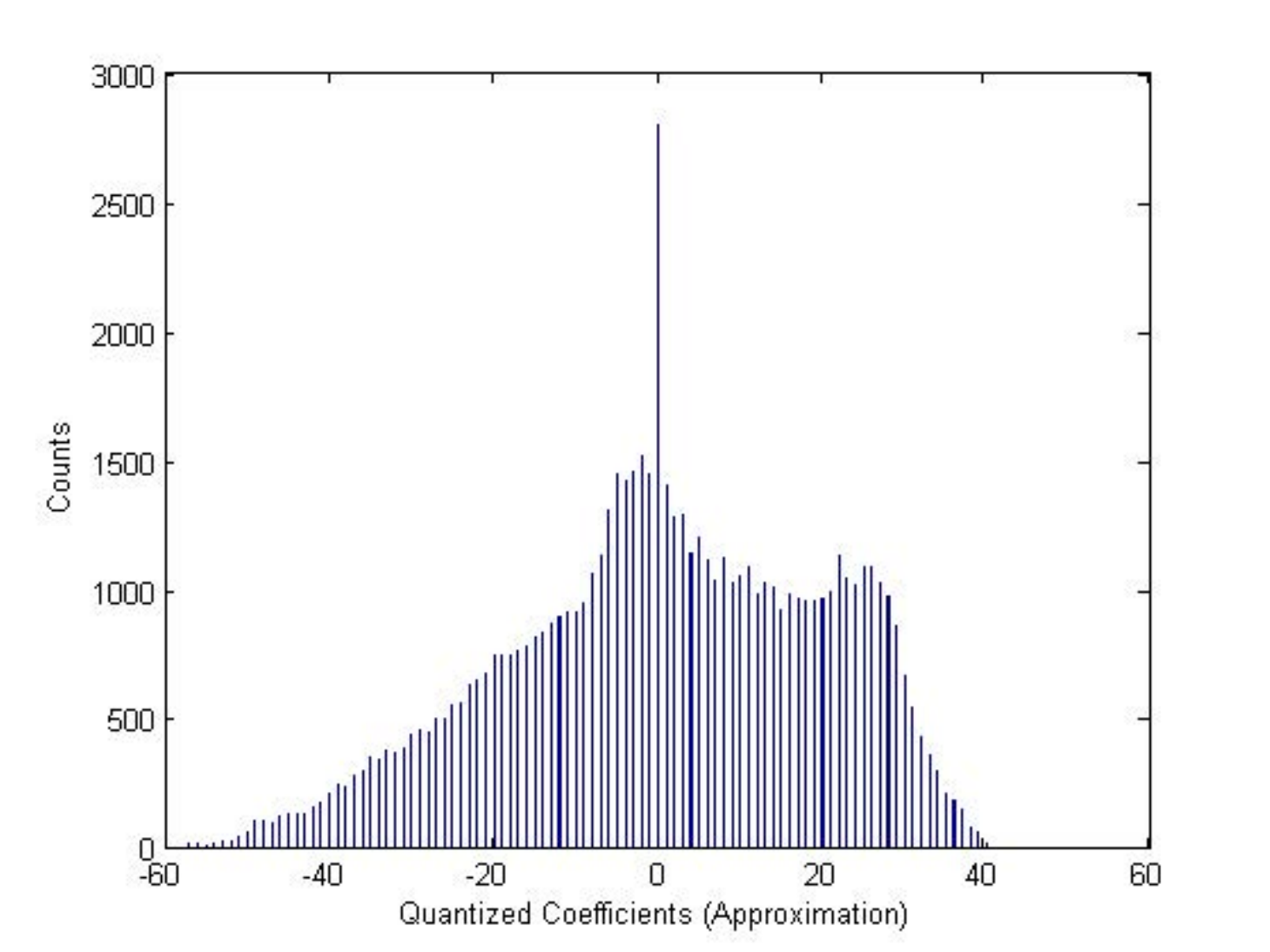}}
  \centerline{(b) UQ, $N_A=104$}\medskip
\end{minipage}
\begin{minipage}[b]{0.24\linewidth}
  \centering
  \centerline{\includegraphics[width=\linewidth]{Approximation_Baboon_e7m8.pdf}}
  \centerline{(c) UQ, $N_A=104$}\medskip
\end{minipage}
\begin{minipage}[b]{0.24\linewidth}
  \centering
  \centerline{\includegraphics[width=\linewidth]{Approximation_Baboon_e7m8.pdf}}
  \centerline{(d) UQ, $N_A=104$}\medskip
\end{minipage}

\begin{minipage}[b]{0.24\linewidth}
  \centering
  \centerline{\includegraphics[width=\linewidth]{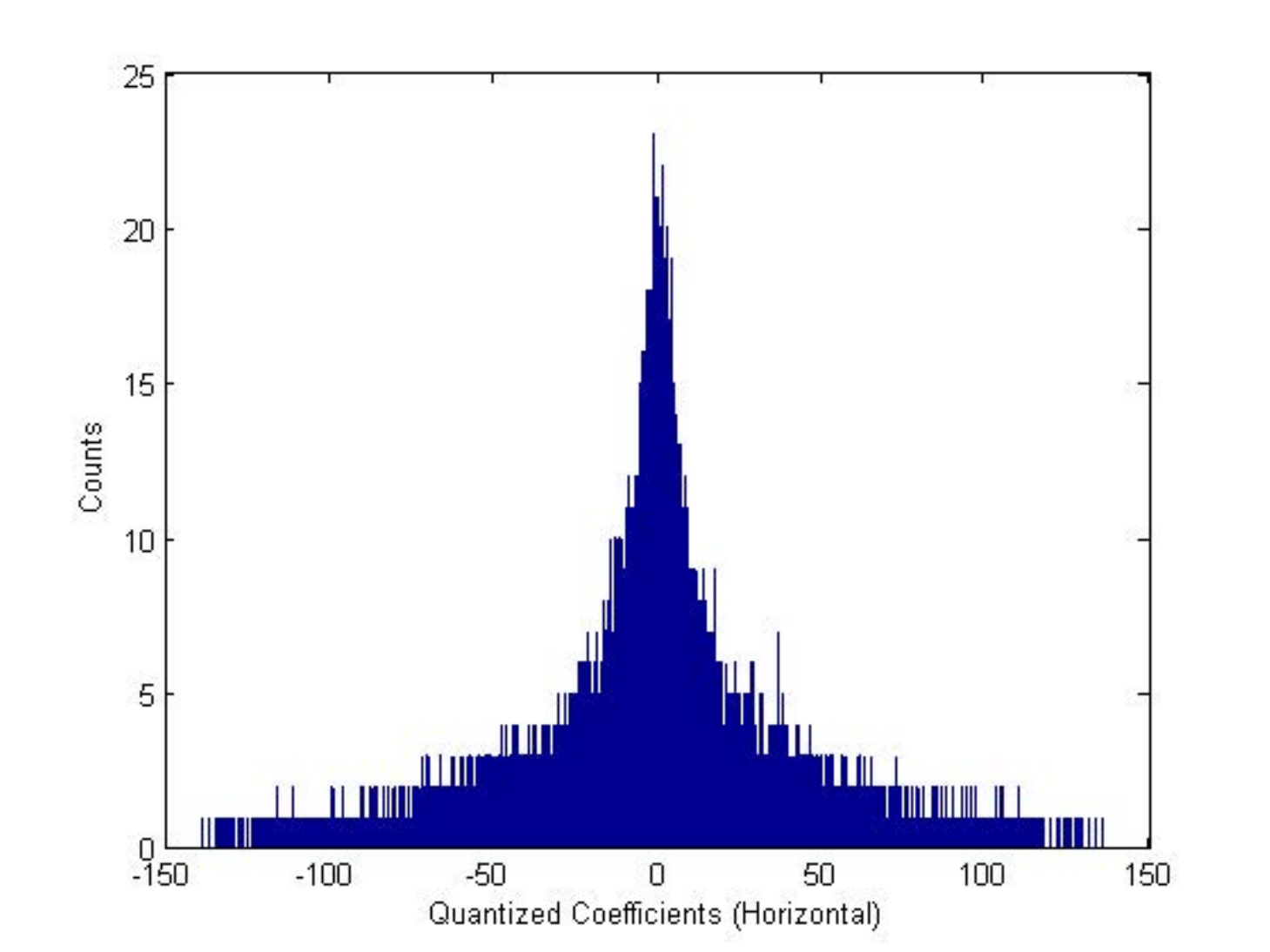}}
  \centerline{(e) Horizontal}\medskip
\end{minipage}
\begin{minipage}[b]{0.24\linewidth}
  \centering
  \centerline{\includegraphics[width=\linewidth]{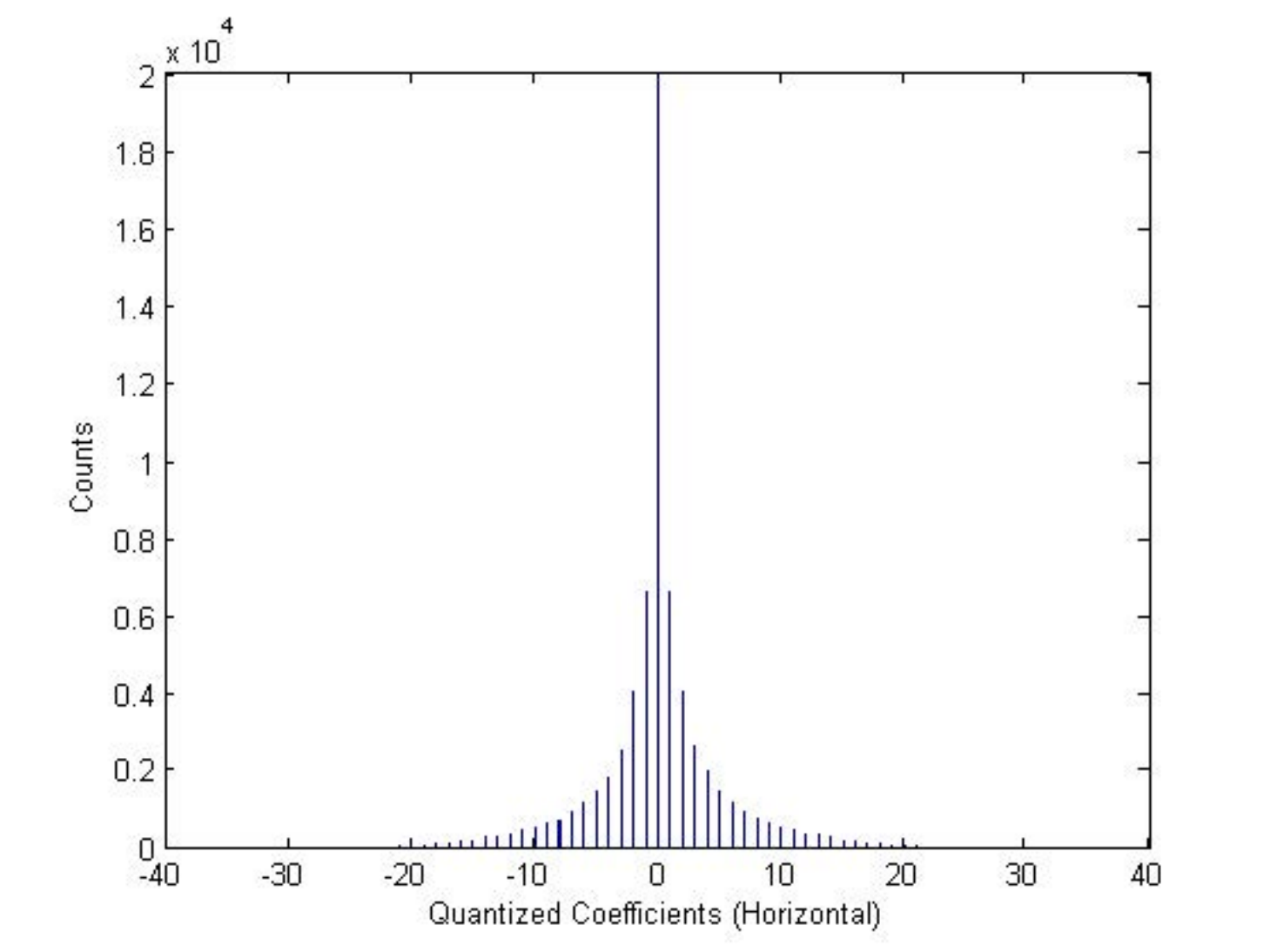}}
  \centerline{(f) UQ, $N_H=68$}\medskip
\end{minipage}
\begin{minipage}[b]{0.24\linewidth}
  \centering
  \centerline{\includegraphics[width=\linewidth]{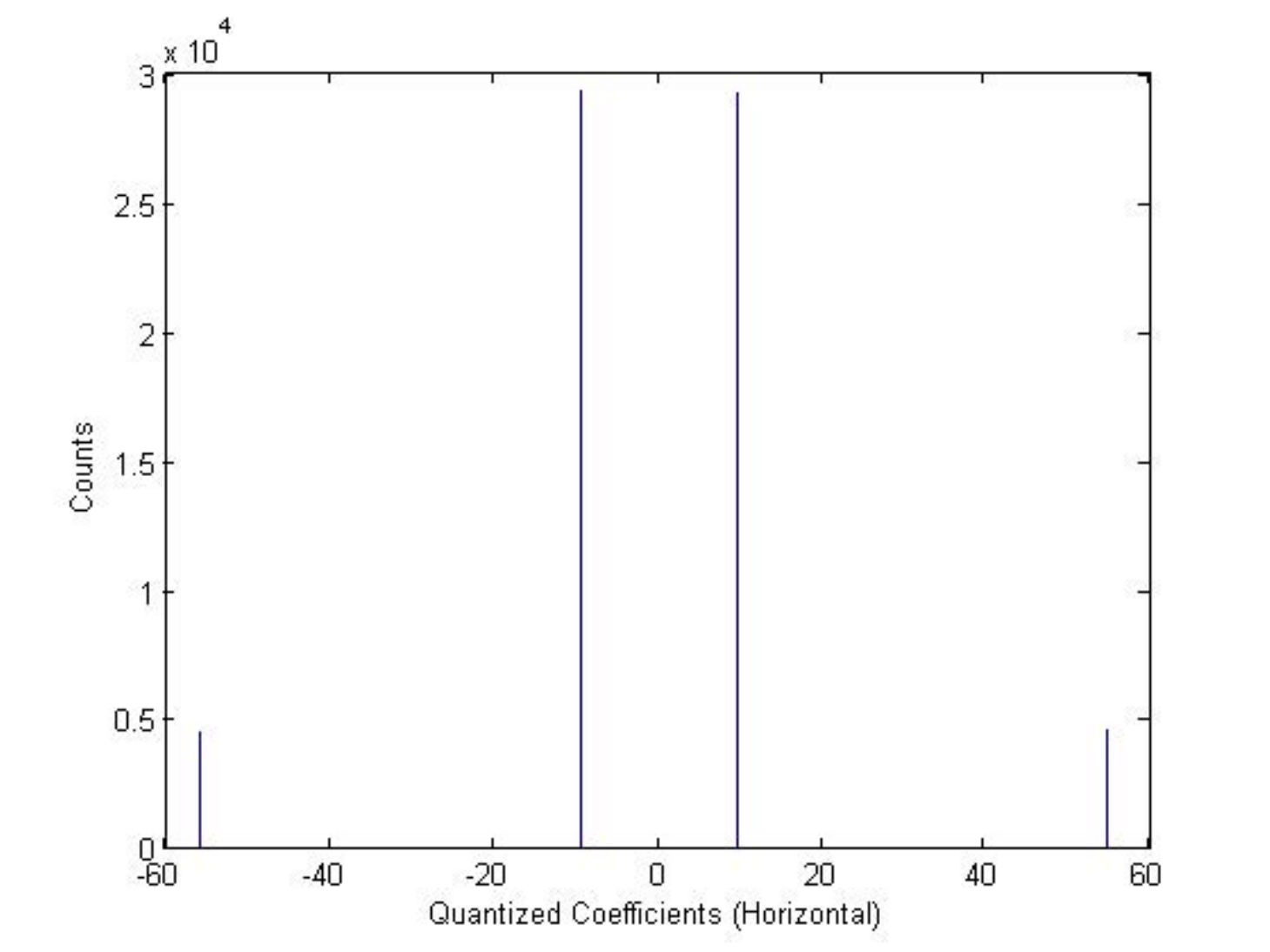}}
  \centerline{(g) NUQ, $N_H=4$}\medskip
\end{minipage}
\begin{minipage}[b]{0.24\linewidth}
  \centering
  \centerline{\includegraphics[width=\linewidth]{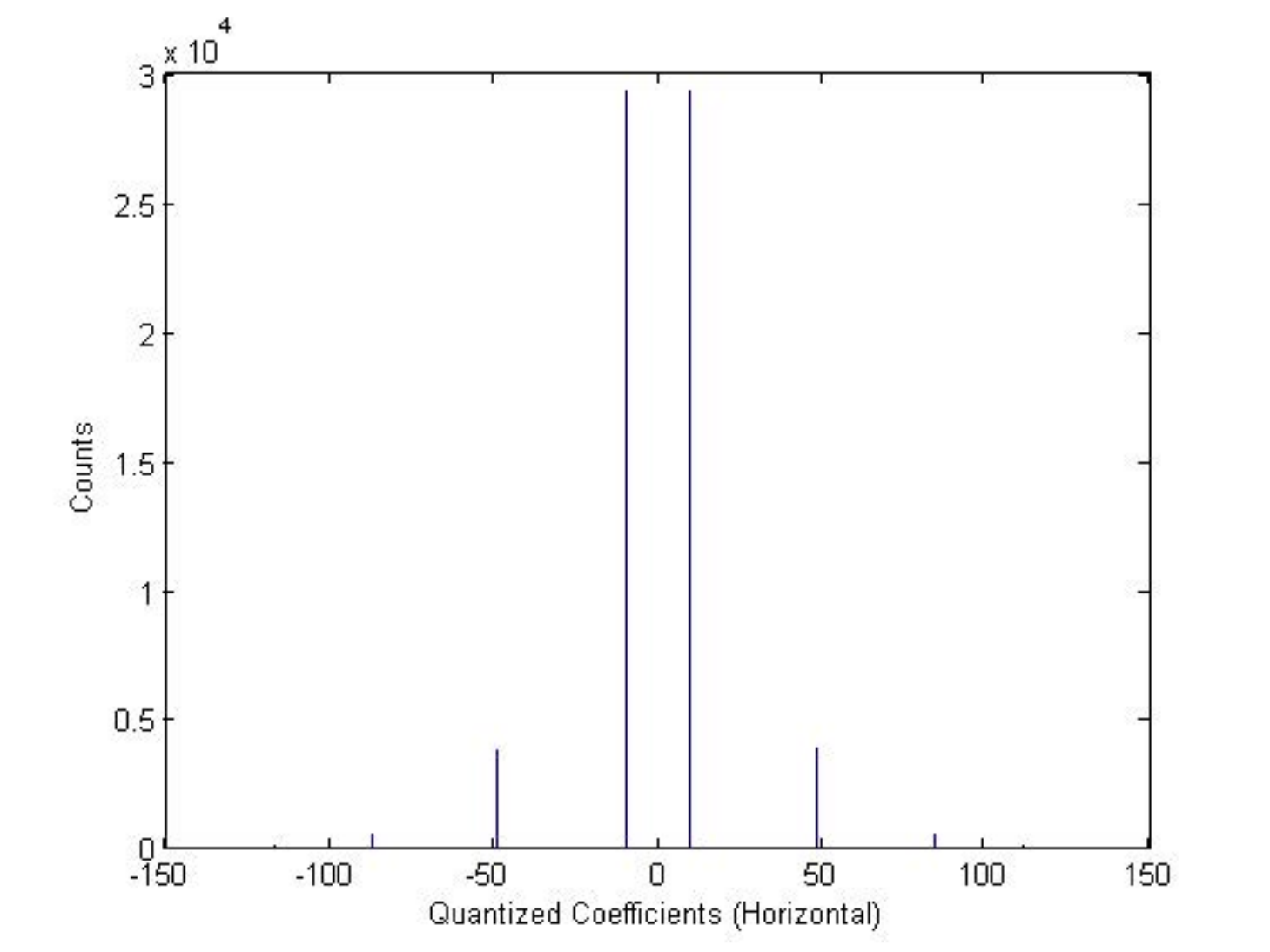}}
  \centerline{(h) NUQ, $N_H=8$}\medskip
\end{minipage}

\begin{minipage}[b]{0.24\linewidth}
  \centering
  \centerline{\includegraphics[width=\linewidth]{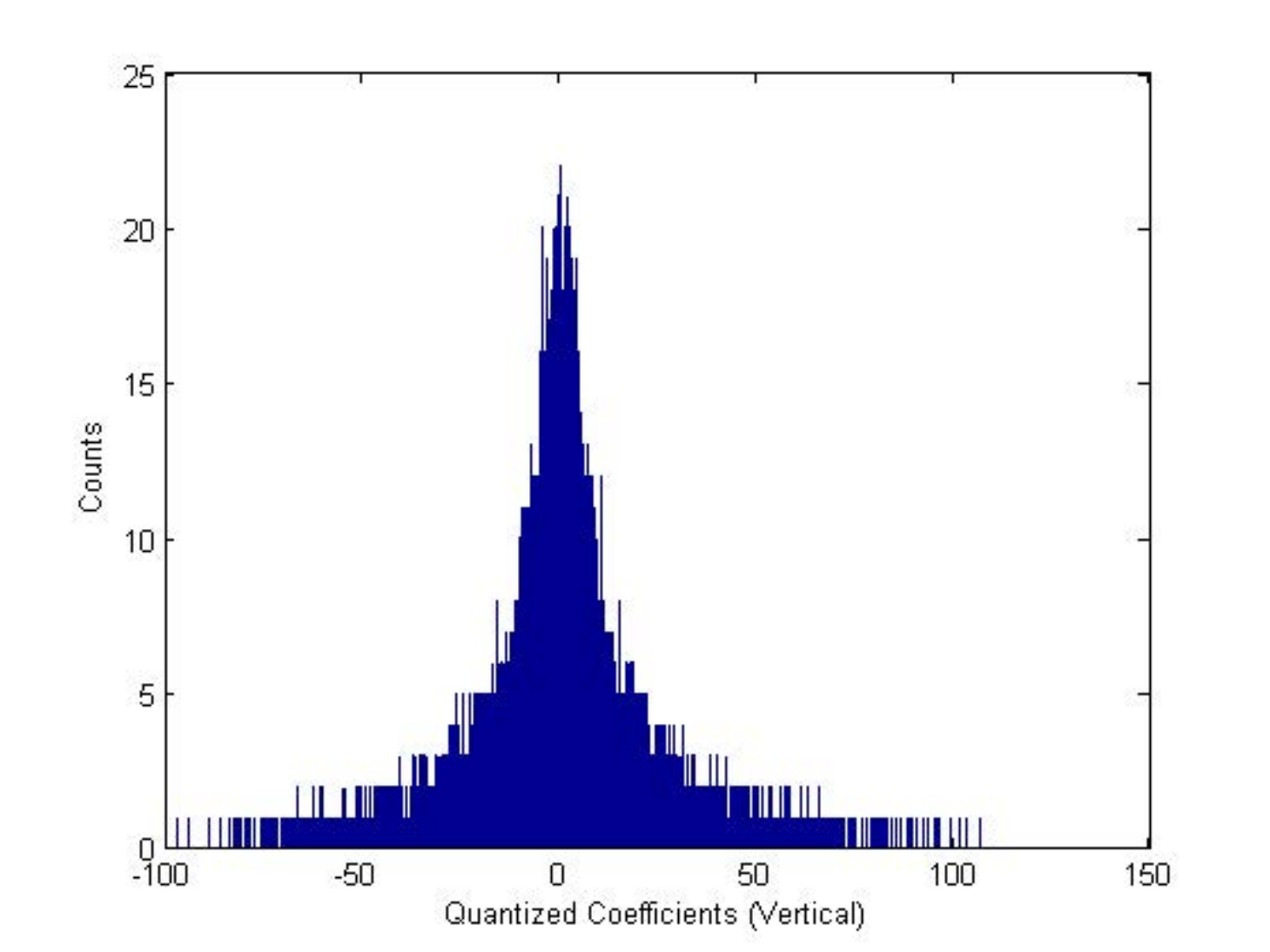}}
  \centerline{(i) Vertical}\medskip
\end{minipage}
\begin{minipage}[b]{0.24\linewidth}
  \centering
  \centerline{\includegraphics[width=\linewidth]{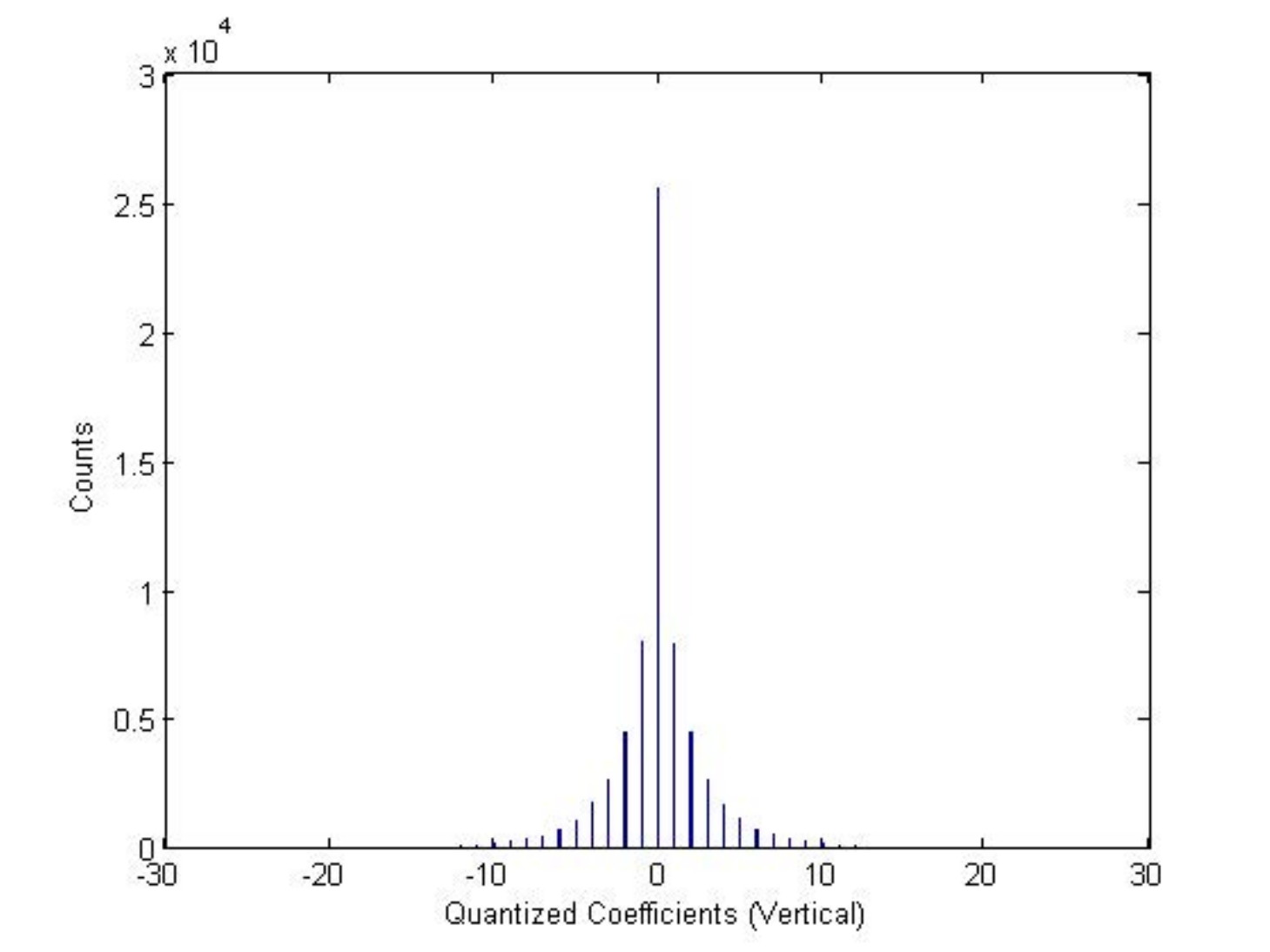}}
  \centerline{(j) UQ, $N_V=51$}\medskip
\end{minipage}
\begin{minipage}[b]{0.24\linewidth}
  \centering
  \centerline{\includegraphics[width=\linewidth]{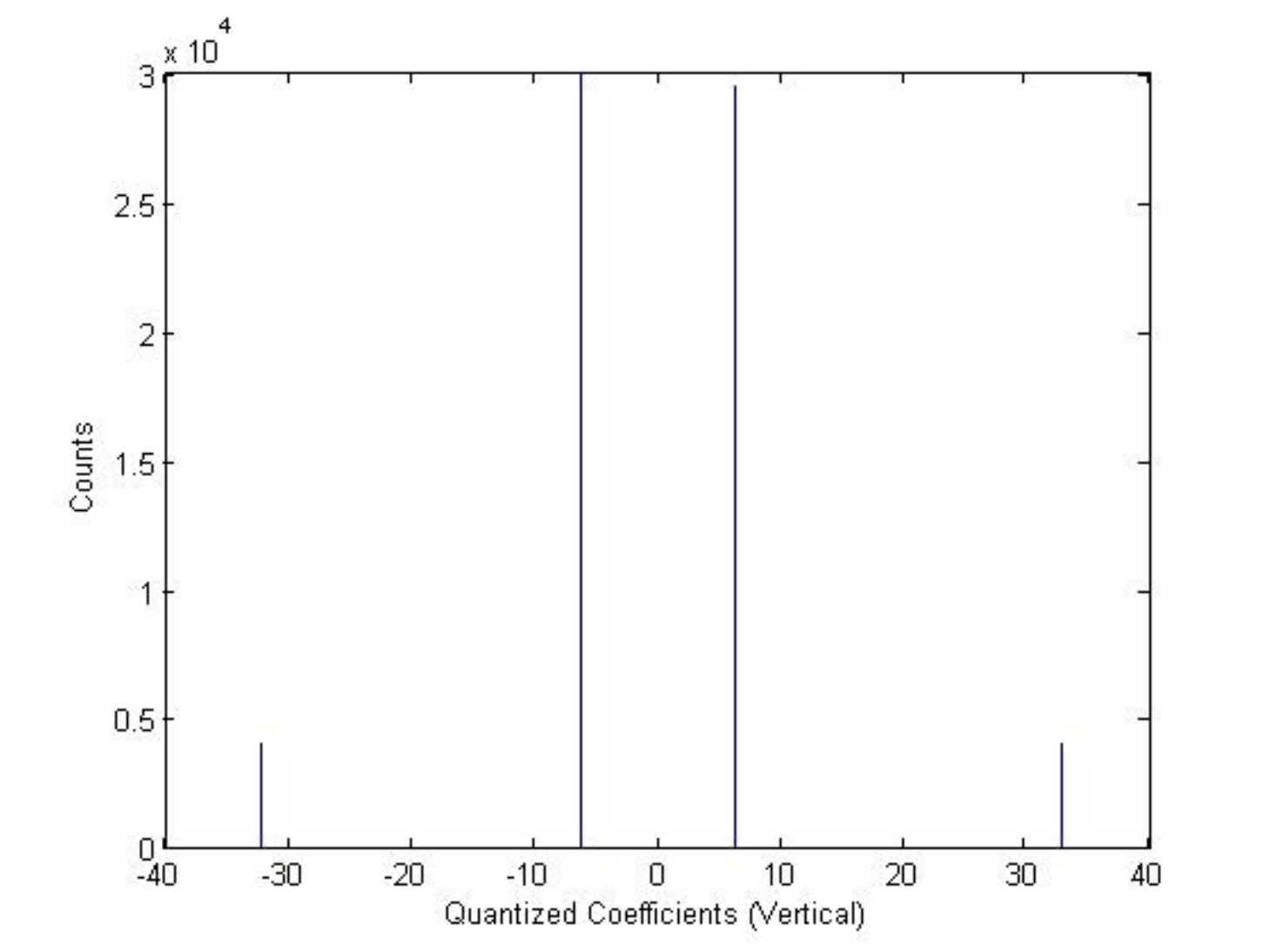}}
  \centerline{(k) NUQ, $N_V=4$}\medskip
\end{minipage}
\begin{minipage}[b]{0.24\linewidth}
  \centering
  \centerline{\includegraphics[width=\linewidth]{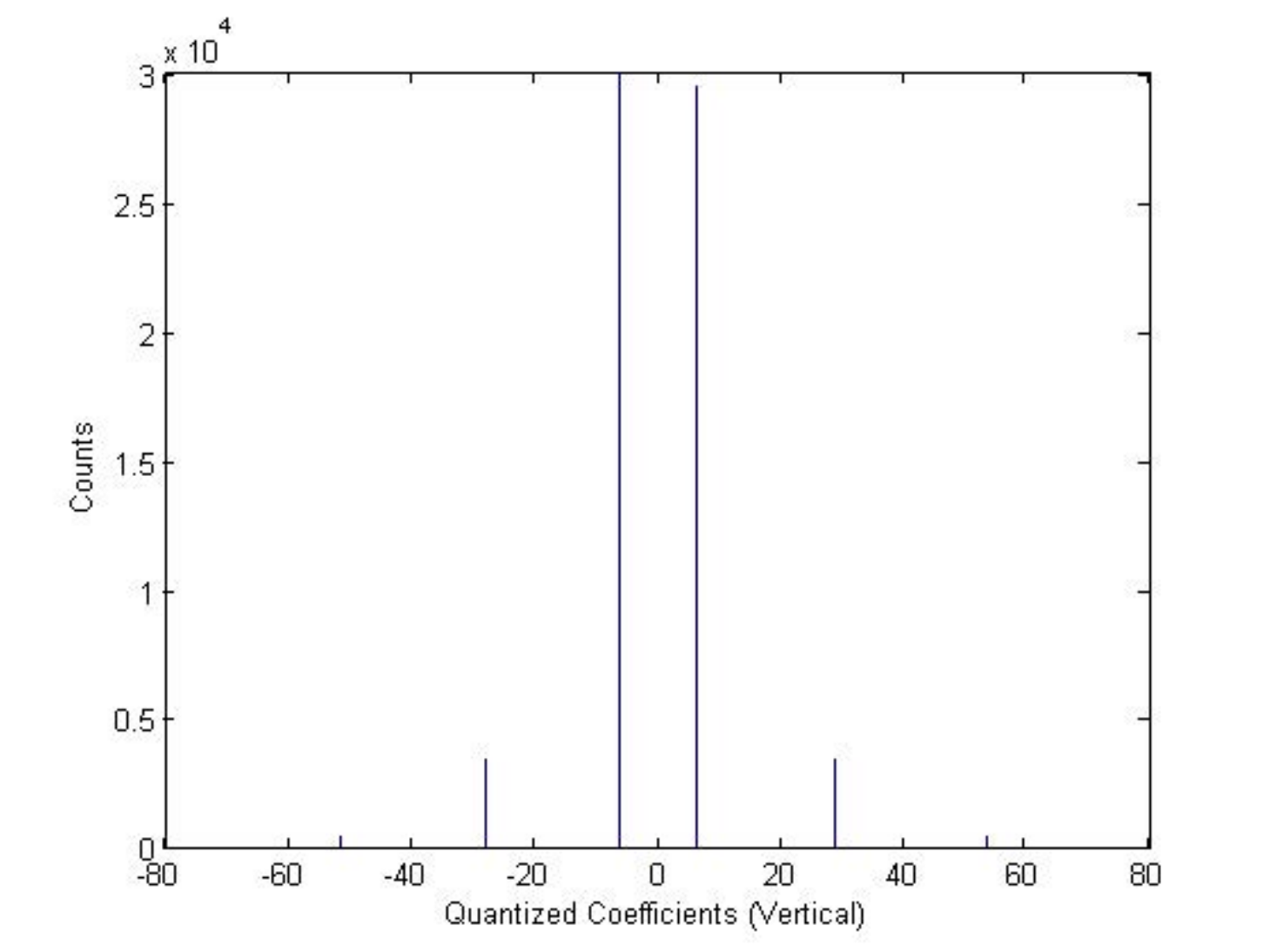}}
  \centerline{(l) NUQ, $N_V=8$}\medskip
\end{minipage}

\begin{minipage}[b]{0.24\linewidth}
  \centering
  \centerline{\includegraphics[width=\linewidth]{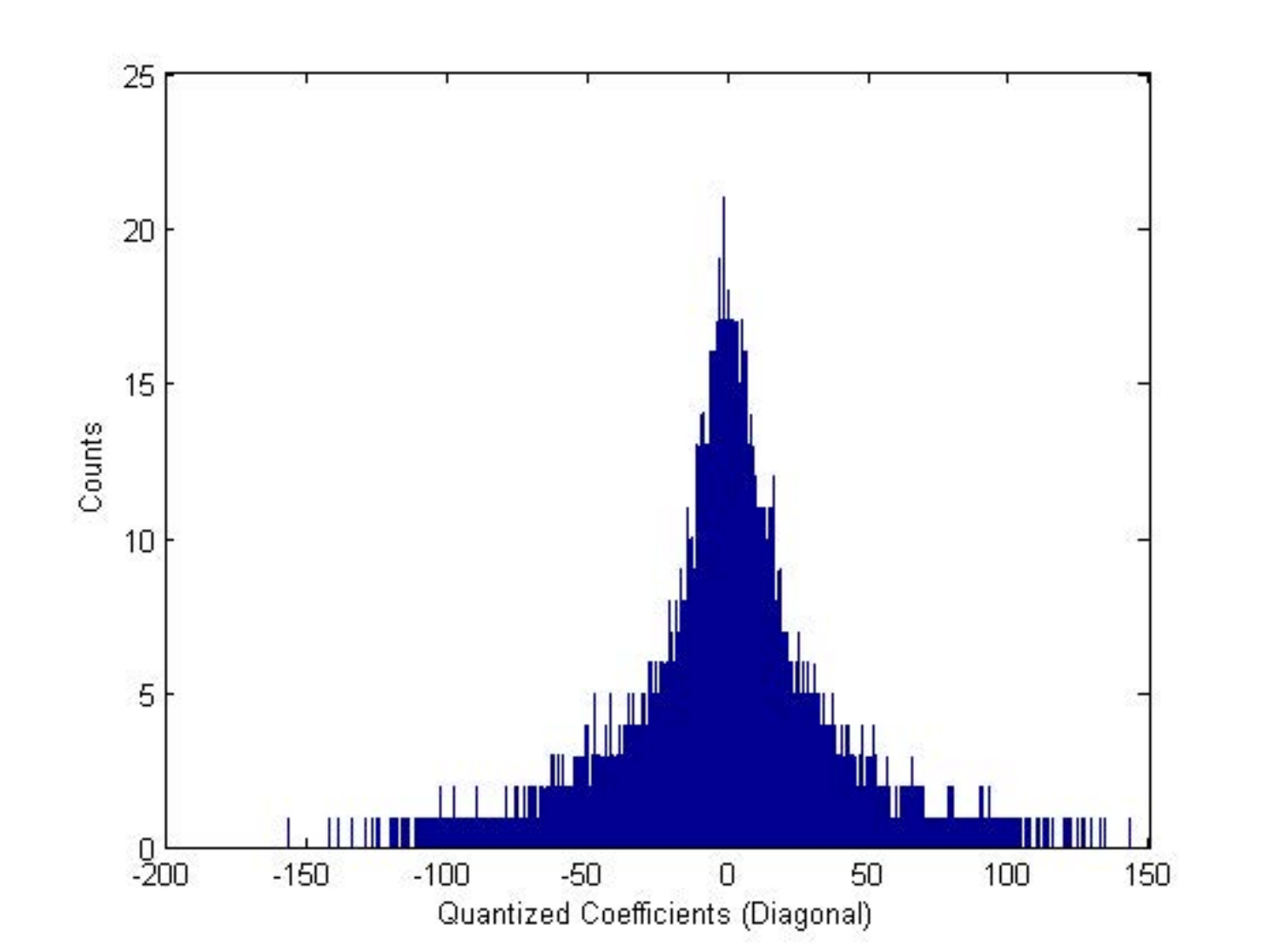}}
  \centerline{(m) Diagonal}\medskip
\end{minipage}
\begin{minipage}[b]{0.24\linewidth}
  \centering
  \centerline{\includegraphics[width=\linewidth]{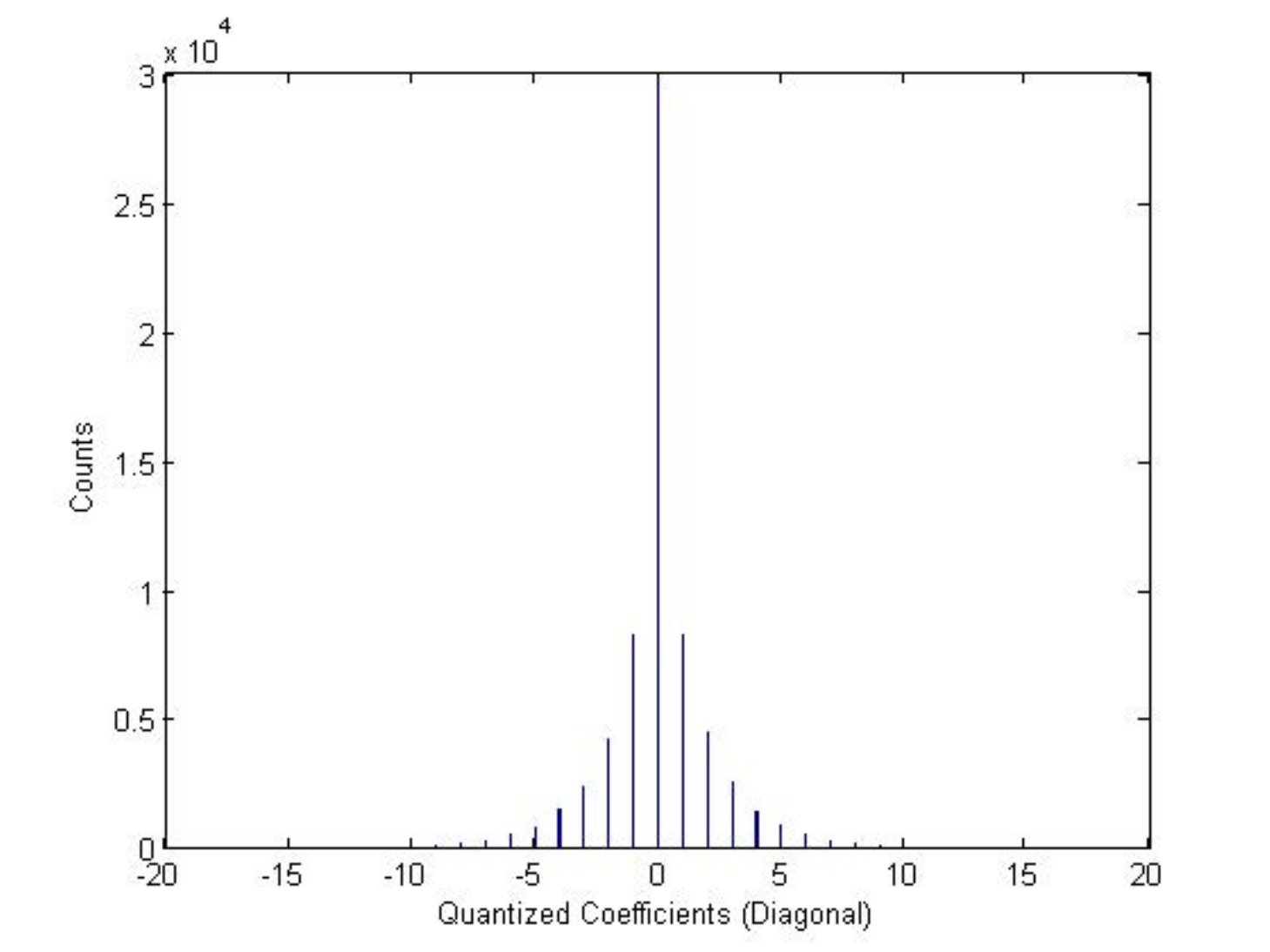}}
  \centerline{(n) UQ, $N_D=36$}\medskip
\end{minipage}
\begin{minipage}[b]{0.24\linewidth}
  \centering
  \centerline{\includegraphics[width=\linewidth]{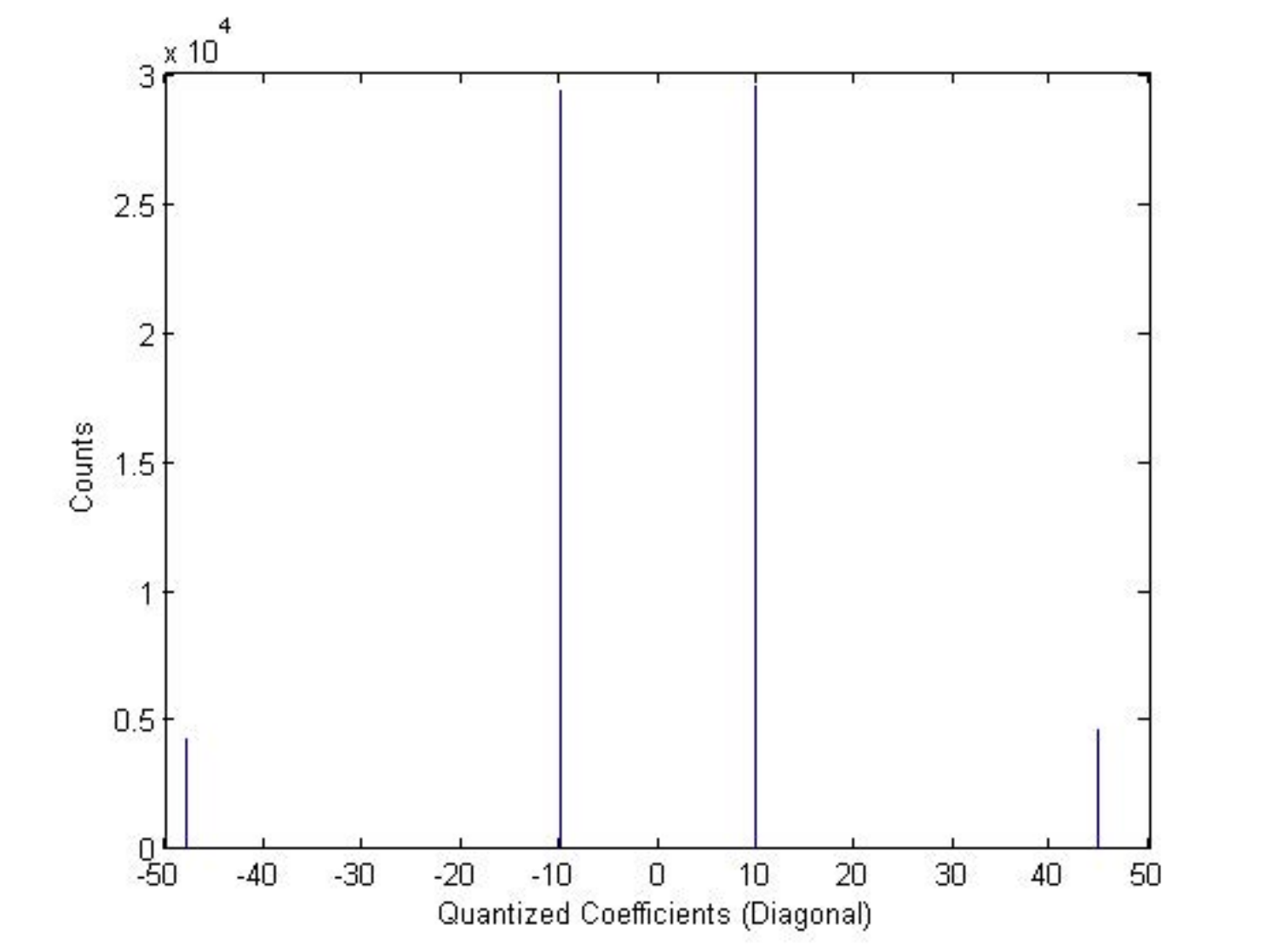}}
  \centerline{(o) NUQ, $N_D=4$}\medskip
\end{minipage}
\begin{minipage}[b]{0.24\linewidth}
  \centering
  \centerline{\includegraphics[width=\linewidth]{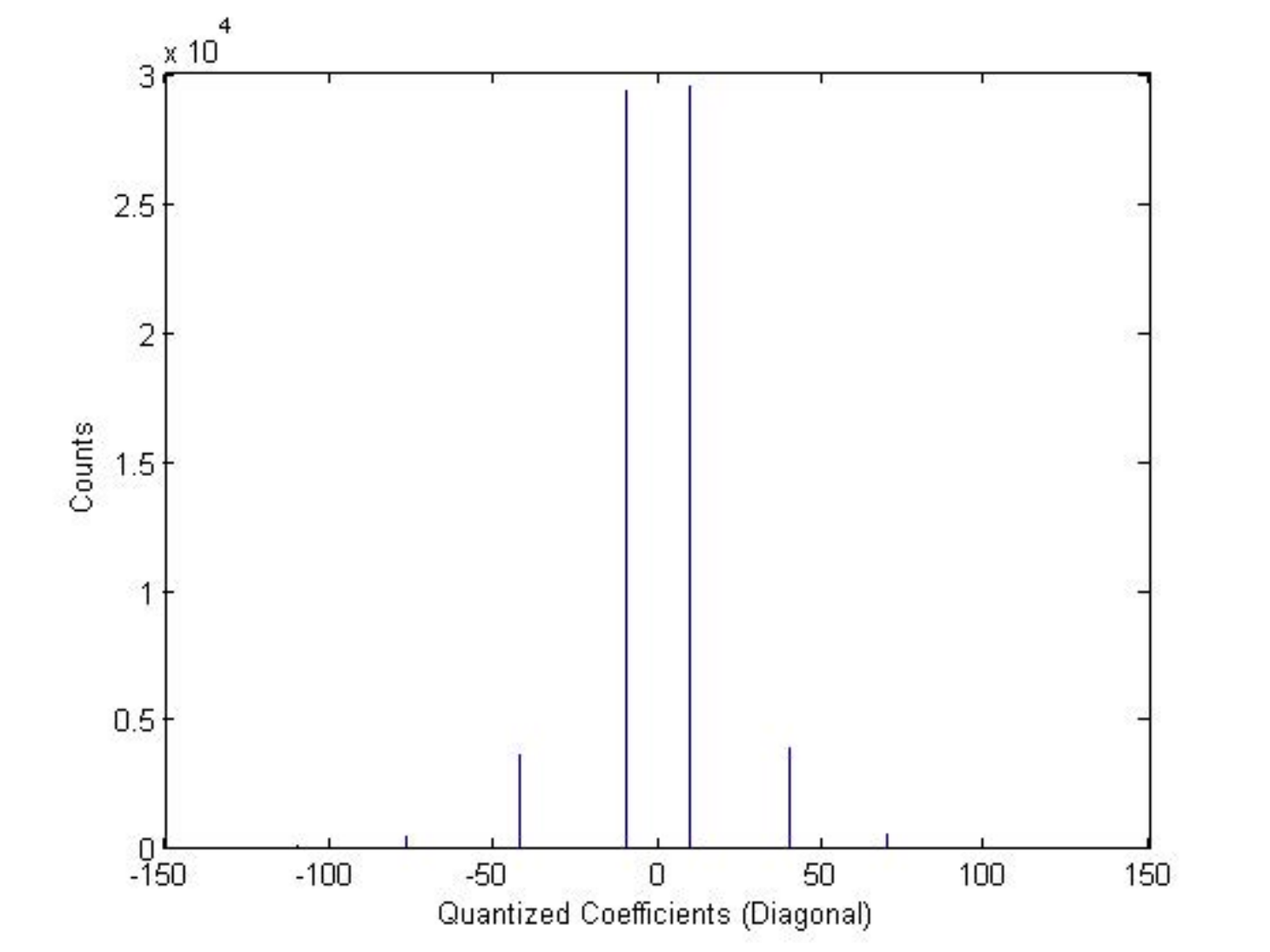}}
  \centerline{(p) NUQ, $N_D=8$}\medskip
\end{minipage}

  \caption{\emph{Baboon} ($512 \times 512$): (a,e,i,m) Histogram plots of the Approximation, Horizontal, Vertical, and Diagonal component, respectively; (b-d) Histogram plots of the Approximation component after the deadzone uniform quantization (UQ) and non-uniform quantization (NUQ); (f-h) Histogram plots of the Horizontal component after UQ and NUQ; (j-l) Histogram plots of the Vertical component after UQ and NUQ; (n-p) Histogram plots of the Diagonal component after UQ and NUQ. \textbf{\emph{N}} is the number of quantized values.}
\end{figure}

Another benefit of the non-uniform quantizer is that it can predetermine the value of $N$, the number of quantized values, which is exactly the same as the number of quantization step sizes; whereas, the deadzone uniform quantizer can only decide on the quantization step size, having no control over the number of quantized values. Moreover, the quantization step size is same for all the Detail components (and also the Approximation component) in the deadzone uniform quantizer. In contrast, the non-uniform quantizer has the flexibility of choosing different number of quantization step sizes (or the quantized values) for each Detail component. See Figs. 3-5 (f,j,n), Figs. 3-5 (g,k,o), and Figs. 3-5 (h,l,p) for comparision.

\subsection{Objective Analysis}

The deadzone uniform quantizer and non-uniform quantizer are objectively evaluated and compared using MSE and MSSIM. MSE provides with the overall quantization error for each component after the quantizer. Only the Detail components are considered for MSE comparison because the two different quantizers are applied on them. As mentioned earlier, the Approximation component is quantized by the deadzone uniform quantizer in both the approaches, and hence, is not considered for MSE results. On the other hand, MSSIM shows the performance of both the quantizers in the overall reconstructed image. MSSIM is better metric than PSNR as former incorporates HVS in measurement \cite{Wang_2004,Wang_2009}. The quantized approximation coefficient values being same, MSSIM reflects the ability of the two quantizers and the extent to which they can visually reproduce the original image. The quantization step sizes for the deadzone uniform quantizer are obtained from equation 3 with $R_1 = 8$, $\mu_1 = 8$ and $\epsilon_1 \in \{4,5,6,7,8,9\}$ that resulted in $N$ quantized values in the increasing order. For the non-uniform quantizer, $N$ values are manually inserted.

MSE and MSSIM are calculated from the following equations,

\begin{equation}
MSE=\frac{1}{L}\sum_{i=1}^{L} [X(i) - \tilde{X}(i)]^2
\end{equation}

\noindent where $L$ is the component length, $X$ is the original component, and $\tilde{X}$ is the quantized component.

\begin{equation}
MSSIM = \frac{1}{M} \displaystyle\sum_{j=1}^{M} \frac{(2\mu_{I_j}\mu_{\tilde{I}_j} + C_1)(2\sigma_{I_{j}\tilde{I}_{j}} + C_2)}{(\mu_{I_j}^2 + \mu_{\tilde{I}_j}^2 + C_1)(\sigma_{I_j}^2 + \sigma_{\tilde{I}_j}^2 + C_2)}
\end{equation}

\noindent where $M$ is the number of local windows of the image, $j$ is the window number, $I$ is the original image, $\tilde{I}$ is the reconstructed image after quantization, $\mu$ is the mean, $\sigma$ is the standard deviation, and $C_1$ and $C_2$ are the arbritary constants to avoid unstable results when $(\mu_{I_j}^2 + \mu_{\tilde{I}_j}^2)$ and $(\sigma_{I_j}^2 + \sigma_{\tilde{I}_j}^2)$ are near zero. For the calculations, $M=1$ (MSSIM of the entire image is considered without windows), $C_1=4$, and $C_2=0.5$.

\begin{table}[ht!]
\begin{center}
\begin{tabular}{|c||c|c|c|c||c|c|c|c||c|c|c|c||}
\hline
Image & \multicolumn{4}{|c||}{Horizontal} & \multicolumn{4}{|c||}{Vertical} & \multicolumn{4}{|c||}{Diagonal}\\
\cline{2-13}
Name & \multicolumn{2}{|c|}{Uniform} & \multicolumn{2}{|c||}{Non-uniform} & \multicolumn{2}{|c|}{Uniform} & \multicolumn{2}{|c||}{Non-uniform} & \multicolumn{2}{|c|}{Uniform} & \multicolumn{2}{|c||}{Non-uniform} \\
\cline{2-13}
 & $N_H$ & MSE & $N_H$ & MSE & $N_V$ & MSE & $N_V$ & MSE & $N_D$ & MSE & $N_D$ & MSE\\
\hline
Lenna & 3 & 16.84 & 4 & 4.05 & 4 & 41.86 & 4 & 10.57 & 2 & 34.07 & 4 & 7.29 \\
($512 \times 512$)	& 6 &	16.46 & 6 & 2.66 & 8 & 39.95 & 6 & 6.07 & 4 & 33.97 & 6 & 5.27 \\ 
	& 12 & 15.12 & 8 & 2.41 & 17 & 35.68 & 8 & 5.55 & 8 & 33.10 & 8 & 4.90 \\
	& 23 & 11.74 & 10 & 2.39 & 34 & 26.96 & 10 & 5.51 & 15 & 29.83 & 10 & 4.84 \\
	& 44 & 5.88 & 12 & 2.38 & 65 & 12.87 & 12 & 5.50 & 29 & 22.58 & 12 & 4.84 \\
	&  84 & 0.30 & 14 & 2.38 & 125 & 0.30 & 14 & 5.50 & 53 & 10.95 & 14 & 4.84 \\
& & & & & & & & & & & &\\
Pepper & 7 & 55.90 & 4 & 22.02 & 7 & 62.08 & 4 & 25.22 & 3 & 93.73 & 4 & 16.24 \\
($512 \times 512$)	& 14 & 53.21 & 6 & 8.91 & 14 & 59.09 & 6 & 10.45 & 5 & 93.38 & 6 & 13.54 \\
	& 27 & 47.32 & 8 & 7.55 & 27 & 52.45 & 8 & 8.86 & 11 & 89.50 & 8 & 12.92 \\
	& 52 & 35.53 & 10 & 7.53 & 53 & 39.30 & 10 & 8.84 & 20 & 78.65 & 10 & 12.78 \\
	& 103 & 16.71 & 12 & 7.53 & 105 & 18.43 & 12 & 8.84 & 37 & 58.79 & 12 & 12.76 \\
	& 194 & 0.31 & 14 & 7.53 & 204 & 0.31 & 14 & 8.84 & 68 & 27.71 & 14 & 12.76 \\
& & & & & & & & & & & &\\
Baboon & 9 & 574.24 & 4 & 108.91 & 7 & 196.88 & 4 & 39.80 & 5 & 445.61 & 4 & 84.56 \\
($512 \times 512$)	& 17 & 539.11 & 6 & 79.09 & 13 & 185.78 & 6 & 27.97 & 9 & 434.24 & 6 & 62.54 \\
	& 34 & 471.13 & 8 & 76.74 & 26 & 162.80 & 8 & 26.72 & 18 & 408.04 & 8 & 60.00 \\
	& 68 & 348.72 & 10 & 76.56 & 51 & 120.99 & 10 & 26.61 & 36 & 356.72 & 10 & 59.74 \\
	& 136 & 158.78 & 12 & 76.56 & 97 & 55.87 & 12 & 26.60 & 70 & 264.27 & 12 & 59.73 \\
	& 264 & 0.32 & 14 & 76.56 & 185 & 0.32 & 14 & 26.60 & 132 & 120.85 & 14 & 59.73 \\
\hline
\end{tabular}
\end{center}
\caption{Comparison of MSE at various quantization levels by the uniform and proposed non-uniform quantizer for the Detail components of the three test images.}
\end{table}

Table 1 displays MSE values obtained from the two quantizers with $N$ quantized values for each Detail component of all the three test images. It can be seen that the non-uniform quantizer outperforms the deadzone uniform quantizer in terms of MSE values. In general, MSE of the non-uniform quantizer is substantially less than the deadzone uniform quantizer, except in some cases when $N$ is very high for the deadzone uniform quantizer. For same $N$, MSE of the deadzone uniform quantizer is many times higher than the non-uniform quantizer. Converse to the previous point, for a particular MSE value, the deadzone uniform quantizer requires far more quantized values than the non-uniform quantizer. In addition, MSE values of the non-uniform quantizer starts saturating at $N=8$, whereas for the deadzone uniform quantizer, it saturates at very high $N$ values ranging from 84 to 264. 

\begin{figure}[th!]
\begin{minipage}[b]{0.5\linewidth}
  \centering
  \centerline{\includegraphics[width=\linewidth]{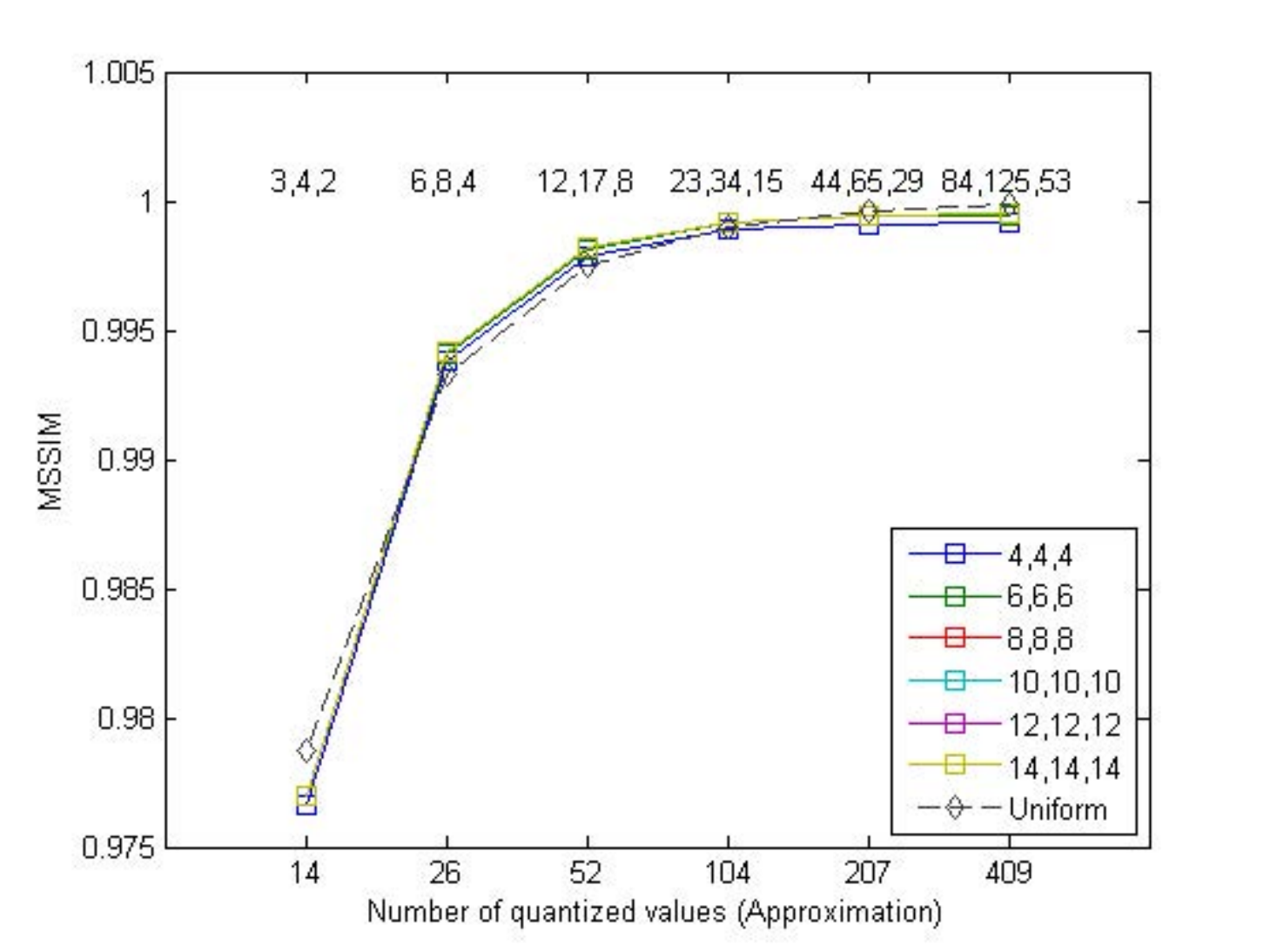}}
  \centerline{(a) Lenna}\medskip
\end{minipage}
\begin{minipage}[b]{0.5\linewidth}
  \centering
  \centerline{\includegraphics[width=\linewidth]{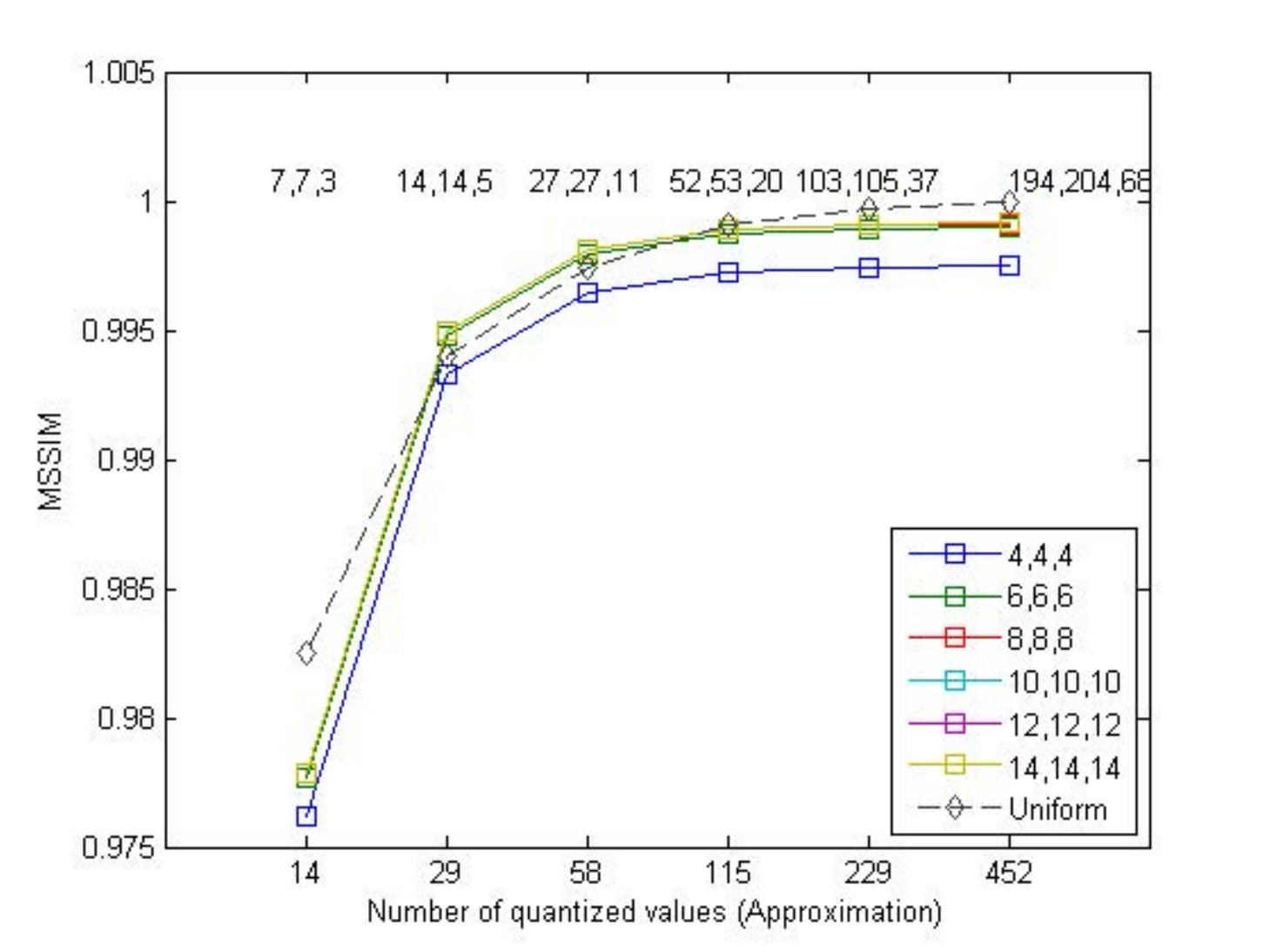}}
  \centerline{(b) Pepper}\medskip
\end{minipage}

\hspace{0.25\linewidth}
\begin{minipage}[b]{0.5\linewidth}
  \centering
  \centerline{\includegraphics[width=\linewidth]{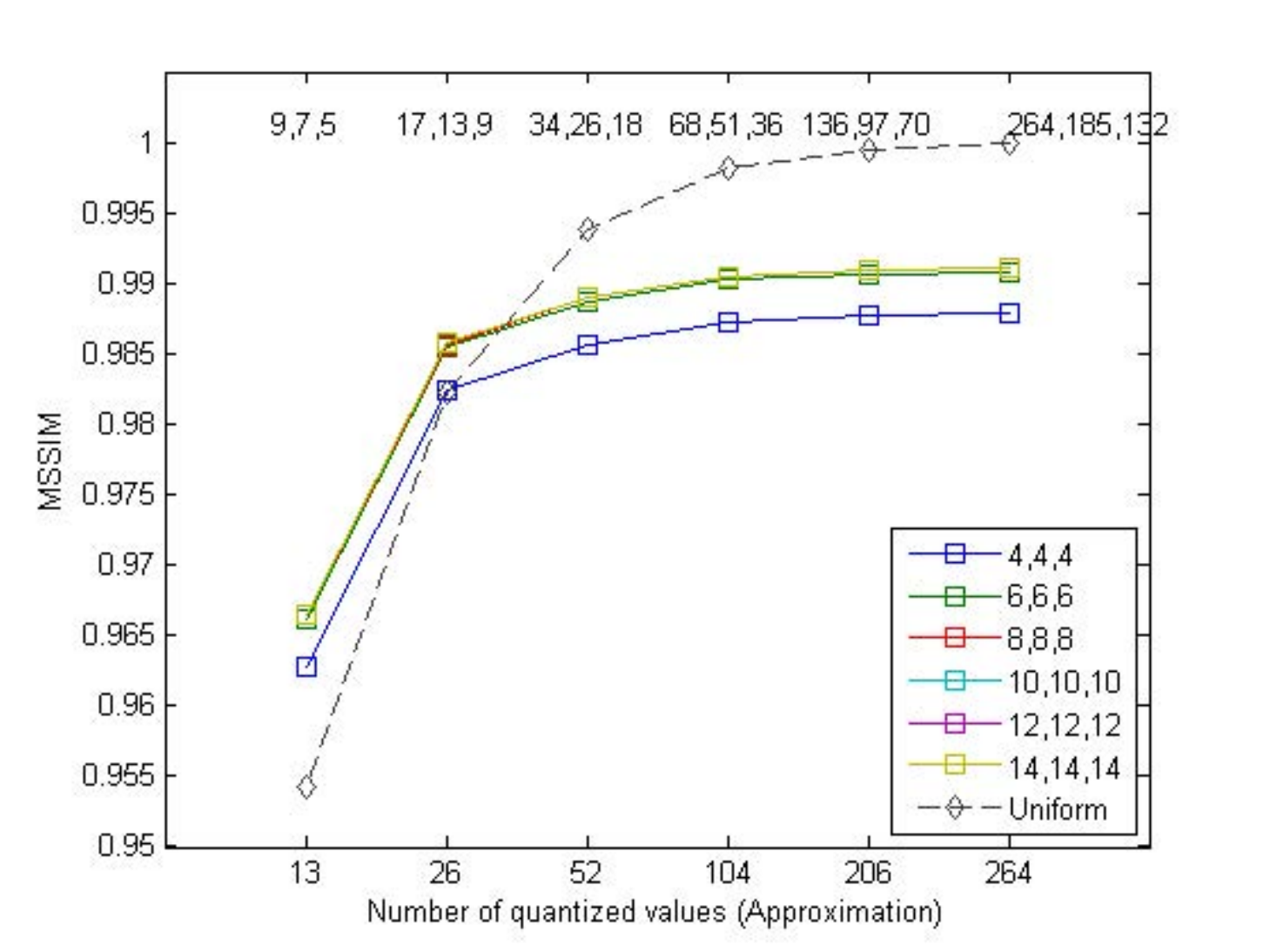}}
  \centerline{(c) Baboon}\medskip
\end{minipage}
 \caption{MSSIM of the test images at different quantized values of the Approximation, Horizontal, Vertical, and Diagonal components, using the deadzone uniform and proposed non-uniform quantizer. The x-axis is the number of quantized values in the Approximation component, while y-axis shows MSSIM value. The legends show the number of quantized values $N_H$, $N_V$, and $N_D$ in the Horizontal, Vertical, and Diagonal component, respectively, for the non-uniform quantizer. $N_H$, $N_V$, and $N_D$ for the deadzone uniform quantizer are provided on the top of the plot at each point.}
\end{figure}

\begin{figure}[ht!]
\begin{minipage}[b]{\linewidth}
  \centering
  \centerline{\includegraphics[width=0.24\linewidth]{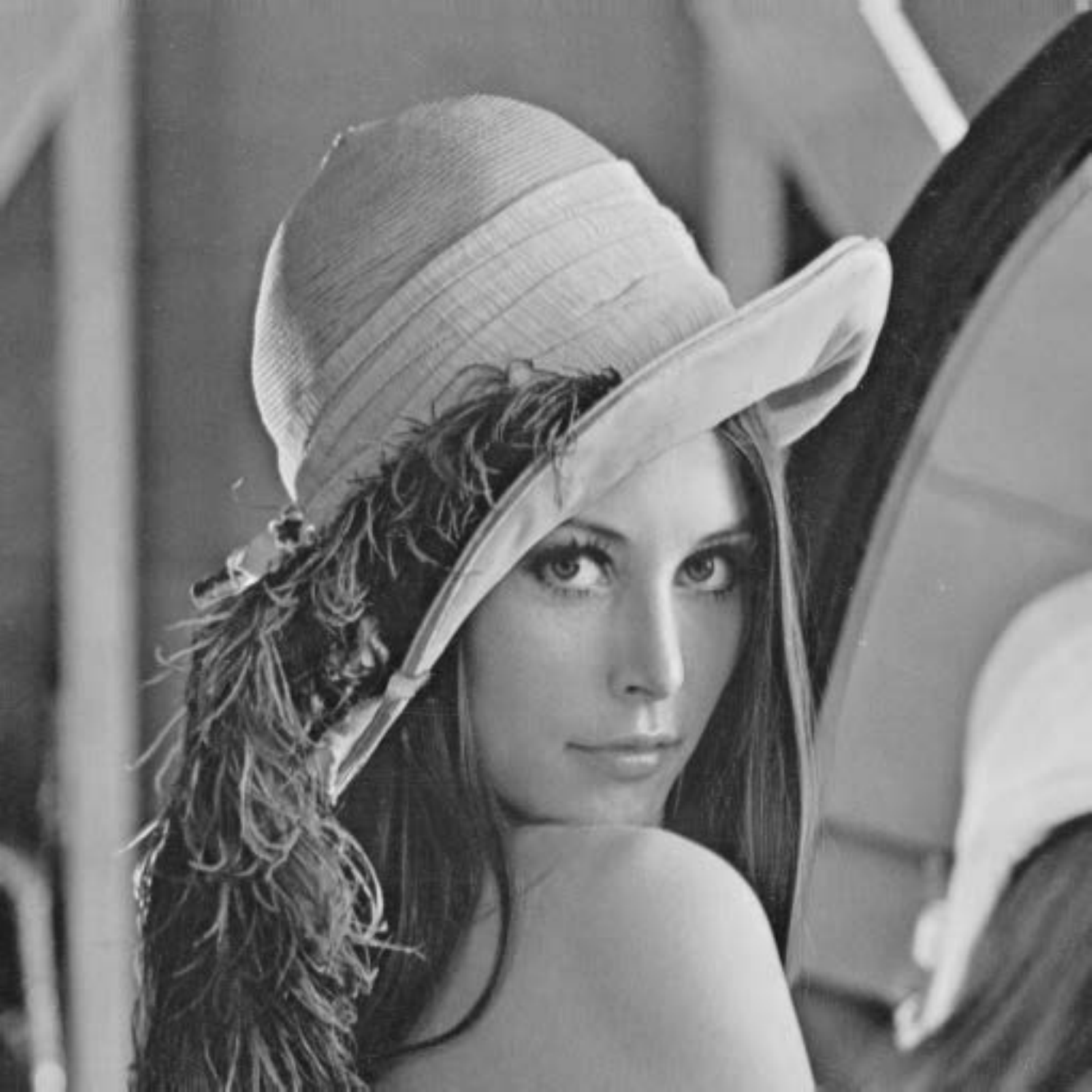}}
  \centerline{(a) Original}\medskip
\end{minipage}

\begin{minipage}[b]{0.24\linewidth}
  \centering
  \centerline{\includegraphics[width=\linewidth]{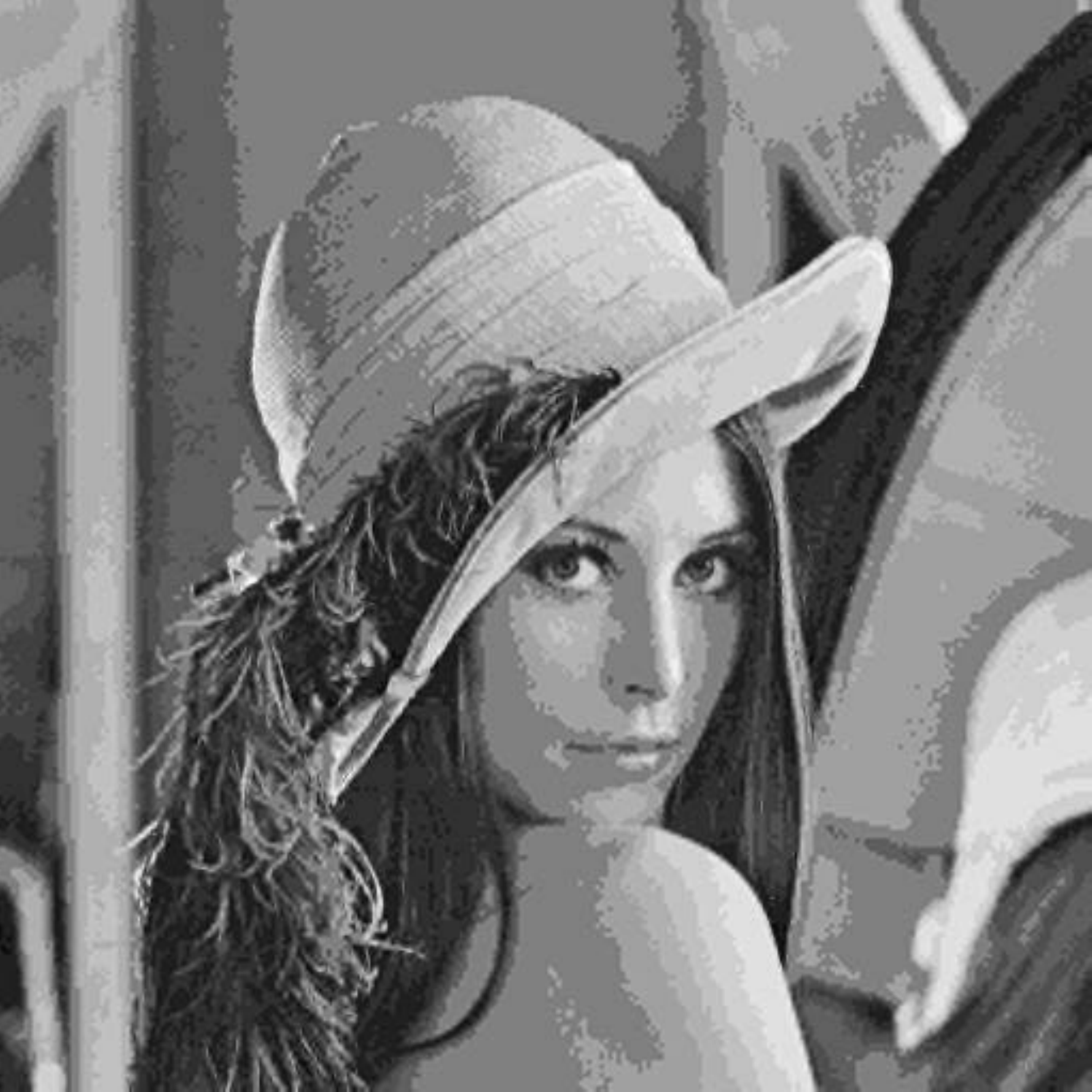}}
  \centerline{(b)$UQ:(14,3,4,2)$}\medskip
\end{minipage}
\begin{minipage}[b]{0.24\linewidth}
  \centering
  \centerline{\includegraphics[width=\linewidth]{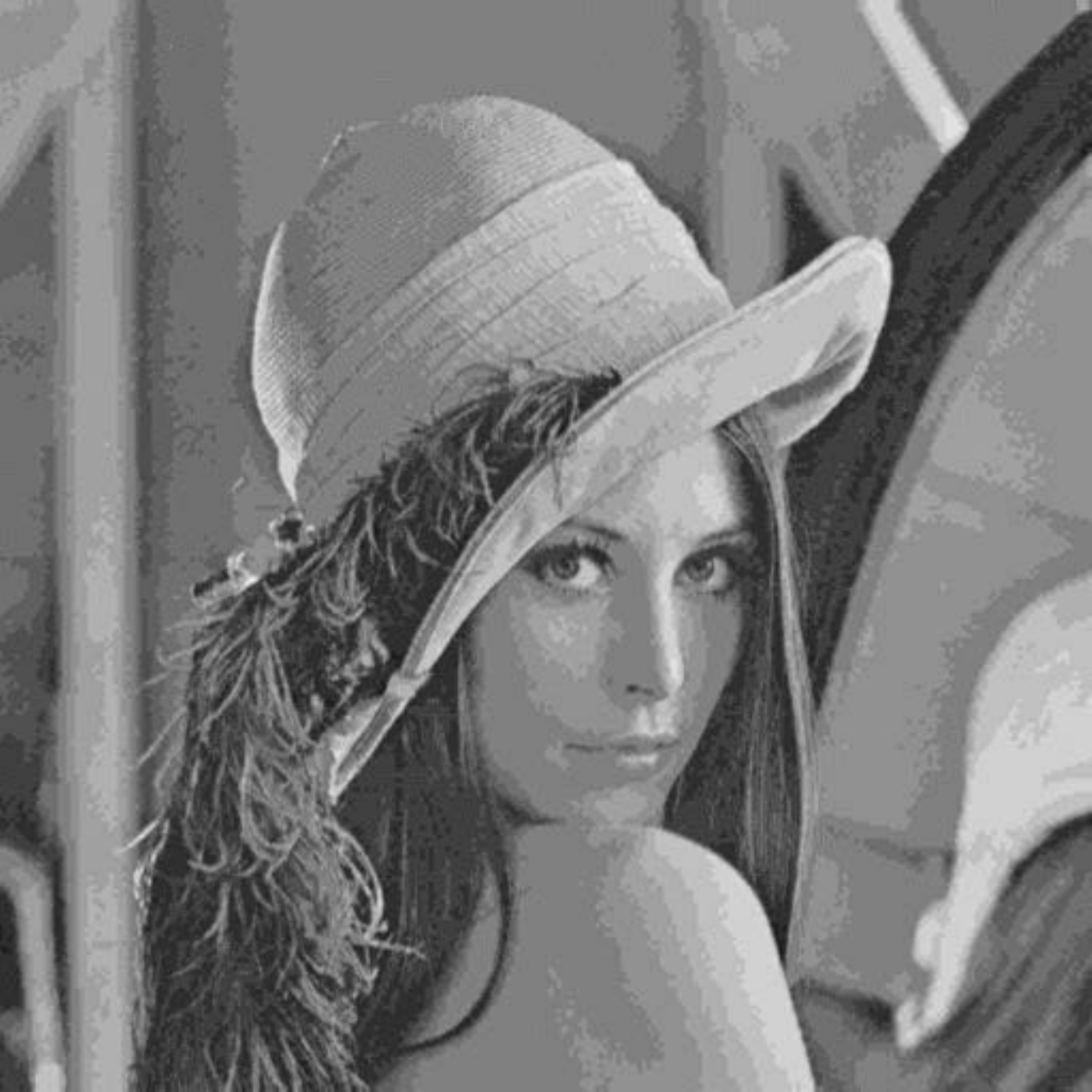}}
  \centerline{(c)$NUQ:(14,4,4,4)$}\medskip
\end{minipage}
\begin{minipage}[b]{0.24\linewidth}
  \centering
  \centerline{\includegraphics[width=\linewidth]{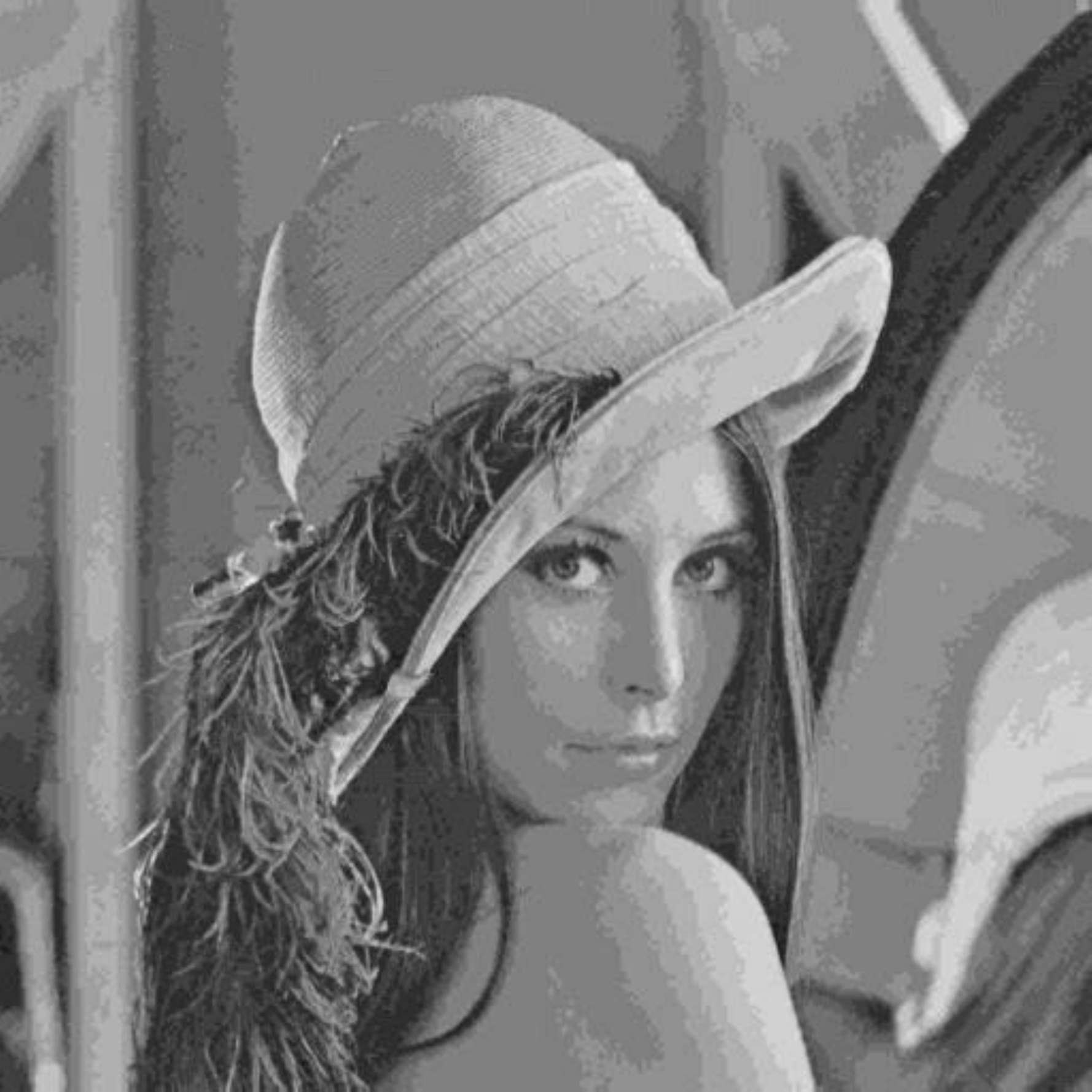}}
  \centerline{(d)$NUQ:(14,8,8,8)$}\medskip
\end{minipage}
\begin{minipage}[b]{0.24\linewidth}
  \centering
  \centerline{\includegraphics[width=\linewidth]{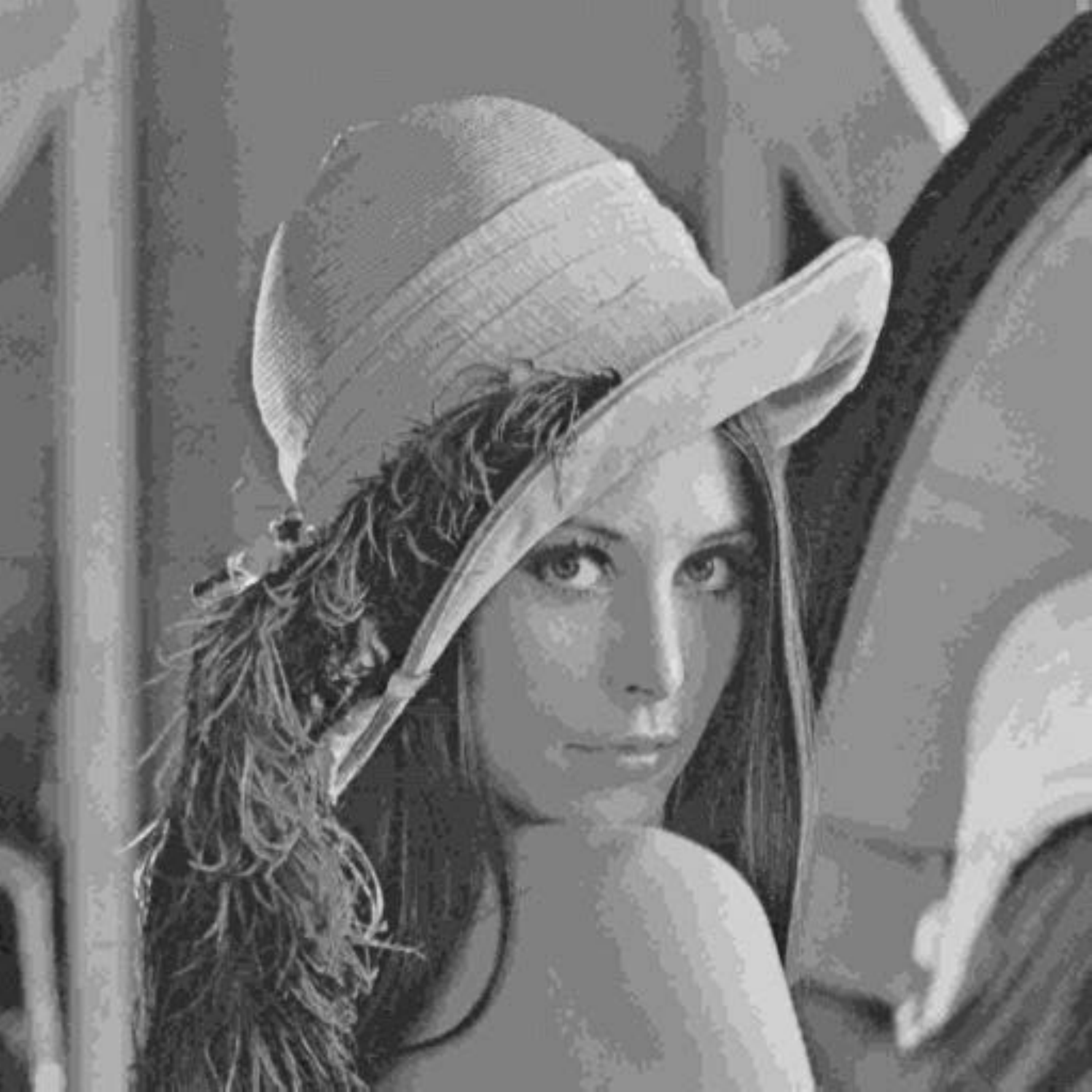}}
  \centerline{(e)$NUQ:(14,14,14,14)$}\medskip
\end{minipage}

\begin{minipage}[b]{0.24\linewidth}
  \centering
  \centerline{\includegraphics[width=\linewidth]{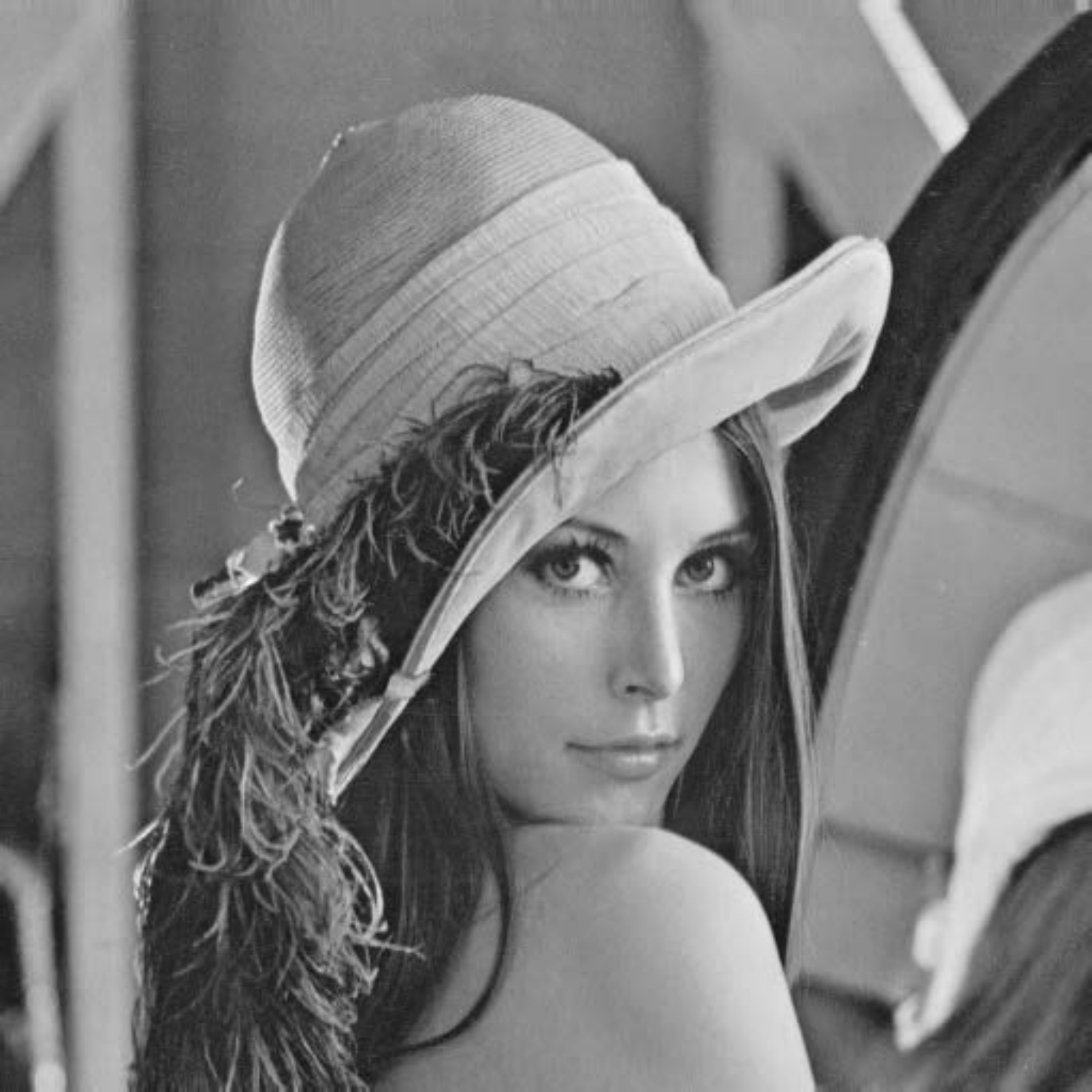}}
  \centerline{(f)$UQ:(104,23,34,15)$}\medskip
\end{minipage}
\begin{minipage}[b]{0.24\linewidth}
  \centering
  \centerline{\includegraphics[width=\linewidth]{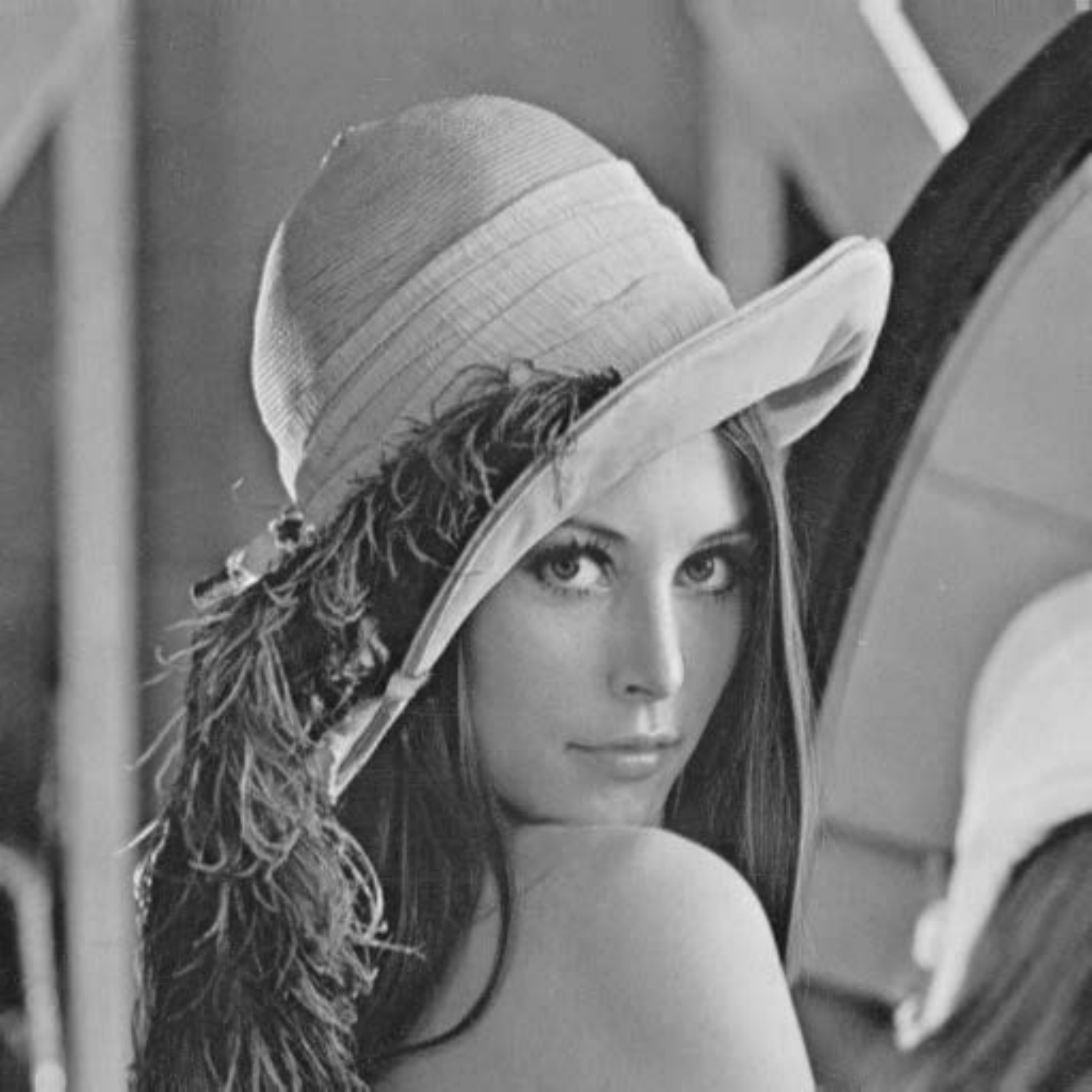}}
  \centerline{(g)$NUQ:(104,4,4,4)$}\medskip
\end{minipage}
\begin{minipage}[b]{0.24\linewidth}
  \centering
  \centerline{\includegraphics[width=\linewidth]{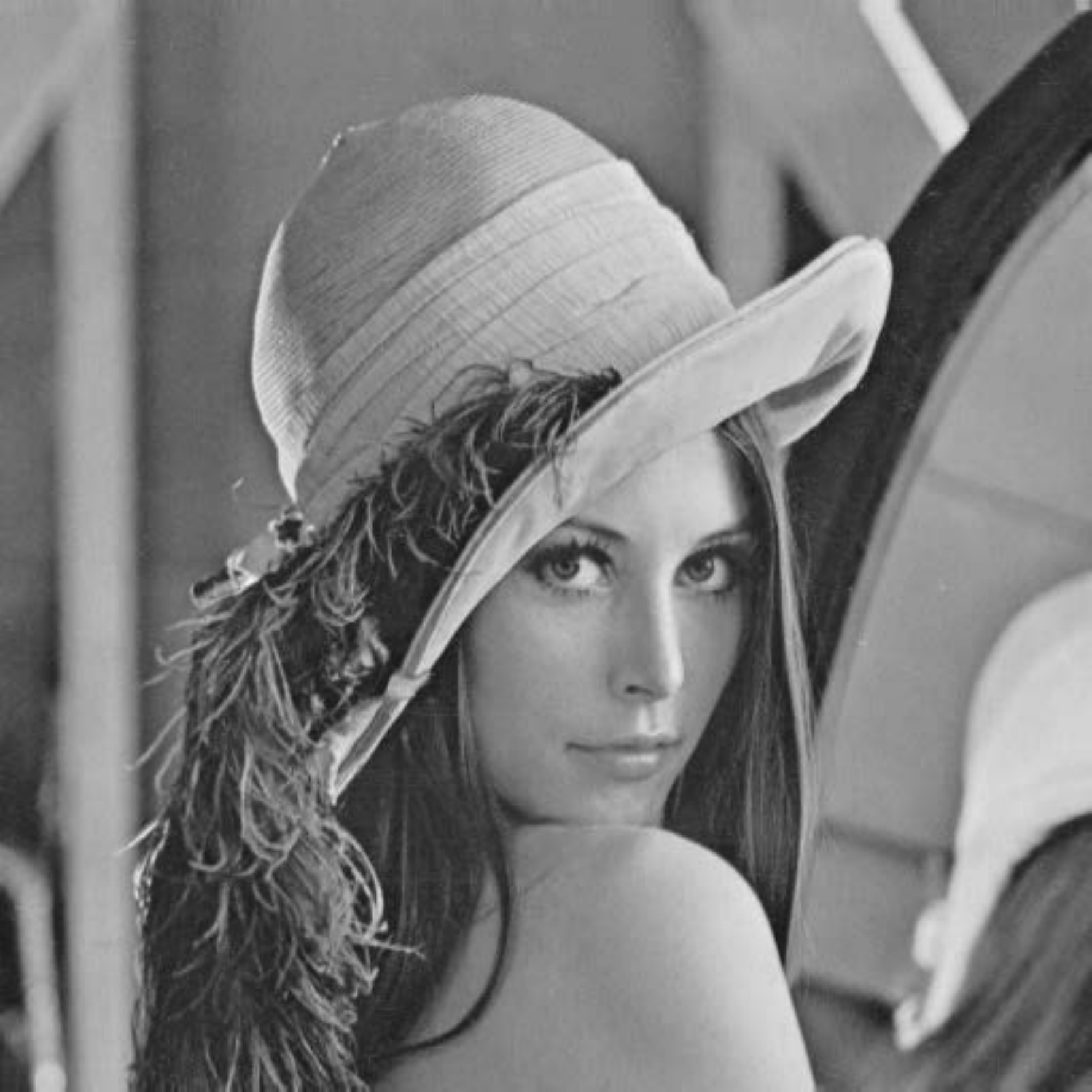}}
  \centerline{(h)$NUQ:(104,8,8,8)$}\medskip
\end{minipage}
\begin{minipage}[b]{0.24\linewidth}
  \centering
  \centerline{\includegraphics[width=\linewidth]{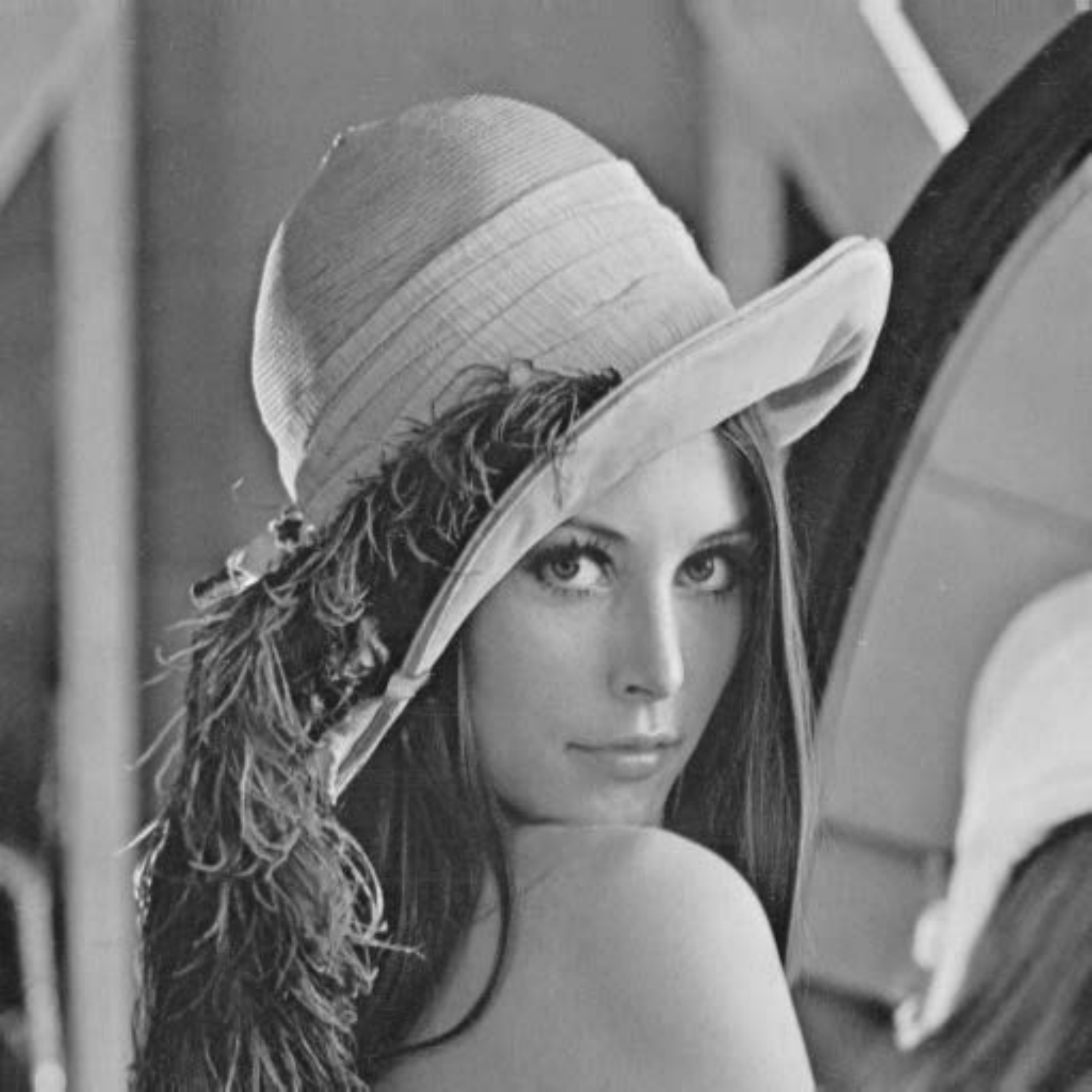}}
  \centerline{(i)$NUQ:(104,14,14,14)$}\medskip
\end{minipage}

 \caption{\emph{Lenna}: Comparison of the reconstructed images produced by the uniform ((b),(f)) and non-uniform quantizer ((c),(d),(e),(g),(h),(i)) at various quantization step sizes. The quantized values for each decomposition component is given in ($N_A$,$N_H$,$N_V$,$N_D$) format, where it represents the Approximation, Horizontal, Vertical, and Diagonal component, respectively.}
\end{figure}

\begin{figure}[ht!]
\begin{minipage}[b]{\linewidth}
  \centering
  \centerline{\includegraphics[width=0.24\linewidth]{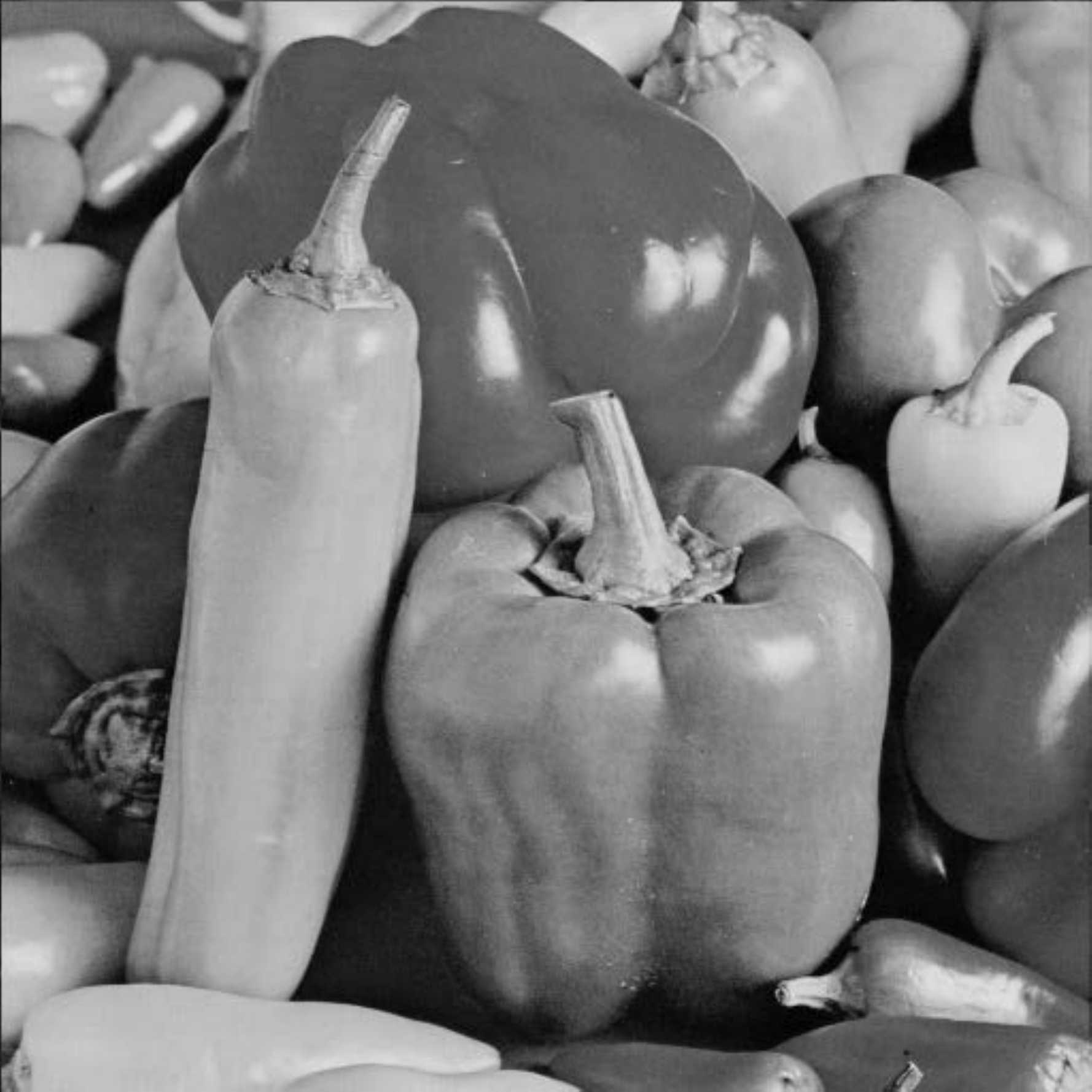}}
  \centerline{(a) Original}\medskip
\end{minipage}

\begin{minipage}[b]{0.24\linewidth}
  \centering
  \centerline{\includegraphics[width=\linewidth]{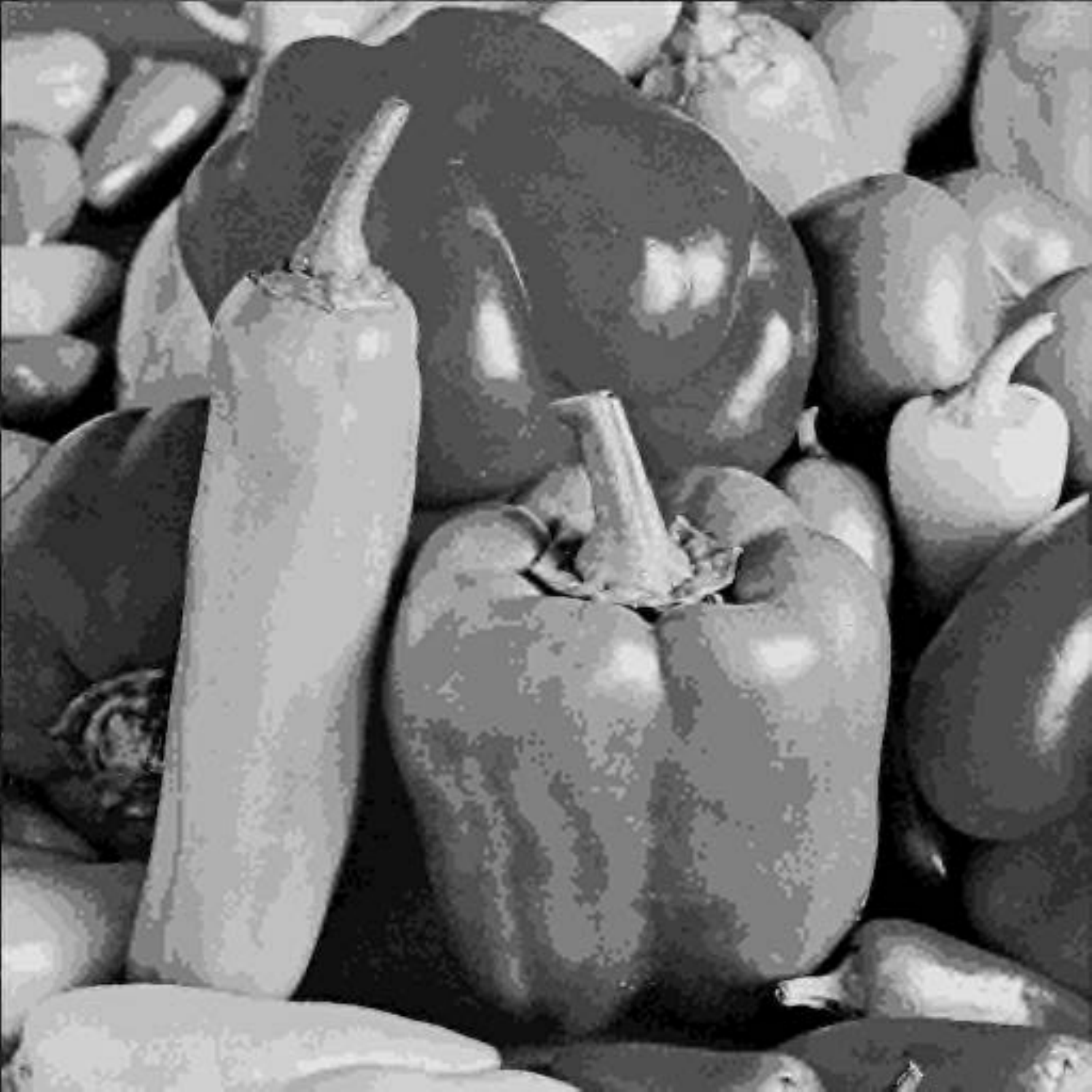}}
  \centerline{(b)$UQ:(14,7,7,3)$}\medskip
\end{minipage}
\begin{minipage}[b]{0.24\linewidth}
  \centering
  \centerline{\includegraphics[width=\linewidth]{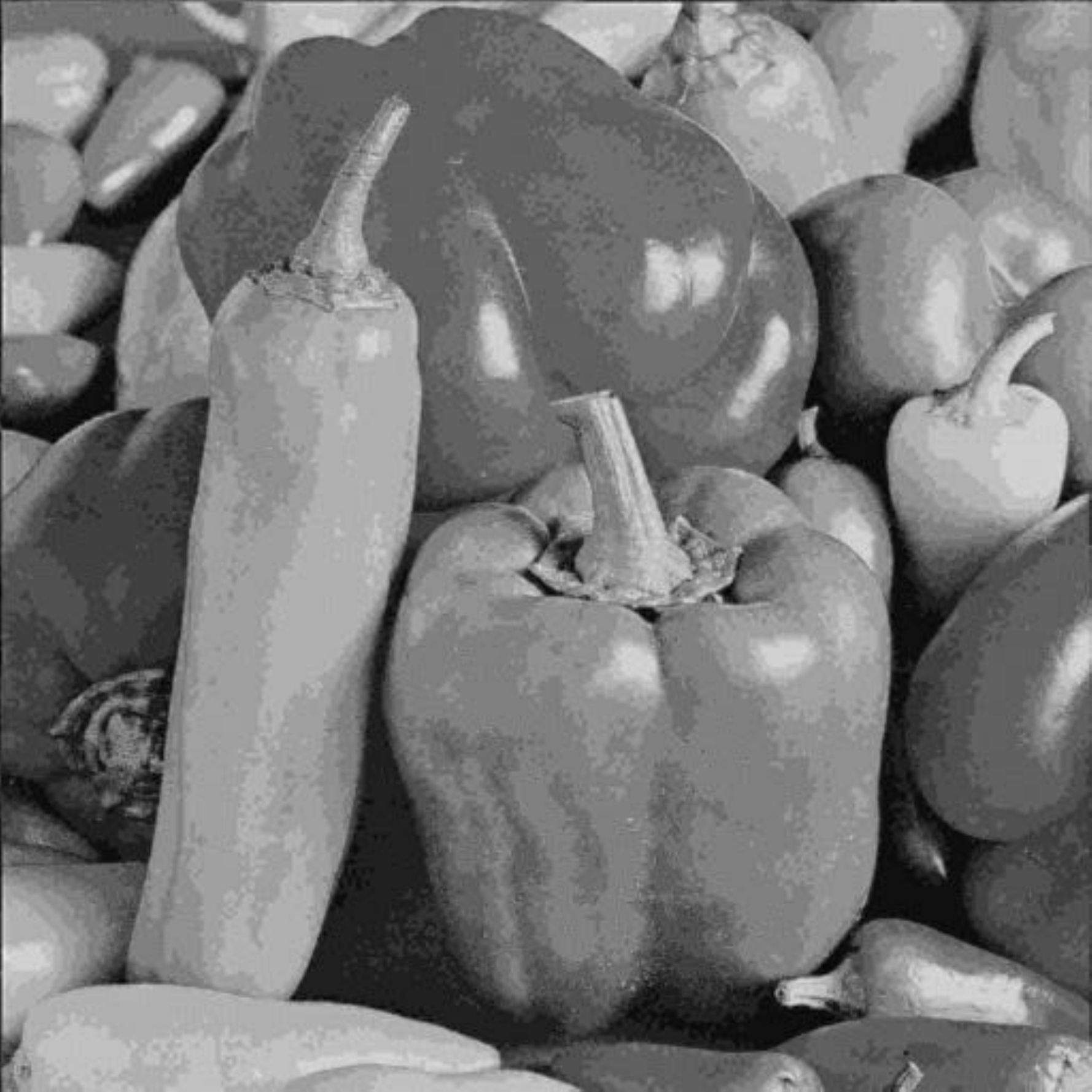}}
  \centerline{(c)$NUQ:(14,4,4,4)$}\medskip
\end{minipage}
\begin{minipage}[b]{0.24\linewidth}
  \centering
  \centerline{\includegraphics[width=\linewidth]{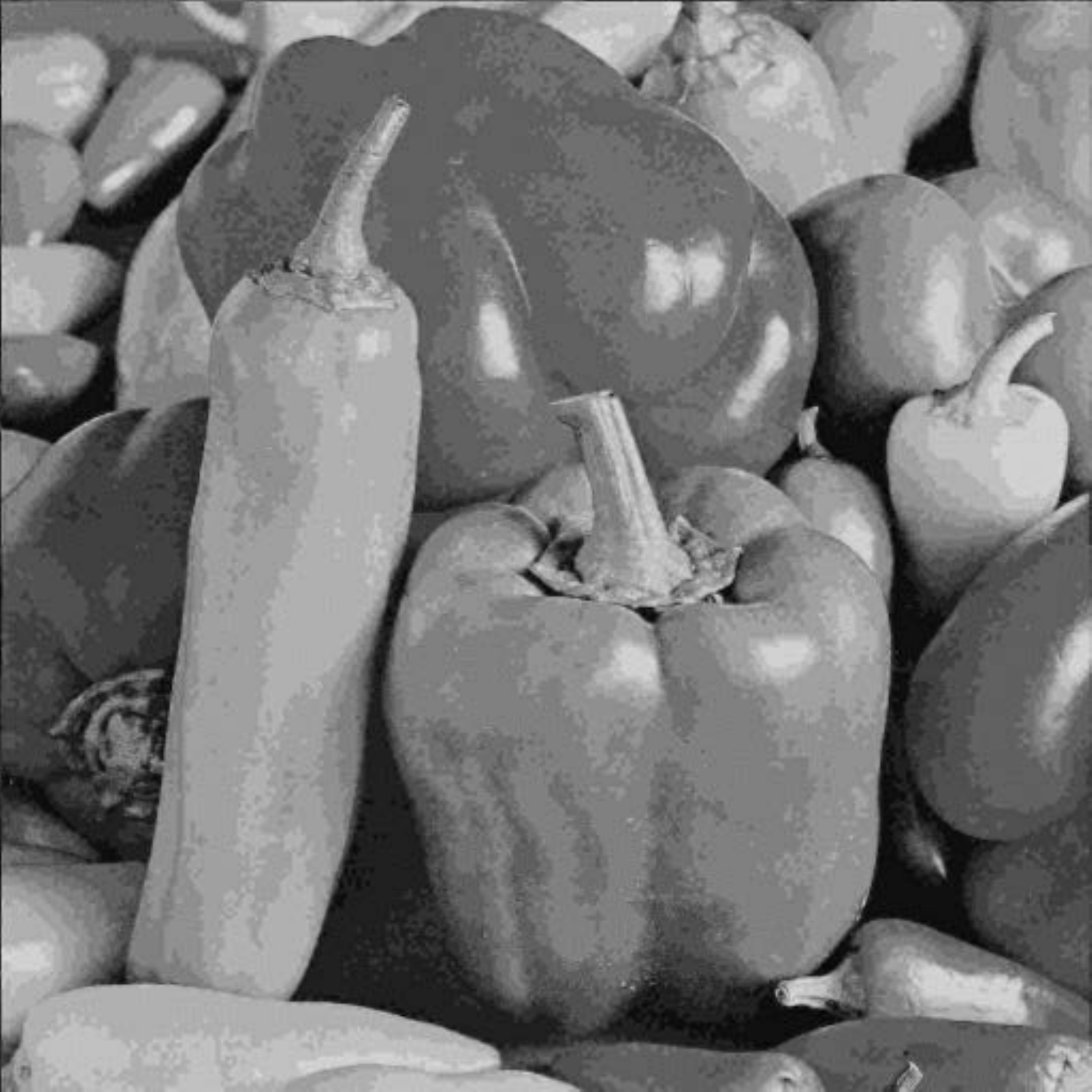}}
  \centerline{(d)$NUQ:(14,8,8,8)$}\medskip
\end{minipage}
\begin{minipage}[b]{0.24\linewidth}
  \centering
  \centerline{\includegraphics[width=\linewidth]{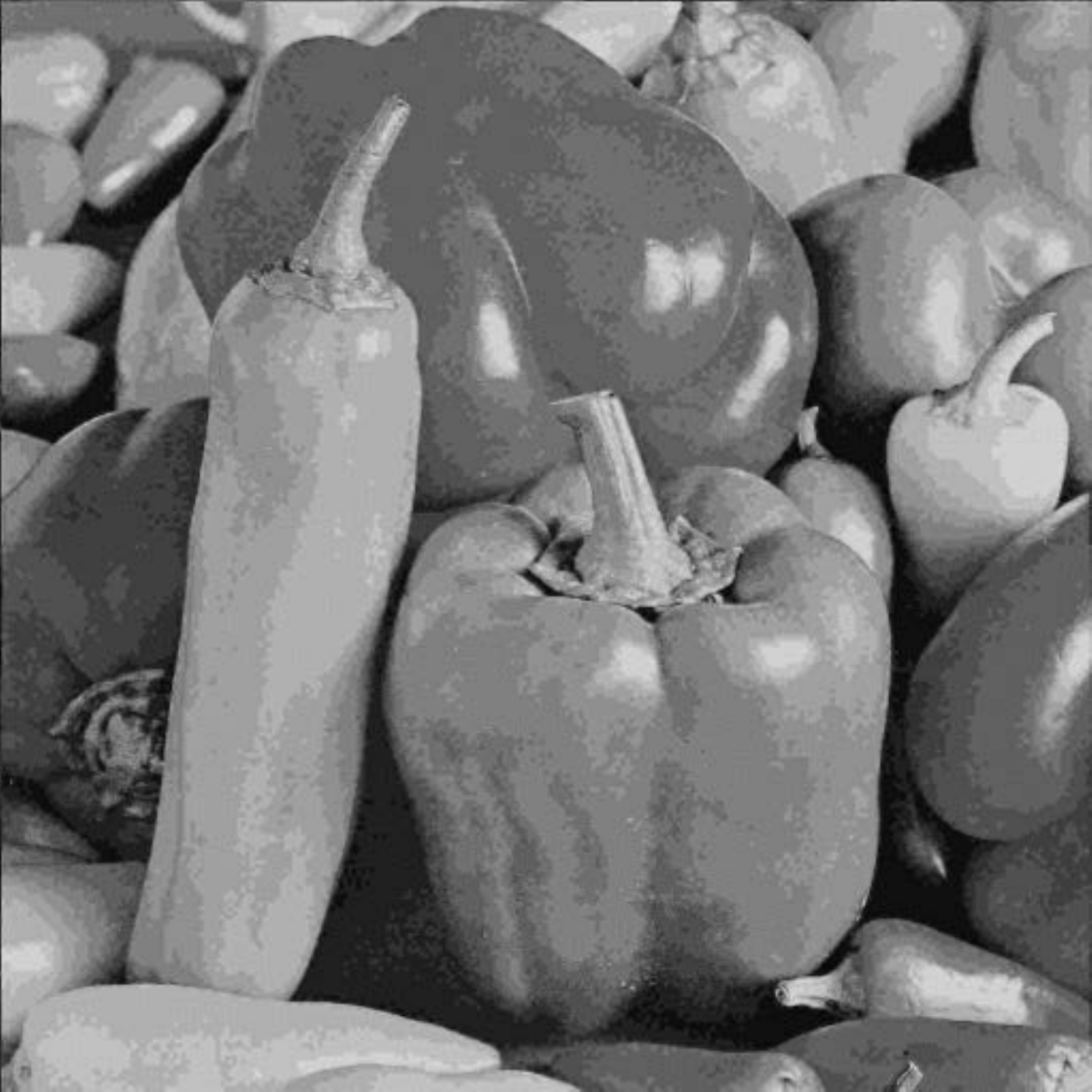}}
  \centerline{(e)$NUQ:(14,14,14,14)$}\medskip
\end{minipage}

\begin{minipage}[b]{0.24\linewidth}
  \centering
  \centerline{\includegraphics[width=\linewidth]{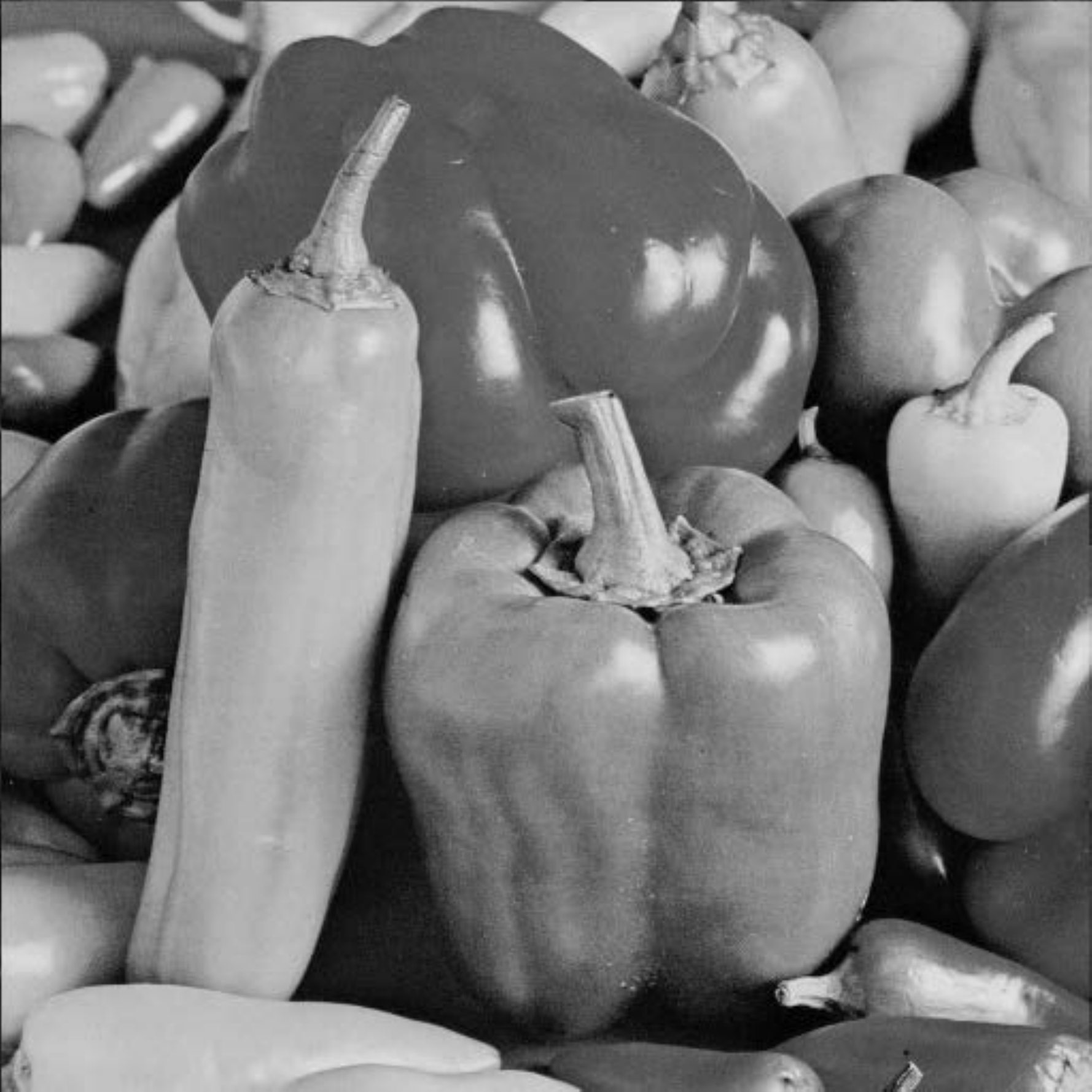}}
  \centerline{(f)$UQ:(115,52,53,20)$}\medskip
\end{minipage}
\begin{minipage}[b]{0.24\linewidth}
  \centering
  \centerline{\includegraphics[width=\linewidth]{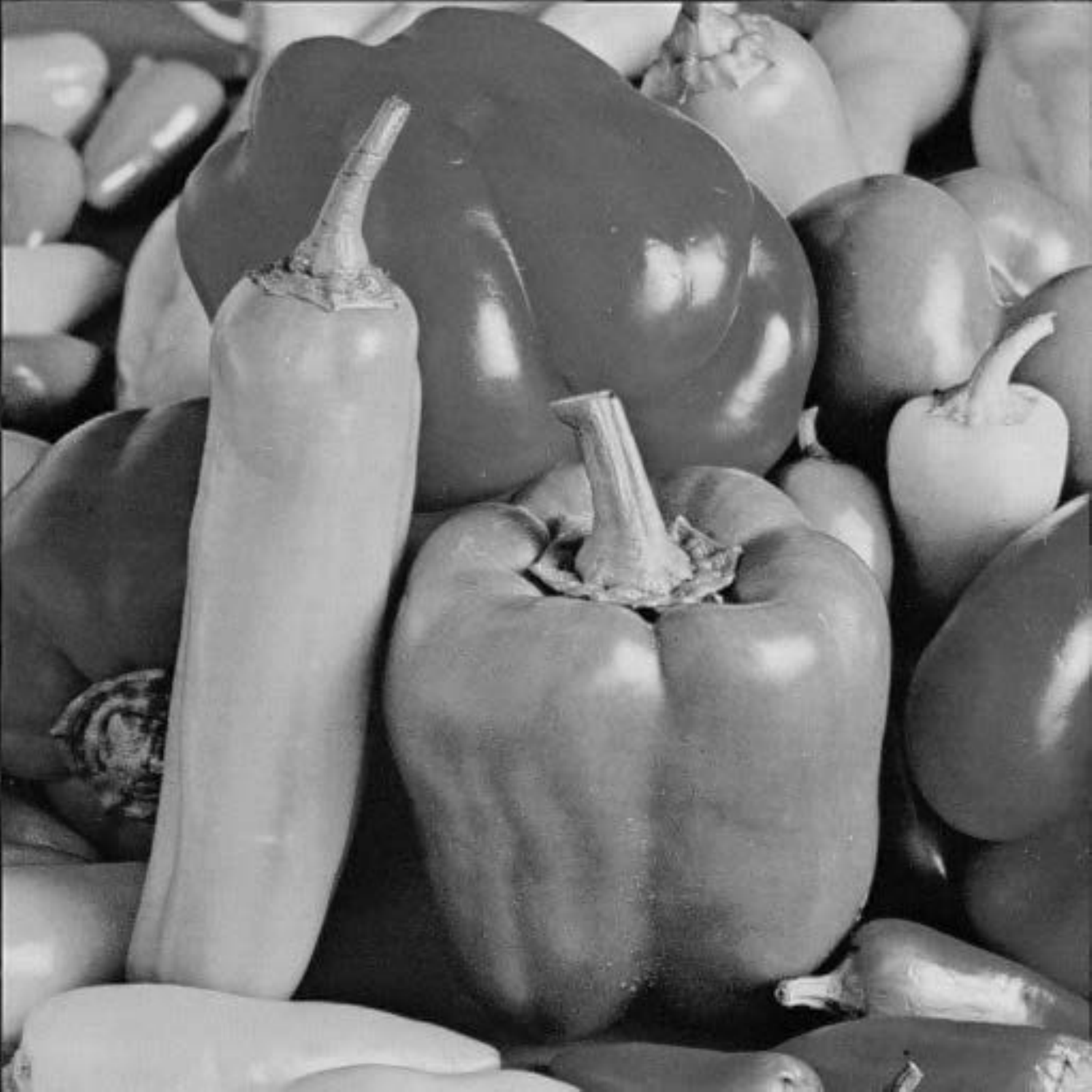}}
  \centerline{(g)$NUQ:(115,4,4,4)$}\medskip
\end{minipage}
\begin{minipage}[b]{0.24\linewidth}
  \centering
  \centerline{\includegraphics[width=\linewidth]{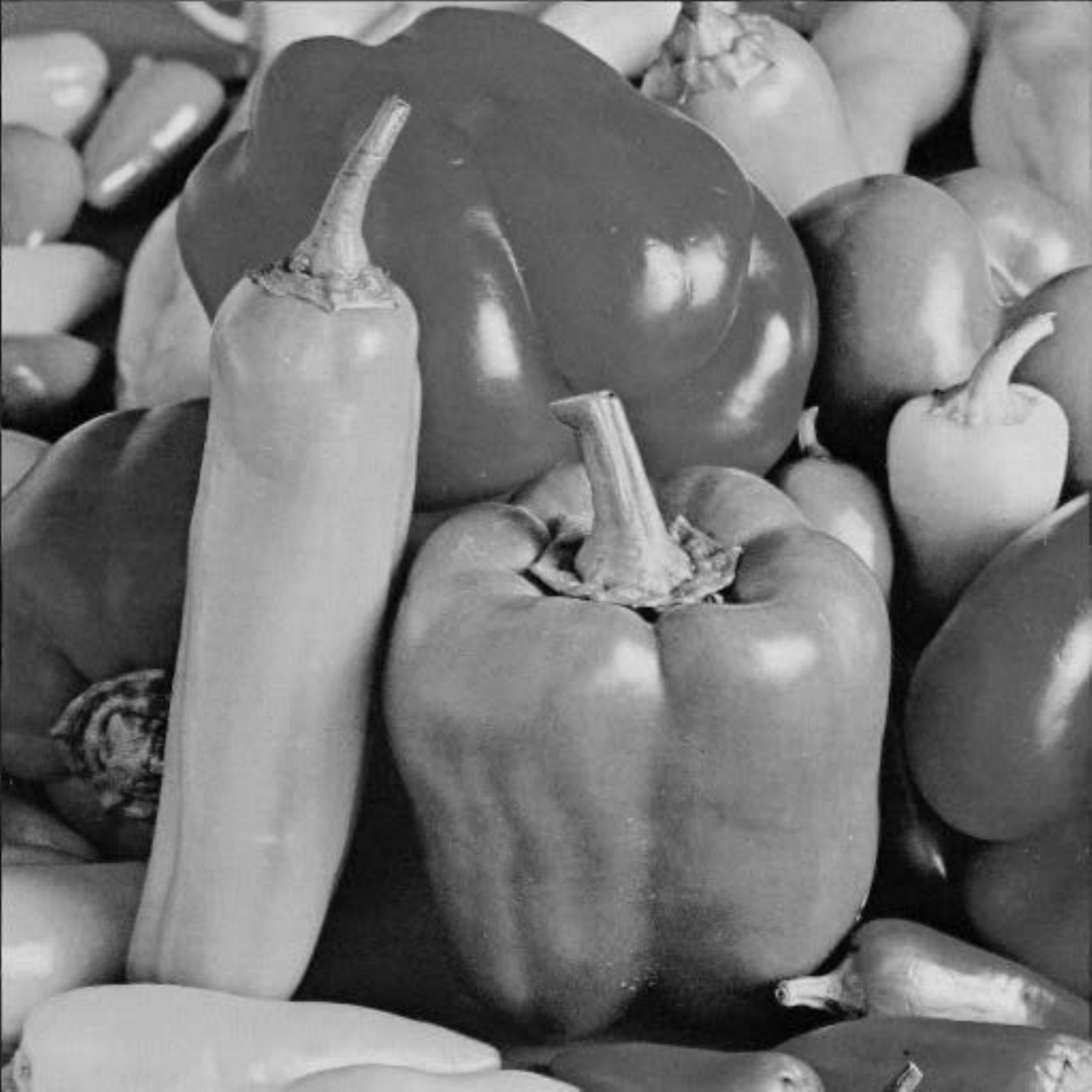}}
  \centerline{(h)$NUQ:(115,8,8,8)$}\medskip
\end{minipage}
\begin{minipage}[b]{0.24\linewidth}
  \centering
  \centerline{\includegraphics[width=\linewidth]{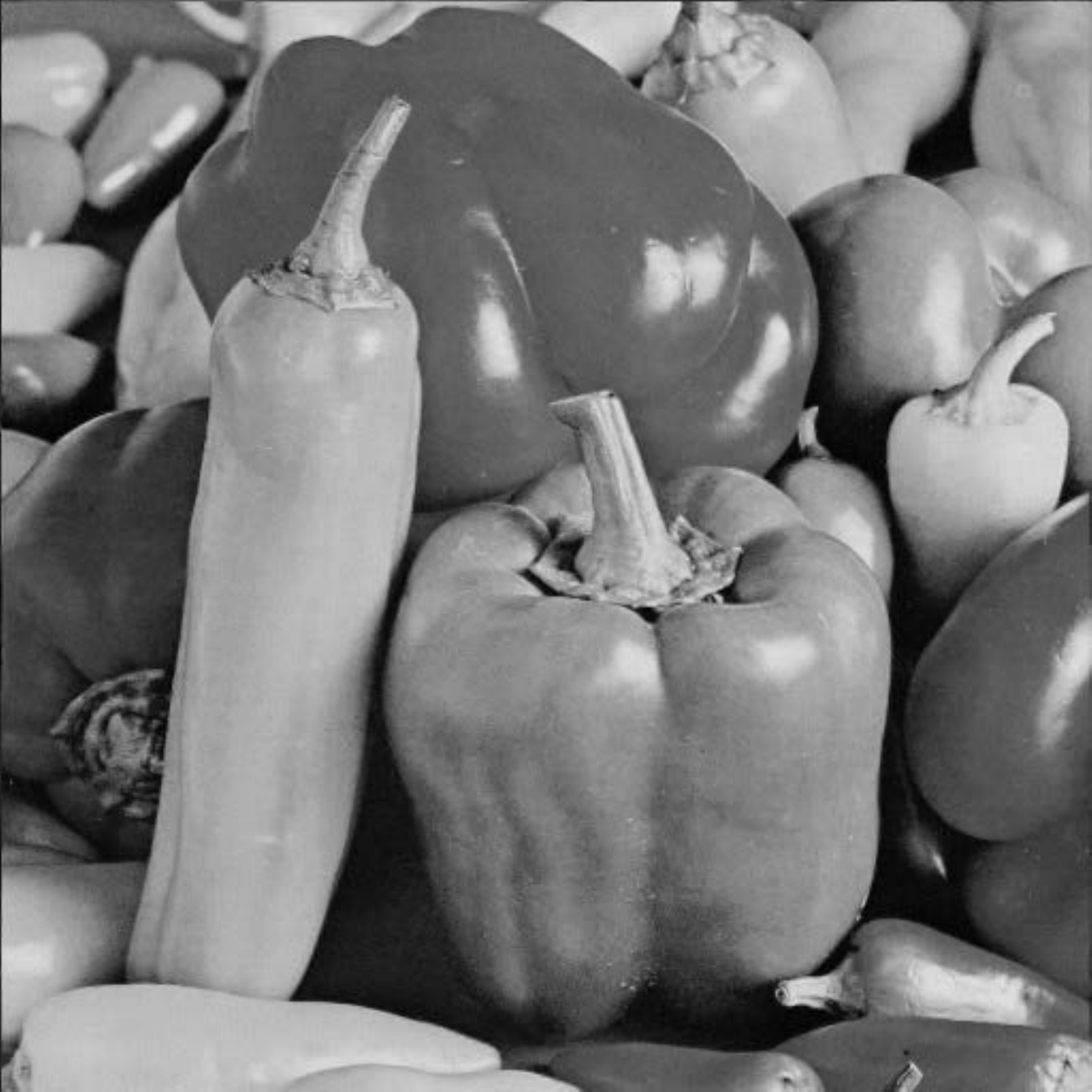}}
  \centerline{(i)$NUQ:(115,14,14,14)$}\medskip
\end{minipage}

 \caption{\emph{Pepper}: Comparison of the reconstructed images produced by the uniform ((b),(f)) and non-uniform quantizer ((c),(d),(e),(g),(h),(i)) at various quantization step sizes. The quantized values for each decomposition component is given in ($N_A$,$N_H$,$N_V$,$N_D$) format, where it represents the Approximation, Horizontal, Vertical, and Diagonal component, respectively.}

\end{figure}

\begin{figure}[ht!]
\begin{minipage}[b]{\linewidth}
  \centering
  \centerline{\includegraphics[width=0.24\linewidth]{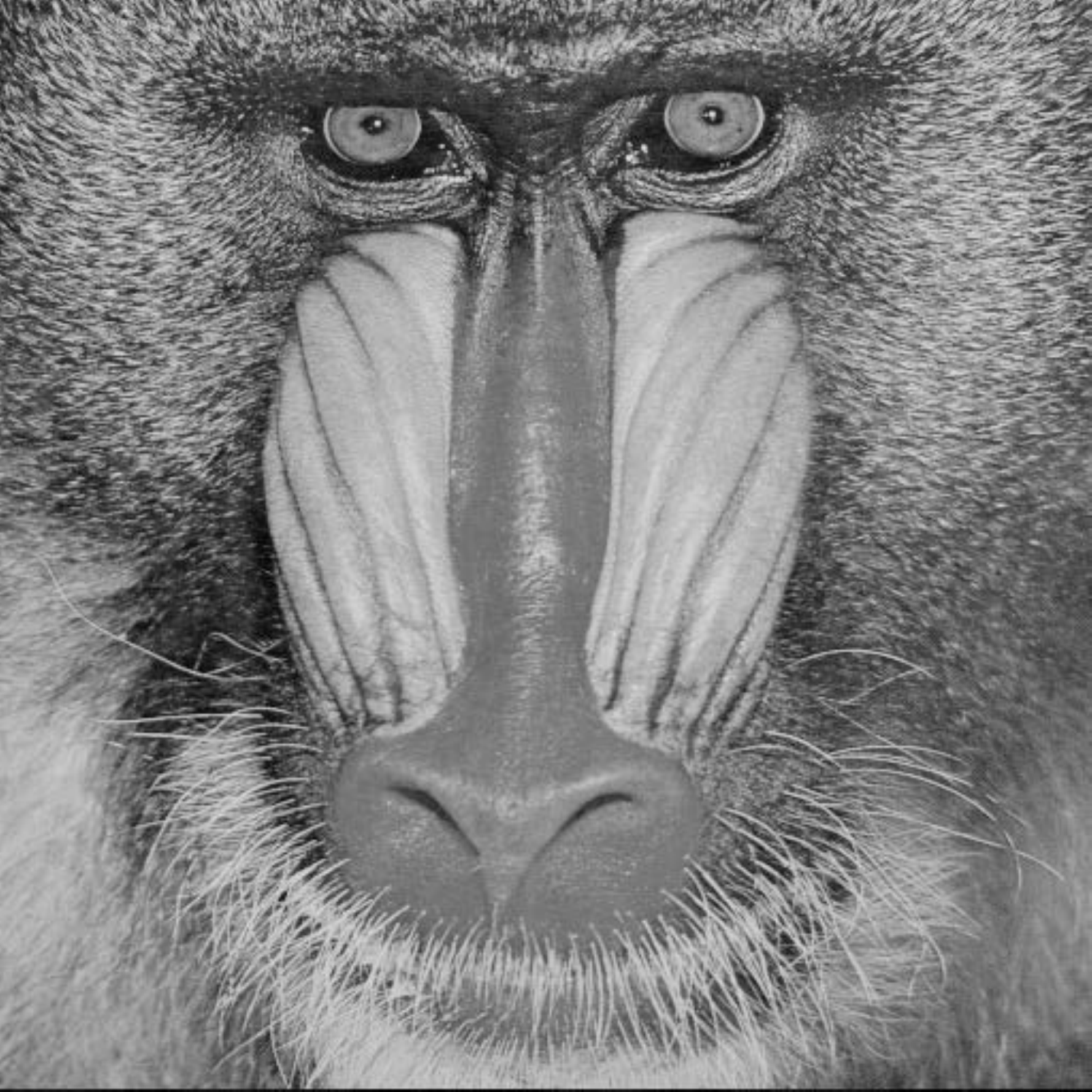}}
  \centerline{(a) Original}\medskip
\end{minipage}

\begin{minipage}[b]{0.24\linewidth}
  \centering
  \centerline{\includegraphics[width=\linewidth]{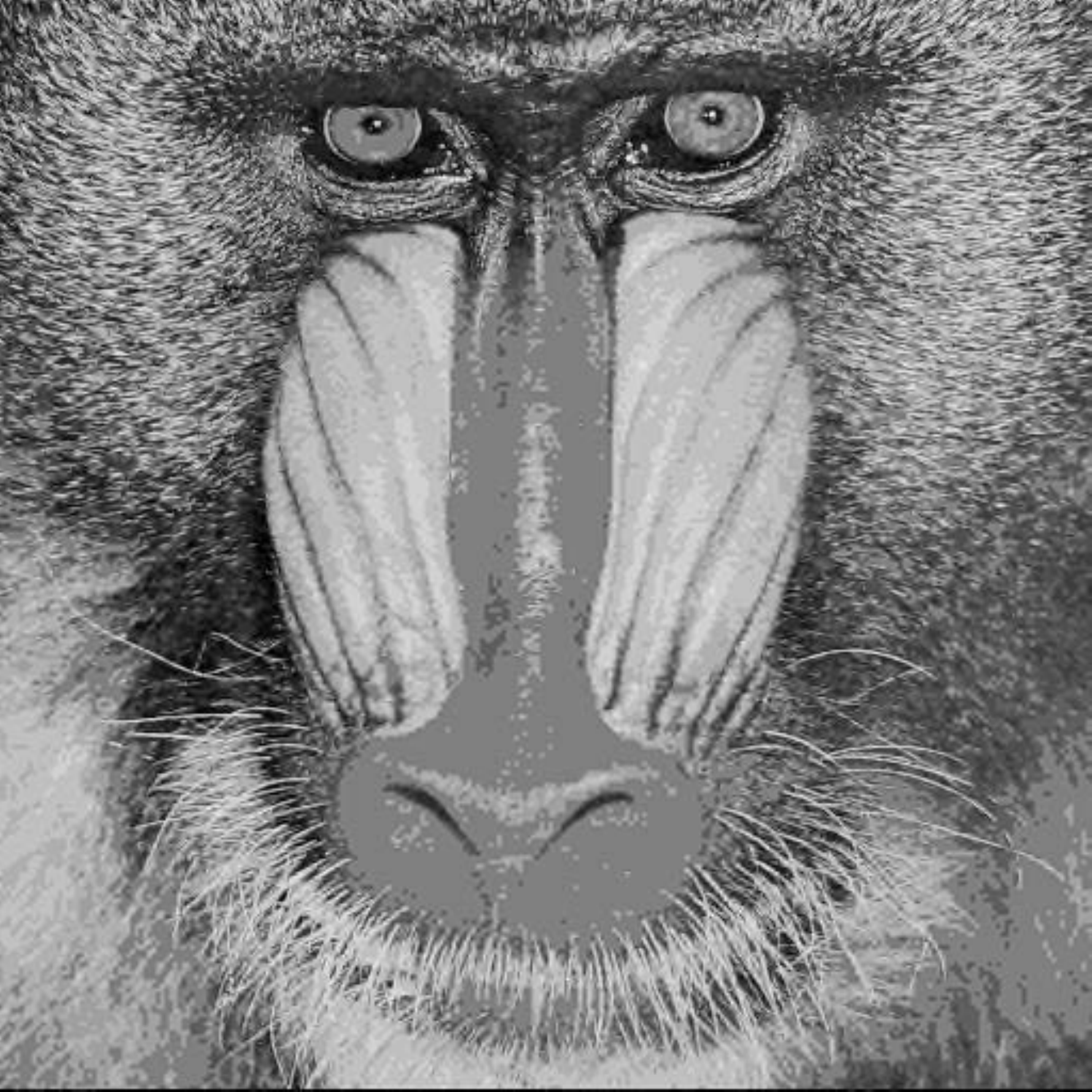}}
  \centerline{(b)$UQ:(13,9,7,5)$}\medskip
\end{minipage}
\begin{minipage}[b]{0.24\linewidth}
  \centering
  \centerline{\includegraphics[width=\linewidth]{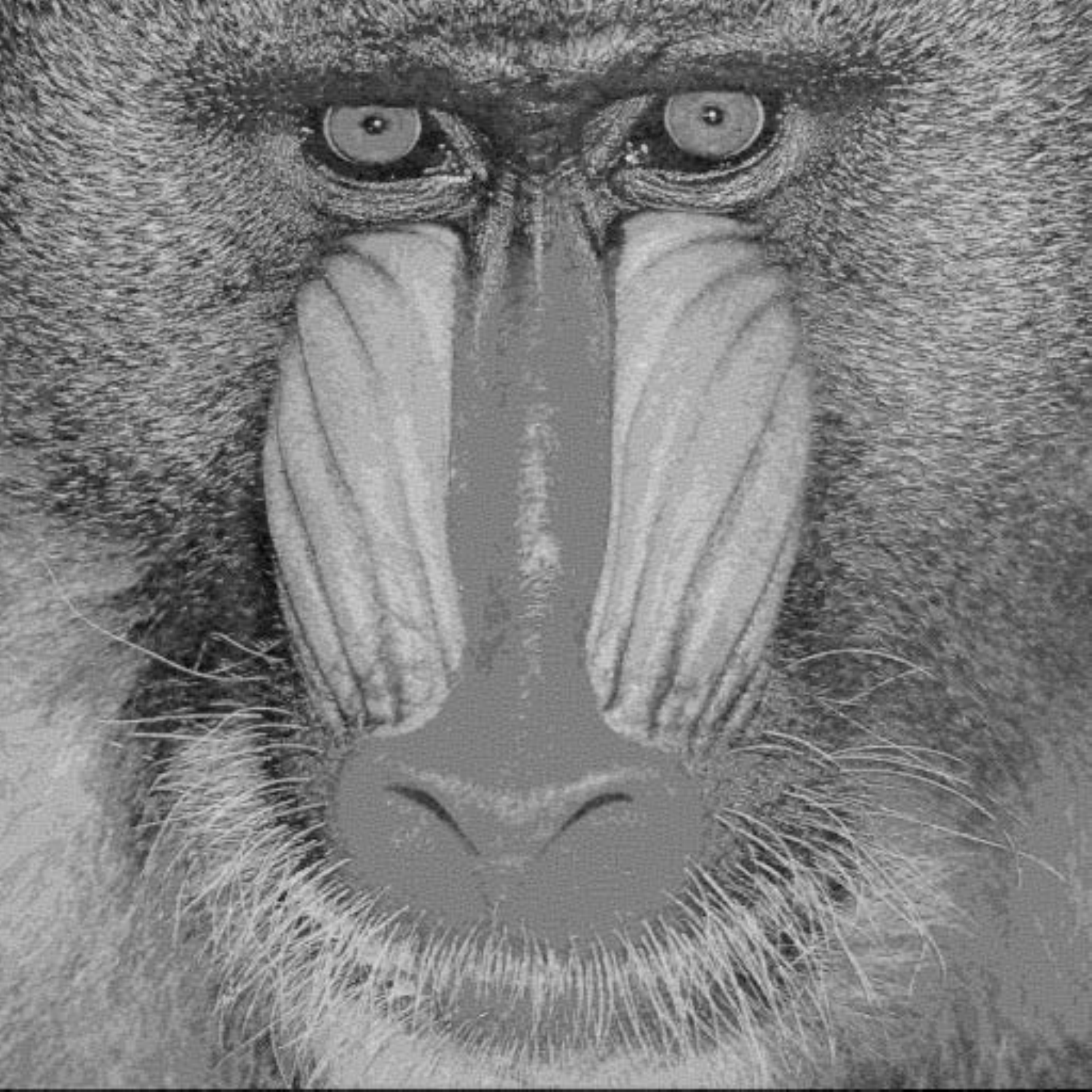}}
  \centerline{(c)$NUQ:(13,4,4,4)$}\medskip
\end{minipage}
\begin{minipage}[b]{0.24\linewidth}
  \centering
  \centerline{\includegraphics[width=\linewidth]{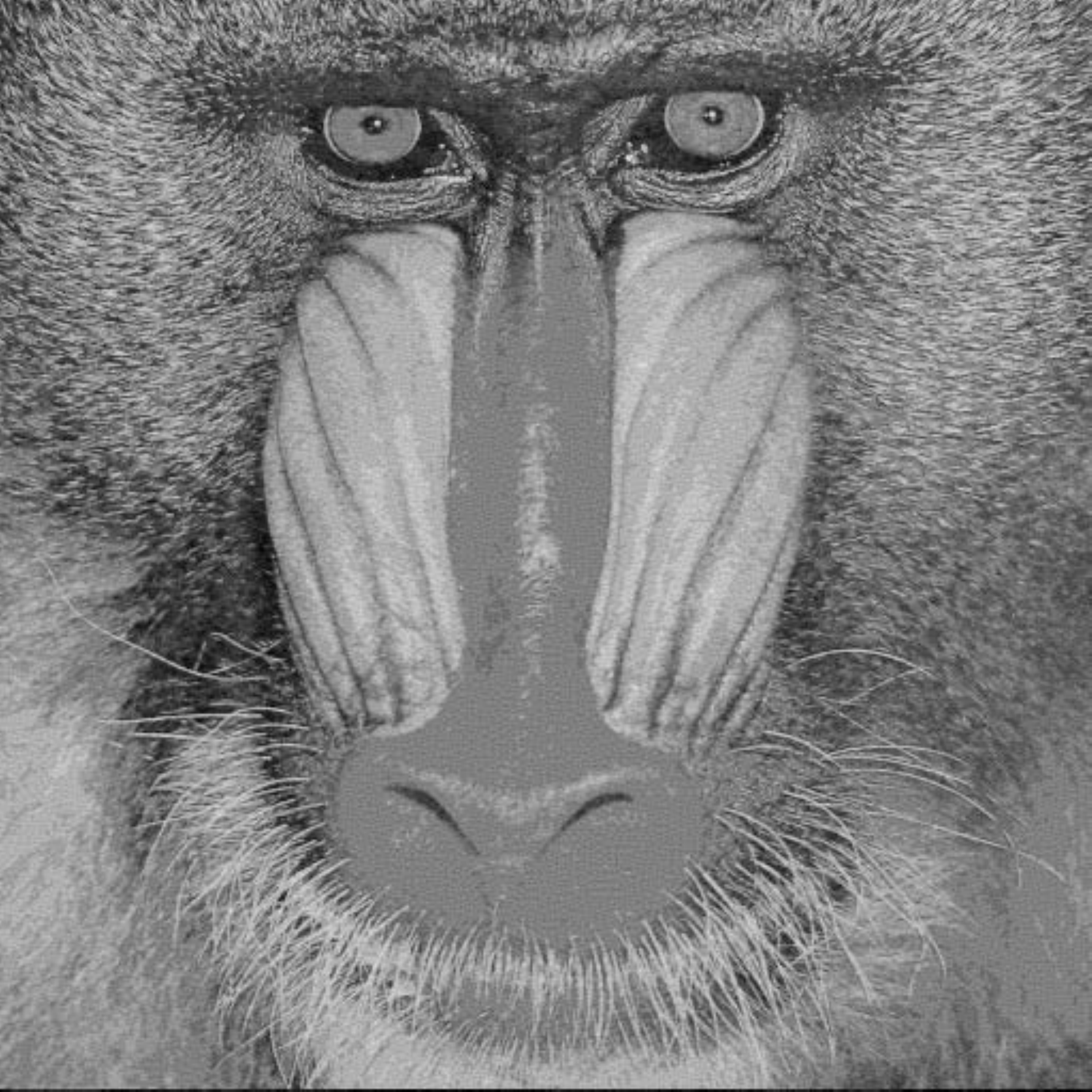}}
  \centerline{(d)$NUQ:(13,8,8,8)$}\medskip
\end{minipage}
\begin{minipage}[b]{0.24\linewidth}
  \centering
  \centerline{\includegraphics[width=\linewidth]{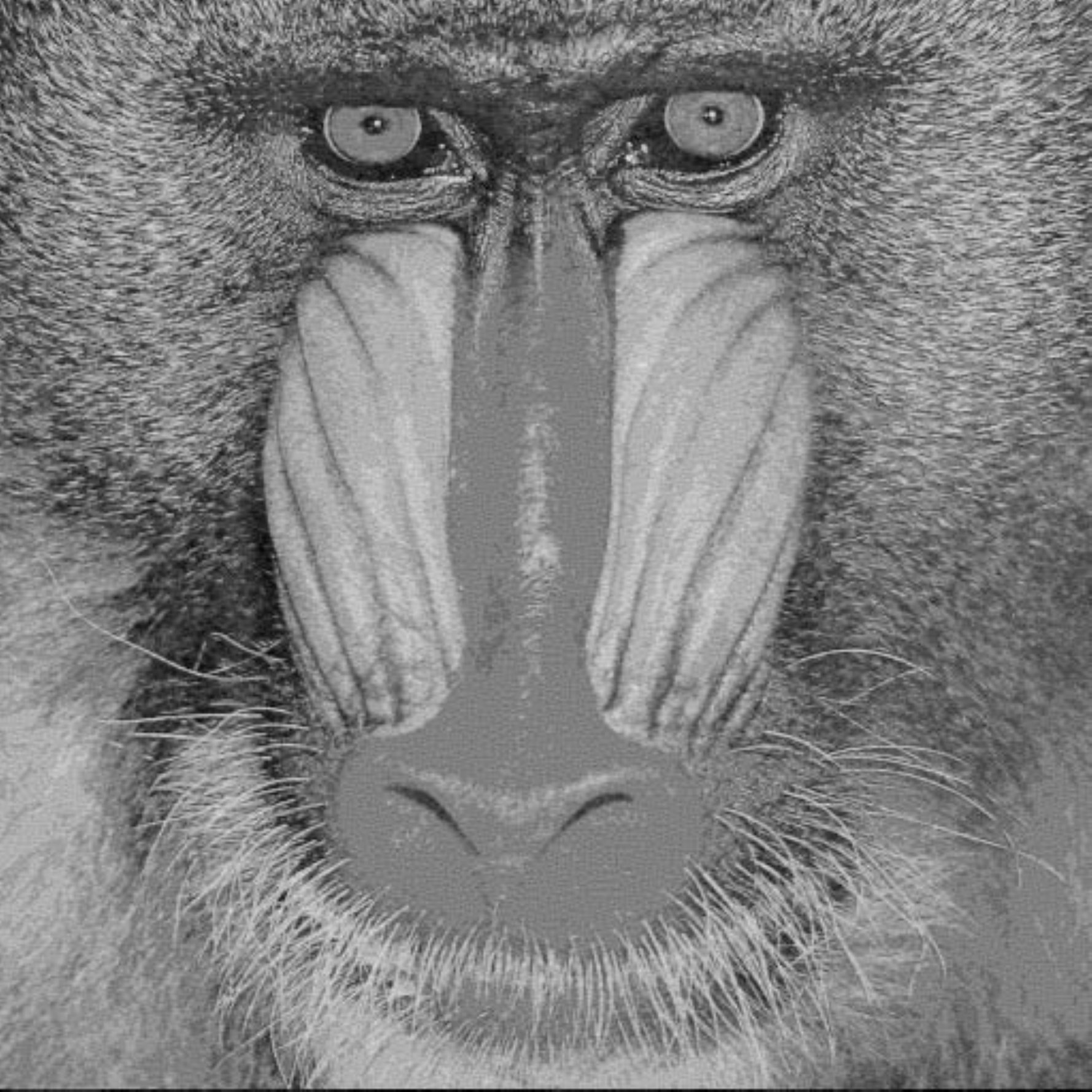}}
  \centerline{(e)$NUQ:(13,14,14,14)$}\medskip
\end{minipage}

\begin{minipage}[b]{0.24\linewidth}
  \centering
  \centerline{\includegraphics[width=\linewidth]{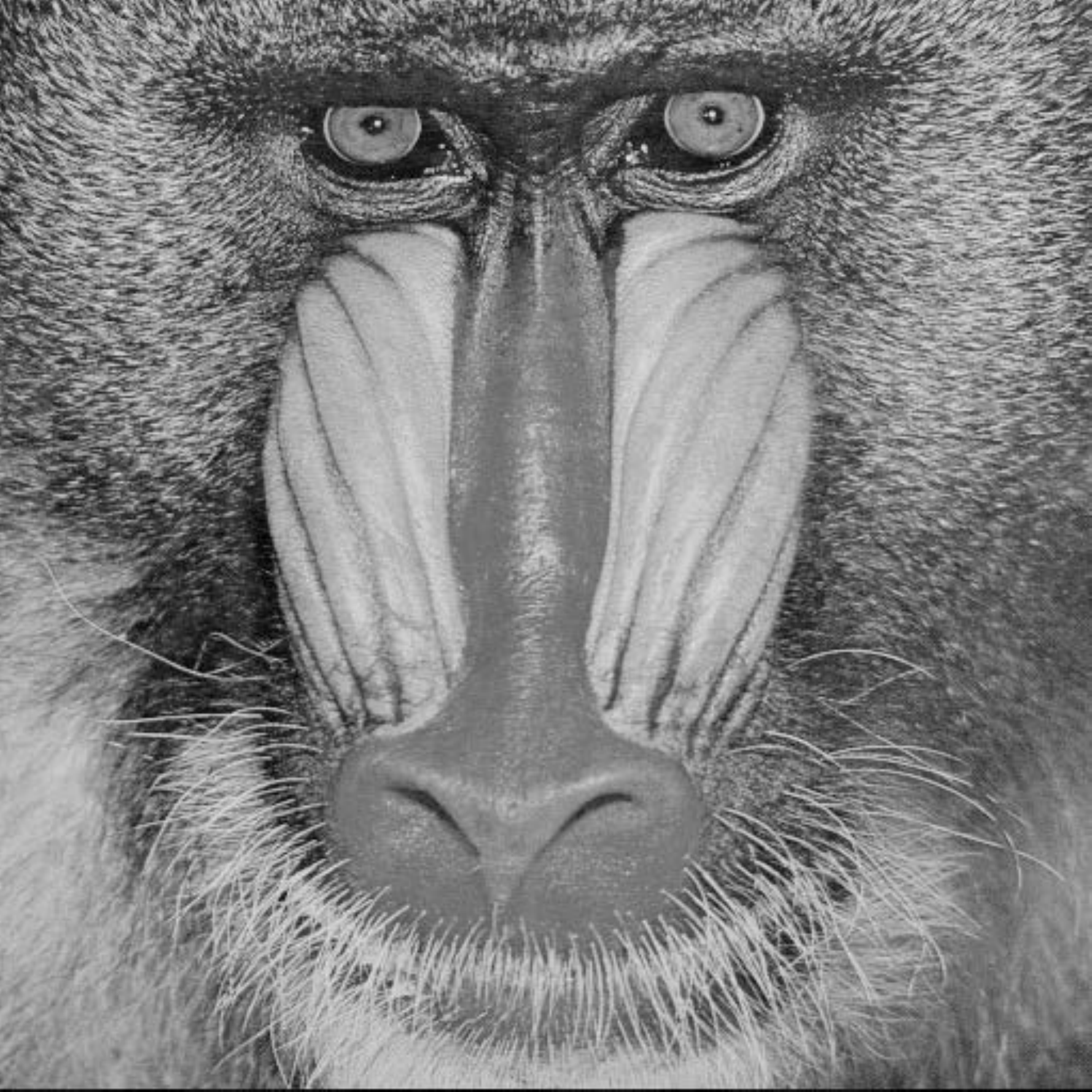}}
  \centerline{(f)$UQ:(104,68,51,36)$}\medskip
\end{minipage}
\begin{minipage}[b]{0.24\linewidth}
  \centering
  \centerline{\includegraphics[width=\linewidth]{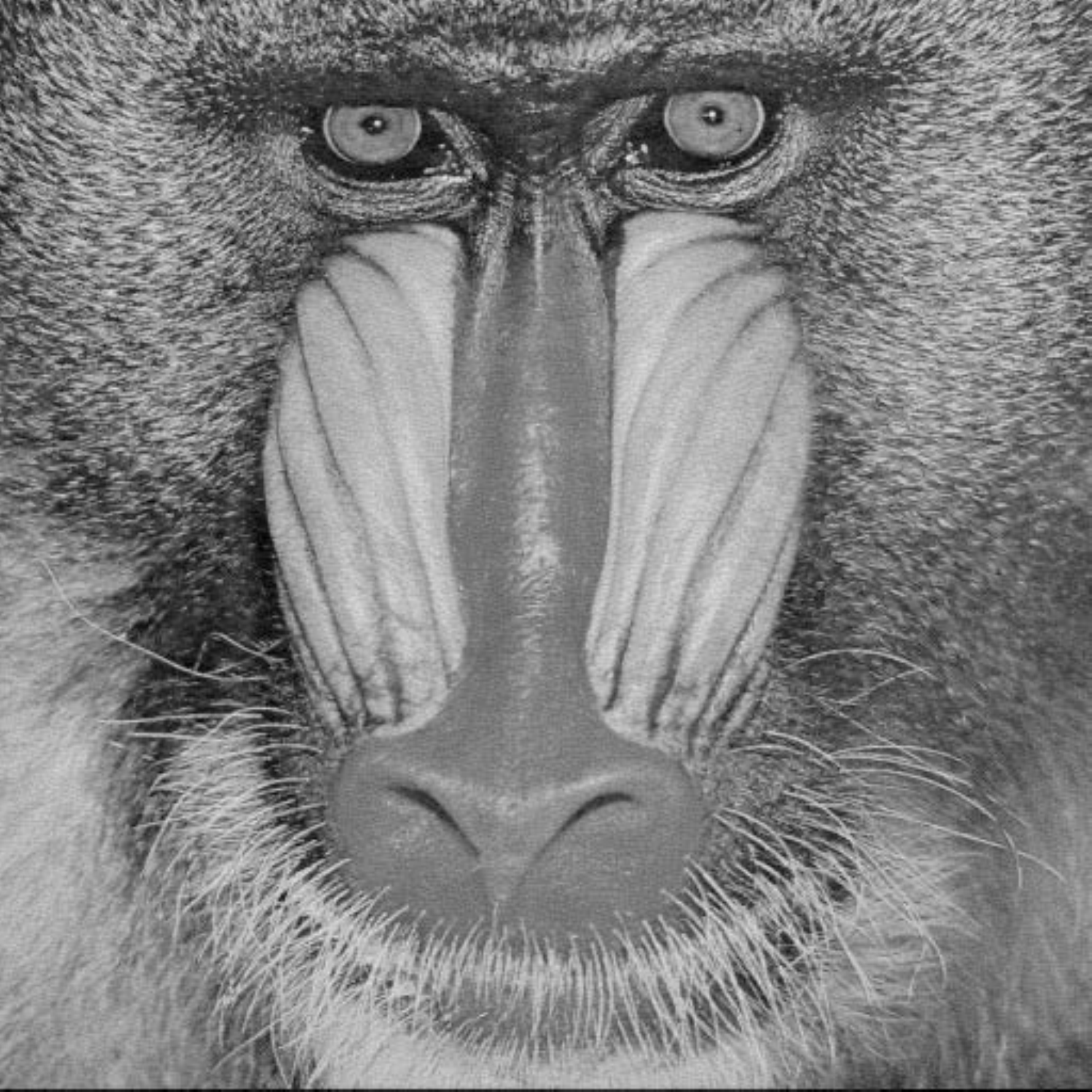}}
  \centerline{(g)$NUQ:(104,4,4,4)$}\medskip
\end{minipage}
\begin{minipage}[b]{0.24\linewidth}
  \centering
  \centerline{\includegraphics[width=\linewidth]{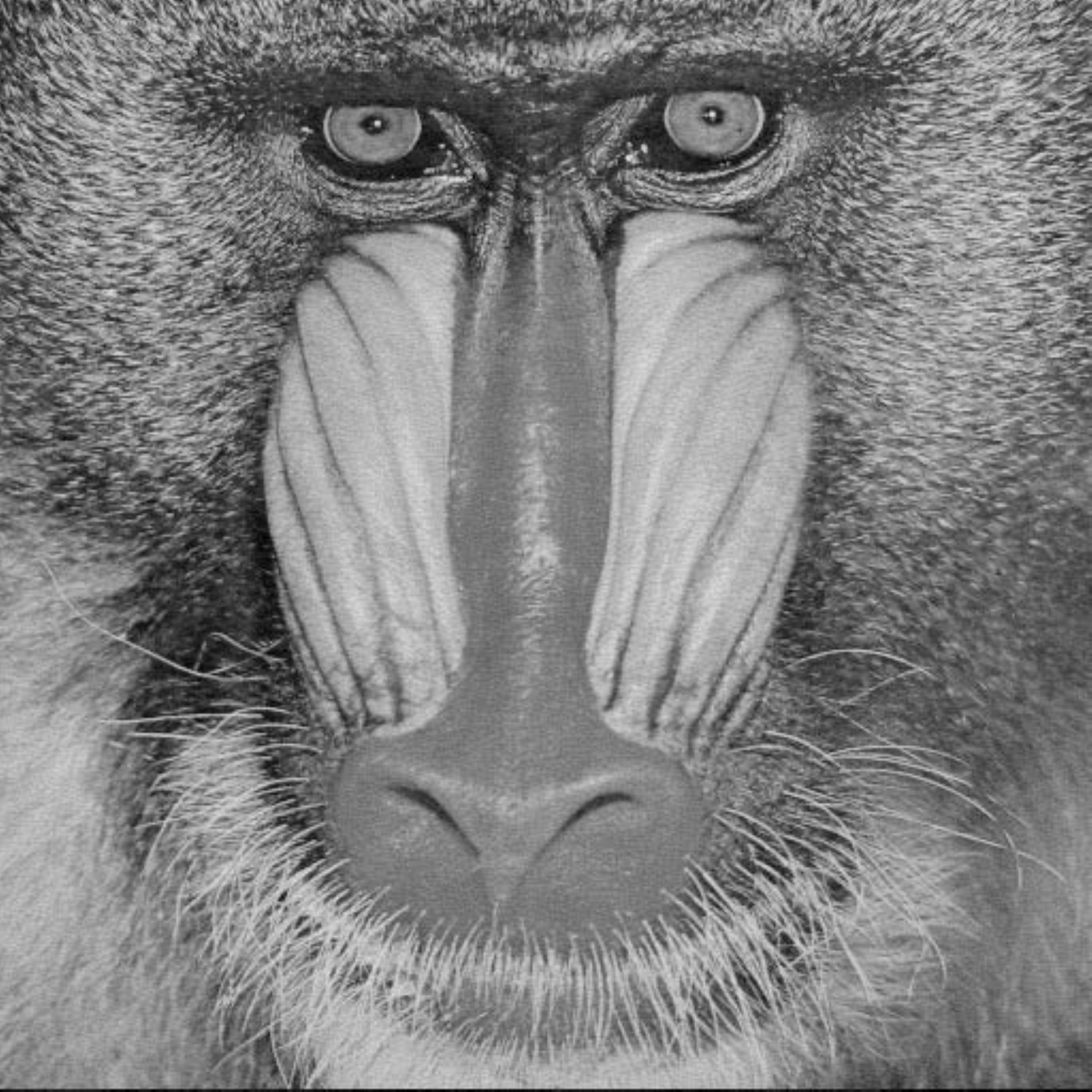}}
  \centerline{(h)$NUQ:(104,8,8,8)$}\medskip
\end{minipage}
\begin{minipage}[b]{0.24\linewidth}
  \centering
  \centerline{\includegraphics[width=\linewidth]{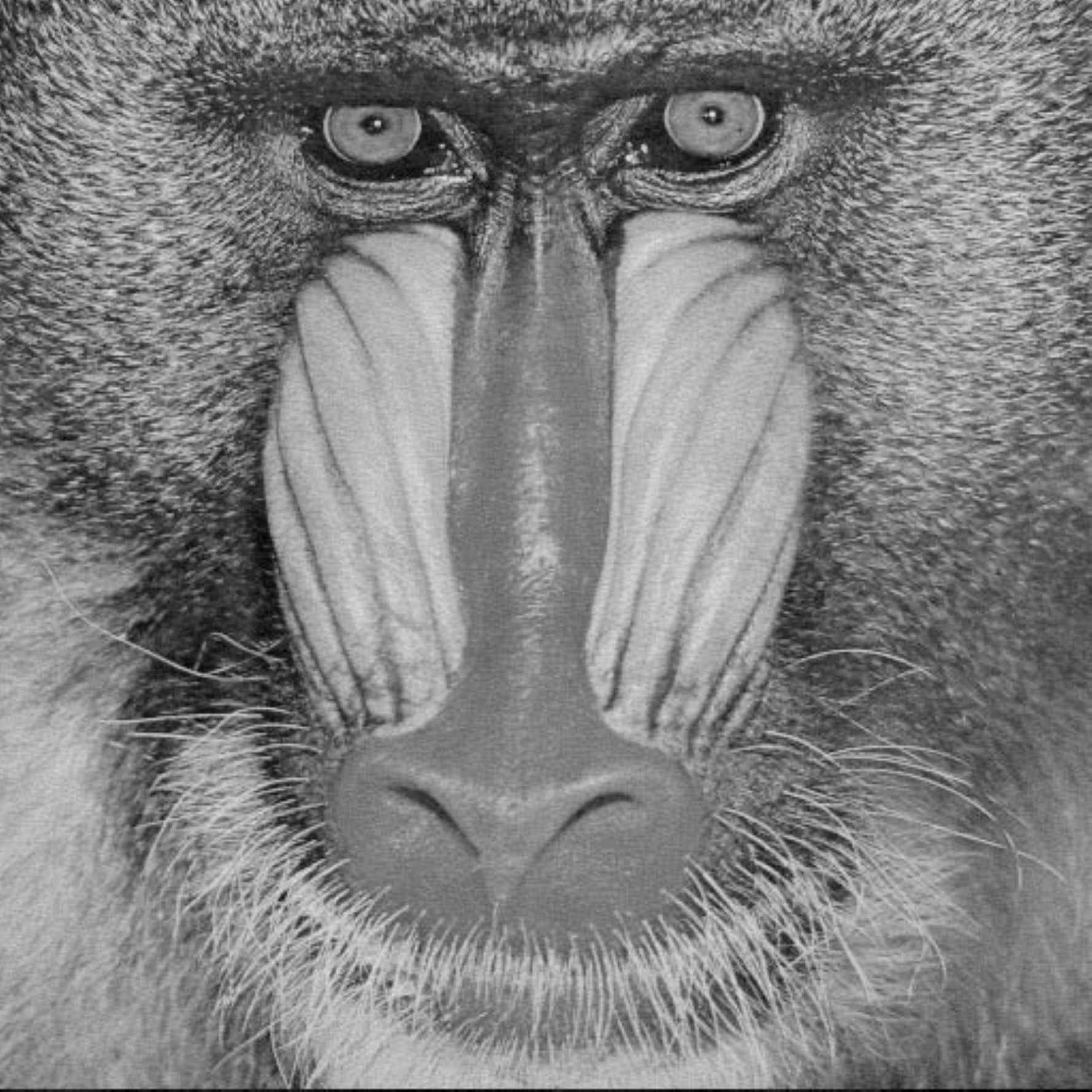}}
  \centerline{(i)$NUQ:(104,14,14,14)$}\medskip
\end{minipage}

 \caption{\emph{Baboon}: Comparison of the reconstructed images produced by the uniform ((b),(f)) and non-uniform quantizer ((c),(d),(e),(g),(h),(i)) at various quantization step sizes. The quantized values for each decomposition component is given in ($N_A$,$N_H$,$N_V$,$N_D$) format, where it represents the Approximation, Horizontal, Vertical, and Diagonal component, respectively.}

\end{figure}

In Fig. 6, MSSIM plots are shown after reconstructing the images from the quantized values of each component. To relate MSSIM values with the number of quantized values associated with the two quantizers, the number of quantized values in the Approximation component is marked as x-axis. It must be recalled that both the quantization schemes have same Approximation component. The number of quantized values for the non-uniform quantizer is shown in the figure legend having sequence in the order of the Horizontal, Vertical, and Diagonal components, while for the deadzone uniform quantizer, they are given at the top of their respective MSSIM value. For the test image \emph{Lenna}, MSSIM values for both the quantizers are very high and close to each other. Within the non-uniform quantizer, MSSIM values among different quantized values are either equal or near equal. Like \emph{Lenna}, the test image \emph{Pepper} has high MSSIM values for both the quantizers, but the number of quantized values in the non-uniform quantizer above 4 are either slightly more or less to MSSIM values of the deadzone uniform quantizer. This shows that the non-uniform quantizer with few quantized values can replicate the results of the deadzone uniform quantizer with many quantized values. 

However, for the test image \emph{Baboon}, MSSIM values are somewhat different for both the quantizers, the non-uniform quantizer performs better when the Approximation component's quantized values are less, whereas the deadzone uniform quantizer does better at more number of quantized Approximation values. This suggests that more number of non-uniform quantized values are needed to reach MSSIM level of the deadzone uniform quantizer. Still, they would be far less than the number of quantized values required by the deadzone uniform quantizer. 

From the above results, it can be seen that the non-uniform quantizer not only captures the information of original coefficients better than (in terms of MSE) or at par with (in terms of MSSIM) the deadzone uniform quantizer but also requires few quantized values to achieve it.

\subsection{Subjective Analysis}

In this section, the reconstructed images are displayed after the application of both the quantizers. Figs. 7-9 exhibit the original and reconstructed images of \emph{Lenna}, \emph{Pepper}, and \emph{Baboon}, respectively. As can be seen, at a few quantized values subfigs. (b-e), the reconstructed images from the non-uniform quantizer are much smoother and perceptually closer to the original image than the deadzone uniform quantizer. Images from the deadzone uniform quantizer are segmented, whereas segmentation in the non-uniformly quantized images is low on contrast. On the other hand, when the Approximation component has large number of quantized values (subfigs. (f-i)), the reconstructed images from both the quantizers are visually the same, in addition to being very close to the original image. Only difference is that the deadzone uniform quantizer uses far more quantized values compared to the non-uniform quantizer in the Detail components. Interestingly, MSSIM values of using the uniform quantizer at those quantized values are higher than the non-uniform quantizer, revealing limitations of objective measurement techniques for HVS. 

Two conclusions can be drawn from the above results: (1) the non-uniform quantizer results in better visual image quality at a particular number of quantized values, and (2) At a perceived image quality, less number of quantized values are required by the non-uniform quantizer.

\section{Discussion}

The non-uniform quantizer preserves the edge information and also allows intrinsic preferential weighting to the high value coefficients by having the quantization step sizes with varying length. It is important to minimize quantization error as the quantization step size approaches the end of the histogram plot of the Detail component to avoid structural deformation. In attempt to preserve the structural information, the deadzone uniform quantizer over preserves the coefficients lying towards zero that represent minimal edge information. This comes at a cost of large number of quantized values (i.e., small quantization step sizes) leading to high bitrates. At low bitrates, the deadzone uniform quantizer is constrained to have large quantization step sizes (see equation 3), distorting edge information. Both of these problems are due to the fixed step sizes. The non-uniform quantizer with variable step size combines the advantages of large and small quantization step sizes and also overcomes their limitations. 

The quantized values assigned in the non-uniform quantizer are the mean of all the coefficients present in their respective quantization step size. This minimizes the quantization error at every step size based on the real statistics of the Detail component for a given image. Unlike the deadzone uniform quantizer's deadzone region, the non-uniform quantizer does not predeterminely assign any value to zero. Embedded coding in the standard assigns equal number of bits to zero as it assigns to the other quantized values, and therefore, replacing zero with a quantized value would not affect the overall compression. 

The number of quantized values required by both the quantizers is in direct correlation with the skewness of each Detail component's histogram plot. Skewer the histogram plot, lesser would be the number of quantized values required. The adaptive approach used by the non-uniform quantizer results in less number of quantization step sizes that are required for the histogram plots with different skewness, compared to the deadzone uniform quantizer, which does not consider coefficients while choosing the step sizes. 

Experimental results show that the non-uniform quantizer produce visually improved lossy image compared to the deadzone uniform quantizer. Due to less number of quantized values in the Detail components and same number of quantized values in the Approximation component, the non-uniform quantizer would also achieve higher image compression. Therefore, its utilization in the standard would result in better image quality as well as higher compression in comparison to the deadzone uniform quantizer. Furthermore, it can easily be embedded in to the standard. Another advantage is that the non-uniform quantizer does not require inverse quantization at the decoder reducing time, resources, and computational complexity.

In case when bitrate is fixed, the use of the non-uniform quantizer would allow more bits to represent the Approximation component. In other words, the difference in bits used by the deadzone uniform and proposed non-uniform quantizer for the Detail components can be allocated to the Approximation component which uses the deadzone uniform quantizer in the proposed scheme. The Approximation component carry more information about the image than the Detail components because the Approximation component is not only the low resolution version of the original image but also represents the low pass frequency coefficients of the image which are of much more equitable significant compared to a Detail component's coefficients.

Apart from the fixed bitrate applications, bits saved from the non-uniform quantizer can be transferred to the Approximation component in the visually lossless compressed images. This would allow HVS algorithms more flexibility in selecting the visibility thresholds to obtain the appropriate quantization step sizes from the deadzone uniform quantizer in the Approximation component. In addition, HVS algorithms can be incorporated in obtaining boundaries for the variable quantization step sizes in the non-uniform quantizer, especially in determining $\kappa$ values in equations 4 and 5.

As the non-uniform quantizer is biased towards the coefficients at the ends of the histogram plot, it is incapable of obtaining the desired thresholds for effective quantization in the Approximation component. The non-uniform quantizer works well for the histogram plots of skewed distributions. In this paper, it is applied on the Detail components at decomposition level 1. However, it can be easily applied to the Detail components at the higher decomposition levels as long as they possess skewed histogram distributions of their respective coefficients.

\section{Future Work}

The future work would require further development of a comprehensive and robust theory for more quantitative analysis. One key challenge is to find the minimum or an optimal number of variable quantization step sizes required by the non-uniform quantizer for each Detail component to achieve a particular bitrate or visual quality. This would also determine the number of bits to be allocated to the Approximation component, eventually determining its quantization step sizes. Another important theoretical challenge would be to find the relationship among the quality factor, compression ratio, required number of step sizes, and impact of varying $\kappa$. Further, including current or developing HVS algorithms to adjust the boundaries of the step sizes in accordance with the human perception would be useful.

\section*{Acknowledgement}

The authors would like to thank Aadhar Jain and Parinita Nene of Cornell University for their useful comments.

\bibliography{refs}

\begin{thebibliography}{10}

\bibitem{Acharya_2005}
T.~Acharya and P.-S. Tsai.
\newblock {\em JPEG2000 Standard for Image Compression: Concepts, Algorithms
  and VLSI Architectures}.
\newblock Wiley-Interscience, 2004.

\bibitem{Skodras_2001}
A.~Skodras, C.~Christopoulos, and T.~Ebrahimi.
\newblock The {JPEG} 2000 still image compression standard.
\newblock {\em IEEE Signal processing Magazine}, 18:36--58, 2001.

\bibitem{Taubman_2002a}
D.S. Taubman and M.W. Marcellin.
\newblock {JPEG2000}: standard for interactive imaging.
\newblock {\em Proceedings of the IEEE}, 90(8):1336--1357, Aug 2002.

\bibitem{Rabbani_2002}
M.~Rabbani and R.~Joshi.
\newblock An overview of the {JPEG} 2000 still image compression standard.
\newblock {\em Signal Processing: Image Communication}, 17(1):3 -- 48, 2002.

\bibitem{Marcellin_2002}
M.W. Marcellin, M.A. Lepley, A.~Bilgin, T.J. Flohr, T.T. Chinen, and J.H.
  Kasner.
\newblock An overview of quantization in {JPEG} 2000.
\newblock {\em Signal Processing: Image Communication}, 17(1):73 -- 84, 2002.

\bibitem{Shannon_1959}
C.E. Shannon.
\newblock Coding theorems for a discrete source with a fidelity criterion.
\newblock In {\em IRE Nat. Conv. Rec., Pt. 4}, pages 142--163. 1959.

\bibitem{Berger_1998}
T.~Berger and J.D. Gibson.
\newblock Lossy source coding.
\newblock {\em IEEE Trans. Inform. Theory}, 44:2693--2723, 1998.

\bibitem{Watson_1993}
A.B. Watson, editor.
\newblock {\em Digital Images and Human Vision}.
\newblock MIT Press, Cambridge, MA, USA, 1993.

\bibitem{Wu_2005}
H.R. Wu and K.R. Rao.
\newblock {\em Digital Video Image Quality and Perceptual Coding (Signal
  Processing and Communications)}.
\newblock CRC Press, Inc., Boca Raton, FL, USA, 2005.

\bibitem{JPEG2000_TechRep}
Information technology -- {JPEG} 2000 image coding system -- part 1: Core
  coding system.
\newblock Technical report, ISO/IEC 15444-1:2000, 2000.

\bibitem{Li_1999}
J.~Li.
\newblock Visual progressive coding.
\newblock In {\em Proc. SPIE}, volume 3653, pages 1143--1154, 1998.

\bibitem{Zeng_2002}
W.~Zeng, S.~Daly, and S.~Lei.
\newblock An overview of the visual optimization tools in {JPEG} 2000.
\newblock {\em Signal Processing: Image Communication}, 17(1):85--104, 2002.

\bibitem{Nadenau_1999}
M.J. Nadenau and J.~Reichel.
\newblock Opponent color, human vision and wavelets for image compression.
\newblock In {\em In Proc. of the 7 tn Color Imaging Conference}, pages
  237--242, 1999.

\bibitem{Taubman_2002b}
D.S. Taubman and M.W. Marcellin.
\newblock {\em {JPEG} 2000: Image Compression Fundamentals, Standards and
  Practice}.
\newblock Kluwer Academic Publishers, Norwell, MA, USA, 2001.

\bibitem{Zeng_2000}
W.~Zeng, S.~Daly, and S.~Lei.
\newblock Point-wise extended visual masking for {JPEG}-2000 image compression.
\newblock In {\em IEEE International Conference on Image Processing}, volume~1,
  pages 657--660, 2000.

\bibitem{Watson_1997}
A.B. Watson, G.Y. Yang, J.A. Solomon, and J.~Villasenor.
\newblock Visibility of wavelet quantization noise.
\newblock {\em IEEE Transactions on Image Processing}, 6(8):1164--1175, Aug
  1997.

\bibitem{Liu_2006}
Z.~Liu, L.J. Karam, and A.B. Watson.
\newblock {JPEG2000} encoding with perceptual distortion control.
\newblock {\em IEEE Transactions on Image Processing}, 15(7):1763--1778, July
  2006.

\bibitem{Larabi_2009}
M.-C. Larabi, P.~Pellegrin, G.~Anciaux, F.-O. Devaux, O.~Tulet, B.~Macq, and
  C.~Fernandez.
\newblock {HVS}-based quantization steps for validation of digital cinema
  extended bitrates.
\newblock In {\em Proc. SPIE}, volume 7240, pages 72400V--72400V--9, 2009.

\bibitem{Ramos_2000}
M.G. Ramos and S.S Hemami.
\newblock Perceptual quantization for wavelet-based image coding.
\newblock In {\em IEEE International Conference on Image Processing}, volume~1,
  pages 645--648, 2000.

\bibitem{Ramos_2001}
M.G. Ramos and S.S. Hemami.
\newblock Suprathreshold wavelet coefficient quantization in complex stimuli:
  psychophysical evaluation and analysis.
\newblock {\em J. Opt. Soc. Am. A}, 18(10):2385--2397, Oct 2001.

\bibitem{Chandler_2005}
D.M. Chandler and S.S Hemami.
\newblock Dynamic contrast-based quantization for lossy wavelet image
  compression.
\newblock {\em IEEE Transactions on Image Processing}, 14(4):397--410, April
  2005.

\bibitem{Gaubatz_2005a}
M.D. Gaubatz, D.M. Chandler, and S.S. Hemami.
\newblock Spatial quantization via local texture masking.
\newblock In {\em Proc. Human Vision and Electronic Imaging}, 2005.

\bibitem{Gaubatz_2005b}
M.D. Gaubatz, D.M. Chandler, and S.S. Hemami.
\newblock Spatially-selective quantization and coding for wavelet-based image
  compression.
\newblock In {\em International Conference on Acoustics, Speech, and Signal
  Processing}, pages 209--212, 2005.

\bibitem{Gaubatz_2006}
M.~Gaubatz, S.~Kwan, B.~Chern, D.~Chandler, and S.S. Hemami.
\newblock Spatially-adaptive wavelet image compression via structural masking.
\newblock In {\em IEEE International Conference on Image Processing}, pages
  1897--1900, Oct 2006.

\bibitem{Albanesi_2003}
M.G. Albanesi and F.~Guerrini.
\newblock An {HVS}-based adaptive coder for perceptually lossy image
  compression.
\newblock {\em Pattern Recognition}, 36(4):997--1007, 2003.

\bibitem{Liu_2008}
K.-C. Liu and C.-H. Chou.
\newblock Locally adaptive perceptual compression for color images.
\newblock {\em IEICE Trans. Fundam. Electron. Commun. Comput. Sci.},
  E91-A(8):2213--2222, August 2008.

\bibitem{Sreelekha_2009}
G.~Sreelekha and P.S. Sathidevi.
\newblock A wavelet-based perceptual image coder incorporating a new model for
  compression of color images.
\newblock {\em International Journal of Wavelets, Multiresolution and
  Information Processing}, 07(05):675--692, 2009.

\bibitem{Sreelekha_2010}
G.~Sreelekha and P.S. Sathidevi.
\newblock An {HVS} based adaptive quantization scheme for the compression of
  color images.
\newblock {\em Digital Signal Processing}, 20(4):1129 -- 1149, 2010.

\bibitem{Wu_2006}
D.~Wu, D.M. Tan, M.~Baird, J.~DeCampo, C.~White, and H.R. Wu.
\newblock Perceptually lossless medical image coding.
\newblock {\em IEEE Transactions on Medical Imaging}, 25(3):335--344, March
  2006.

\bibitem{Wu_2010}
D.~Wu, D.M. Tan, and H.R. Wu.
\newblock Perceptual coding at the threshold level for the digital cinema
  system specification.
\newblock In {\em 2010 IEEE International Conference on Multimedia and Expo
  (ICME)}, pages 796--801, July 2010.

\bibitem{Oh_2013}
H.~Oh, A.~Bilgin, and M.W. Marcellin.
\newblock Visually lossless encoding for {JPEG2000}.
\newblock {\em IEEE Transactions on Image Processing}, 22(1):189--201, Jan
  2013.

\bibitem{Reichel_2001}
J.~Reichel, G.~Menegaz, M.J. Nadenau, and M.~Kunt.
\newblock Integer wavelet transform for embedded lossy to lossless image
  compression.
\newblock {\em IEEE Transactions on Image Processing}, 10(3):383--392, Mar
  2001.

\bibitem{Long_2002}
M.~Long, H.-M. Tai, and S.~Yang.
\newblock Quantisation step selection schemes in {JPEG2000}.
\newblock {\em Electronics Letters}, 38(12):547--549, Jun 2002.

\bibitem{Srivastava_2013b}
M.~Srivastava, S.K. Singh, and P.K. Panigrahi.
\newblock A semi-automated statistical algorithm for object separation.
\newblock {\em Circuits, Systems, and Signal Processing}, 32(6):3059--3078,
  2013.

\bibitem{Srivastava_2013a}
M.~Srivastava and P.K. Panigrahi.
\newblock Non-uniform quantization of detail components in wavelet transformed
  image for lossy {JPEG2000} compression.
\newblock In {\em ICPRAM}, pages 604--607, 2013.

\bibitem{Wang_2004}
Z.~Wang, A.C. Bovik, H.R. Sheikh, and E.P Simoncelli.
\newblock Image quality assessment: from error visibility to structural
  similarity.
\newblock {\em IEEE Transactions on Image Processing}, 13(4):600--612, April
  2004.

\bibitem{Gonzalez_2003}
R.C. Gonzalez, R.E. Woods, and S.L. Eddins.
\newblock {\em Digital Image Processing Using MATLAB}.
\newblock Prentice-Hall, Inc., Upper Saddle River, NJ, USA, 2003.

\bibitem{Wang_2009}
Z.~Wang and A.C. Bovik.
\newblock Mean squared error: Love it or leave it? a new look at signal
  fidelity measures.
\newblock {\em IEEE Signal Processing Magazine}, 26(1):98--117, Jan 2009.

\end{thebibliography}
\bibliographystyle{unsrt}

\end{document}